\RequirePackage{fix-cm}

\documentclass[natbib,smallextended,final]{svjour3}
\usepackage{color}
\usepackage{float}
\usepackage{graphicx,longtable}
\usepackage[rflt]{floatflt}
\usepackage{latexsym}
\usepackage{bm,array}
\usepackage{amsfonts}
\usepackage{amssymb}
\usepackage{amsmath}
\usepackage{mathptmx}

\newcommand{\rosat}{{\it ROSAT}}

\newcommand{\chandra}{{\it Chandra}}
\newcommand{\suzaku}{{\it Suzaku}}
\newcommand{\xmm}{{\it XMM-Newton}}
\newcommand{\ace}{{\it ACE}}
\newcommand{\wind}{{\it Wind}}
\newcommand{\soho}{{\it SOHO}}
\newcommand{\stereo}{{\it STEREO}}
\newcommand{\ulys}{{\it Ulysses}}

\newcommand{\oqkev}{$1\over4$~keV}

\newcommand\arcdeg{\mbox{$^\circ$}}% 
\newcommand\fdg{\mbox{$.\!\!^\circ$}}% 
\newcommand\sun{\odot}% 
\newcommand\er{R$_E$}%

\journalname{The Astronomy and Astrophysics Reviews}

\begin{document}
\title{Solar Wind Charge Exchange: An Astrophysical Nuisance}
\titlerunning{Solar Wind Charge Exchange}
\author{K. D. Kuntz}
\thanks{This work was supported by the XMM Guest Observer Facility at NASA GSFC.}

\institute{K. D. Kuntz \at
The Henry A. Rowland Department of Physics and Astronomy, \\
Johns Hopkins University, 3701 San Martin Drive, Baltimore, MD, 21218 \\
Tel.: +001-410-516-5628\\
\email{kkuntz1@jhu.edu}}
\date{Received: date / Accepted: date}
\maketitle

\begin{abstract}
Solar Wind Charge-Exchange (SWCX) emission is present in every X-ray observation of an astrophysical object. The emission is problematic when one cannot remove the foreground by the simultaneous measurement of a nearby field. SWCX emission is a serious impediment to the study of the diffuse hot ISM, including the Galactic halo, as its contribution to diagnostic emission lines is temporally variable. Modeling the SWCX emission, in order to remove it from our observations, has proven to be more difficult than originally anticipated. This work reviews our current understanding of SWCX emission, with special attention to all of the components required for future modeling tools. Since, in the absence of such a tool, observing programs can still be constructed to minimize the effect of SWCX, mitigation strategies are discussed. Although some aspects of SWCX will be very difficult to characterize, progress continues on many fronts.
\keywords{
X-rays: Diffuse Background}
\end{abstract}

\clearpage
\section{Motivations and Caveats \label{sec:motiv}}

Solar wind charge exchange (SWCX) was first clearly recognized to be a significant issue for the study of the hot Galactic ISM at the {\it Local Bubble and Beyond} conference in Garching bei M{\"u}nchen in 1997. In following years, the existence of the issue was well understood but its scope was not. It was clear neither whether anything could be done vis-{\'a}-vis astrophysical observations nor whether anything significant needed to be done. This uncertainty was due, in part, to the interdisciplinary nature of the problem, involving heliophysics\footnote{Disciplinary boundaries are usually poorly defined, and discipline labels are not always a good fit. ``Heliophysics'', as defined by NASA, is the study of the heliosphere, excluding planets.}, space physics, planetary physics, and laboratory astrophysics. Since then, the necessary interdisciplinary collaborations have grown; astrophysicists have sought heliophysicists for assistance, and heliophysicists have discovered SWCX to be a powerful new tool for exploring the Earth's magnetosheath. Both have turned to laboratory astrophysics for vital atomic data.

Although it was not recognized as such, SWCX was first detected in the \rosat\ All-Sky Survey (RASS) as a temporally variable signal, the Long-Term Enhancements (LTEs). These enhancements had time scales from hours to days. They were soon recognized as a local signal because the LTE rate during an observation of the Moon was consistent with the count rate towards the unilluminated portion of the Moon \citep{rosat_moon}. \citet{freyberg_1994} had noted the correlation of LTEs with geomagnetic storms and variations in the solar wind before recognizing the importance of the \rosat\ detection of Comet Hyakutake \citep{lisse_etal_1996}. Taking up the suggestion by \citet{cravens_1997} that X-rays could be produced by charge exchange between neutral atoms and high state ions in the solar wind, he suggested that the Earth is surrounded by a bright X-ray emitting region \citep{freyberg_1998}. \citet{cox_1998}, at the same meeting, suggested the neutral interstellar medium (ISM) that flows through the heliosphere could also interact with the solar wind, producing a relatively uniform X-ray emission. 

\begin{figure*}
\center{\includegraphics[width=3.75cm,angle=90]{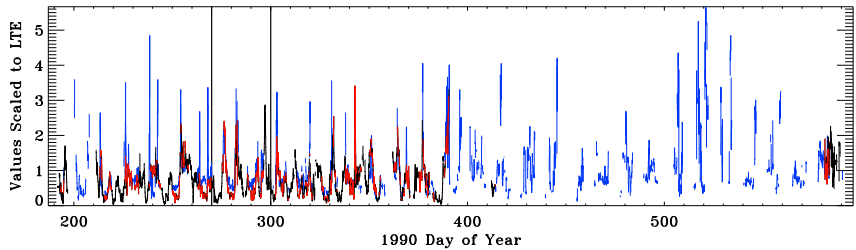}}
\caption{Comparison of the \rosat\ \oqkev\ Long-Term Enhancement (LTE) rates and the solar wind flux over the period of the \rosat\ All-Sky Survey. The solar wind flux as measured by {\it IMP-8}, shown in {\it blue} has been scaled to the LTE rate {\it red}. The {\it black} is the LTE rate when {\it IMP-8} data are not available. The two vertical lines denote the period originally studied by \citet{cravens_etal_2001}.}
\label{fig:c_lc}
\end{figure*}

We now recognize two regimes of SWCX based on the target neutrals. The {\it magnetospheric}\footnote{In this case ``magnetospheric'' is a  misnomer. The magnetosphere is the region containing the Earth's magnetic field. The solar wind does not enter this region, and thus the magnetosphere is free of SWCX emission. The magnetosheath is the region between the magnetopause (the boundary of the magnetosphere) and the bow shock, where the solar wind can interact with the neutral exosphere. However, ``magnetosheathic'' is a bit cumbersome, so common practice has been to use ``magnetospheric'' even though it is not correct.} or {\it geocoronal} emission is due to the solar wind interacting with the outer reaches of the Earth's atmosphere (the exosphere). Because the supersonic solar wind is nearly fully ionized and carries with it the interplanetary magnetic field, it cannot flow directly into the Earth's magnetosphere. Thus a shock exists upwind of the magnetopause that slows and deflects the solar wind. As the solar wind flux increases, the shock and the solar wind plasma move into denser parts the exosphere, meaning that the strength and location of this emission strongly depends upon the strength of the solar wind. The observed strength of the emission is also very sensitive to the location of the observer and the look direction. The {\it heliospheric} emission is due to the solar wind interacting with the neutral ISM everywhere within the heliopause. Local contributions from the heliospheric emission can be relatively strong and temporally variable. The contribution from the remainder of the line of sight should show only slow, low-amplitude variation because it is due to integration over $\sim$100 au and because the solar wind density decreases with distance from the Sun. The solar wind is strongly structured, which means that the heliospheric emission is direction dependent and, more importantly, the amplitude of the variation is also direction dependent.

There remains a certain amount of ambiguity as to which SWCX source dominates the emission contaminating particular astrophysical observations. The first unambiguous SWCX detection with \xmm\ has been modeled both as heliospheric emission \citep{koutroumpa_etal_2006} and as more local magnetospheric emission \citep{snowden_etal_2004}. There remain difficulties reconciling magnetospheric models with observations (see \S\S\ref{sec:hist_chan} and \ref{sec:comp_ms} for references). There are also difficulties with the heliospheric modeling of individual SWCX events, mostly because the solar wind data required for careful analysis often do not exist. Nevertheless, there have been several efforts to build models to provide some measure of the probability of strong SWCX emission for a given observation.

Such efforts have not yet borne fruit, in part because the systems being modeled are much more complex than was originally recognized by astrophysicists. To model the heliospheric emission one needs to model the flow of the neutral ISM through the heliosphere, a topic of active research, as well as the outward flow of the solar wind, another not entirely resolved issue that requires data outside the scope of any planned mission.  Modeling the magnetospheric emission requires three inputs: first, a good measure of the input solar wind properties (to some extent measured by upstream monitors); second, a model of the Earth's exosphere (for which we are currently relying on results from the 1990s that have only been partially validated); and third, magnetohydrodynamic (MHD) models of the interaction of the solar wind with the Earth's magnetic field (which carry a number of their own hotly contested controversies). Astrophysicists have often assumed that because SWCX is a local phenomenon, all of the required science would be well understood, but now find themselves collaborating with space physicists to elucidate our very local universe. 

This review organizes our current understanding of SWCX as a contaminating foreground component of astrophysical observations. A complementary review of SWCX as a means of studying the magnetopause, magnetosheath, and cusps can be found in \citet{sibeck_ssr_2018}. What is known is not synonymous with that which has been published. Although the bulk of this review covers results in the literature, there have been a number of unpublished results from ``back-of-the-envelope'' calculations. I include my own versions here, not because they are particularly the best, but because they are useful for understanding SWCX emission. Although there are many interesting results from charge-exchange observations of planets and comets, they are included here only when they have bearing on the problem of understanding SWCX emission as an astrophysical annoyance.

%ROADMAP

The first section sketches the difficulties that SWCX poses to X-ray astrophysics, as well as a short history of how we have arrived at the current impasse. The second section constructs the mathematical formalism required to model the SWCX emission as a way of introducing all of the relevant parameters. It then briefly describes the relevant issues of both the magnetospheric and heliospheric SWCX. The third section is a discussion of the solar wind, reviewing our understanding of the solar wind, the available data, and introducing the relevant models. The solar wind interaction with the Earth is a special case, with its own set of MHD models, which are discussed in \S\ref{sec:sw_ms}. The neutral atom distributions within the heliosphere and the exosphere are reviewed in \S\ref{sec:neut_hs} and \S\ref{sec:neut_ms} respectively, while the required atomic data are discussed in \S\ref{sec:atom}. Section~\ref{sec:play} covers some useful results from back-of-the-envelope type calculations while the following section reviews the confrontation between data and models. Although the models are promising, perplexing issues remain. Even if we may not be ready to model the SWCX emission, the results of \S\S\ref{sec:play} and \ref{sec:comp} can be used to devise strategies for mitigating the effects of SWCX emission on X-ray observations through the more carefully constructed observational strategies discussed in \S\ref{sec:miti}. 

%What is known is not synonymous with that which has been published. Although the bulk of this review covers results in the literature, there have been a number of unpublished results from ``back-of-the-envelope'' calculations that are useful for understanding SWCX emission. 

\subsection{The Chaos Created by SWCX \label{sec:contro}}

The SWCX emission contaminates astrophysical observations, modifying line ratios, changing the derived temperature of astrophysical plasmas and, at times, mimicking soft emission components. 

{\it Local Hot Bubble: } By far the most serious controversy caused by SWCX emission has been its implications for the existence of the Local Hot Bubble (LHB). Once it was suggested that heliospheric SWCX emission could contribute some of the X-ray emission attributed to the LHB, it became apparent that reducing its emissivity would reduce its pressure and potentially resolve the problem that the pressure in the LHB is four times greater than in the Local Interstellar Cloud. \citet{lallement_2004} made the first model of heliospheric SWCX emission in the \rosat\ All-Sky Survey. Since many cross sections were unknown, she could calculate the angular distribution but not the normalization. The largest possible normalization, 100\% of the observed emission in the faintest direction, removed nearly all of the LHB emission in the Galactic plane, but left a significant amount at the poles, sparking rumors of the LHB's demise. However, it is often overlooked that she proposed a less extreme scaling, leaving significant LHB emission in the Galactic plane, which would make the volume occupied by the LHB more consistent with the boundary of the Local Cavity. 

Further modeling of the heliospheric emission using the contemporaneous cross sections yielded mixed results. \citet{koutroumpa_lbb} found that nearly all of the $\frac{1}{4}$ keV LHB emission in the Galactic plane could be accounted for by SWCX, while \citet{robertson_lbb}, at the same conference, found that only half of the LHB emission in the Galactic plane could be due to SWCX. All models agreed that the bulk of the observed $\frac{3}{4}$ keV LHB emission was due to SWCX. This controversy has been partially resolved only recently by \citet{galeazzi_etal_2014} and \citet{snowden_etal_2014}, by measuring the broad-band SWCX emissivity in the helium focussing cone\footnote{The ``helium focussing cone'' is the region of the solar system with a higher density of neutral helium due to gravitational focussing by the sun of the neutral helium from the ISM (see \S\ref{sec:enlil}). It would be better to call it the ``focussed helium cone'', but the current usage is well established.}. There remain, however, lingering issues.
%\textcolor{red}{DS suggests actually talking about how this was done.}

{\it The Galactic Halo: } Isolating the emission due to the Galactic halo from the foreground emission is done with shadowing studies which observe a cloud with a known column density that absorbs more distant emission. With a sufficient signal-to-noise ratio and range of absorbing column densities, one can fit 
\begin{equation}
I_{total}=I_{local}+I_{distant} e^{-n_H \sigma_{eff}}
\end{equation}
where $n_H$ is the column density of hydrogen nucleons, and $\sigma_{eff}$ is the effective cross section per hydrogen nucleon. For broad bands, such as the \rosat\ R12 band, it should be noted that $\sigma_{eff}$ is a function of both the absorbing column density {\it and} the spectral shape of the absorbed emission \citep[see][for a demonstration]{kuntz_snowden_2000}. Fitting this equation produces the emission strength of both the local and the distant components and, if the distance to the absorbing cloud is known, a lower limit to the distance to the more distant emission, and an upper limit to the path length through the foreground emission. In the \rosat\ era, shadowing experiments were rather straight-forward since the absorbed region and the unabsorbed region were generally in the same field of view (FOV). Then, even if there were a time-variable foreground, the measurement of the distant component is still secure.

In the \chandra /\xmm\ era, shadowing studies became more difficult. First, neither of those observatories can observe below $\sim$0.35 keV, so they can observe only those shadows produced by relatively high column densities. Second, the FOV is sufficiently small that it is nearly impossible to observe the ``on-cloud'' and ``off-cloud'' regions simultaneously. The danger of not observing them simultaneously was shown by \citet{henley_shelton_2008}, who compared observations of an absorbing cloud taken with \suzaku\ with earlier observations of the same directions taken with \xmm\ \citep{henley_etal_2007}. The solar wind flux during the \xmm\ observations was a relatively steady $1.76\times10^8$ cm$^{-2}$ s$^{-1}$ (35th percentile). The observations did have a substantial pathlength through the magnetosheath, but only through the flanks where the emission is relatively low. Further, the two observations (just 2 degrees apart) were executed over a span of 16 hours, with a 5 hour gap between them. Conversely, for the Suzaku observations, the solar wind flux was stronger and more variable, and the line of sight through the flanks of the magnetosheath was fairly short. However, \citet{henley_shelton_2008} found stronger low energy emission in the \xmm\ observations that were missing from the \suzaku\ observations. They attributed the difference to SWCX, and showed that the \suzaku\ observations derived a much higher halo emission measure as a result.

Other shadowing targets have had similar issues. MBM12, since it was key in setting the size of the Local Hot Bubble (LHB), has had a long history of observations by multiple groups and instruments \citep[see][and references therein]{koutroumpa_etal_2011,smith_etal_2007}; the SWCX contributions have been highly variable. Here, at least, there have been a sufficient number of observations with sufficiently well-behaved solar wind that by plotting the measured values of the O VII and O VIII emission against the model predictions, one can fit a straight line to determine the value of the line emission when there is zero SWCX contribution \citep{koutroumpa_etal_2011}. Studies that include only one on-cloud/off-cloud pair, while sometimes producing interesting, suggestive, or even disturbing results, are always subject to doubt given the possibility (or even probability) that one or the other of the observations were affected by heliospheric SWCX which may not be reflected in upstream monitor data. Extended, multi-year observation programs, such as Galeazzi's as yet unpublished large Suzaku program, provide a much more secure measure as well as an estimate of what the uncertainty due to SWCX actually is.

{\it Extended Diffuse Emission: } Diffuse emission that nearly fills the FOV is particularly problematic. A single observation will have an undetermined amount of SWCX. For the soft emission due to the hot ISM in the Galaxy, or the soft emission on the outskirts of large galaxy clusters, this can be disastrous. The classic example is the detection of WHIM emission in only one of several fields on the outskirts of the Coma cluster \citep{finoguenov_etal_2003} which was later shown to be due to SWCX emission \citep{takei_etal_2008}. Of course, these results were published well before SWCX emission had been noticed in either \xmm\ or \chandra .

SWCX emission is a time-variable foreground. Currently, its strength, for a given target, can be estimated only from multiple observations of the same target. If there is a sufficient number of observations with sufficiently well-behaved solar wind conditions (and we will see what that means in a later section) then one might be able to extrapolate to SWCX-less conditions. Modeling the SWCX emission for an arbitrary time and look direction is the ideal, but as this review will show, that goal is still far in the future.

\subsection{A Short History of SWCX}

\subsubsection{The \rosat\ Era}

\begin{figure}
\center{\includegraphics[width=8cm,angle=0]{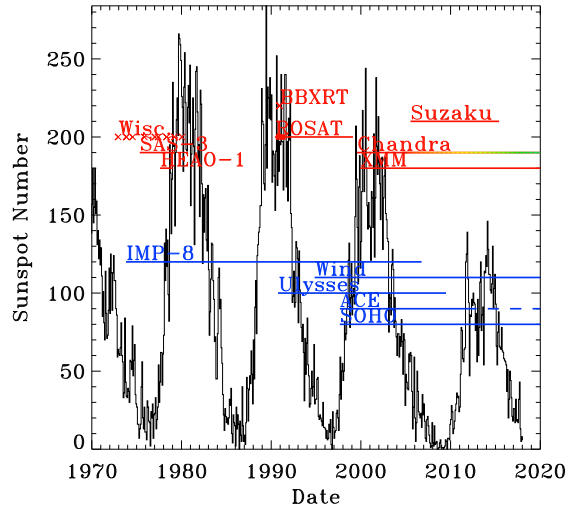}}
\caption{Sunspot number in the X-ray era. The horizontal lines indicate the duration of the various missions. The {\it red} bars show X-ray astrophysical missions while the {\it blue} bars are solar wind missions.The wide part of the \rosat\ line indicates the time over which the All-Sky Survey was executed. The dashed part of the \ace\ line indicates the time during which it was no longer producing extensive ion state data. The change of color of the \chandra\ line is to remind the reader that the soft response faded after launch. The Wisconsin survey was composed of individual launches, as shown. The sunspot data is from the SILSO dataset, Royal Observatory of Belgium, Brussels.}
\label{fig:sunspot}
\end{figure}

As the opening description of the advent of charge exchange in astrophysics implied, SWCX has required a multi-disciplinary approach. The observation of LTEs would probably have remained a curiosity had \rosat\ not observed Comet Hyakutake\footnote{Incidently, there are $\sim$6 comets detected in the \rosat\ All-Sky Survey, so it is likely that someone would have realized that comets emit X-rays had Lisse not done so spectacularly, but it probably would have taken much longer!}. Hyakutake was observed by Lisse for reasons far removed from charge exchange. Hyakutake was an extremely dusty comet that would pass very close to the Earth. He remembered a paper that argued that dust released by comets on highly non-Keplerian orbits should impact zodiacal cloud dust at 10 to 100 km s$^{-1}$, vaporizing in the process, and creating glowing plasma clouds that would emit X-rays \citep{ibadov_1990}. Lisse proposed to observe this effect, or at least put upper bounds on it (Lisse, private communication). Instead, Hyakutake was dazzlingly bright in the X-ray band \citep{lisse_etal_1996}.

After \citet{cravens_1997} had identified solar wind charge exchange as the likely mechanism to explain cometary X-ray emission, that explanation was quickly applied to LTEs \citep[i.e.,][]{freyberg_1998}. Cravens, primarily a space scientist, found the problem sufficiently interesting that he crossed disciplinary boundaries to work with Snowden to correlate the \rosat\ $\frac{1}{4}$ keV LTE rate with the solar wind flux (see Figure~\ref{fig:c_lc}). They argued that the tight correlation between the LTE rate and the solar wind flux measured at the Earth indicated a local, geocoronal source of the emission. The heliospheric SWCX emission is due to emission between the observer and the heliopause (at $>120$ au) and thus should show variability only at much longer time scales. This analysis was revisited by \citet{kuntz_etal_2015}. They repeated the $\frac{1}{4}$ keV band analysis with a much larger data set and showed that that the $\frac{1}{4}$ keV LTE rate was closely correlated with the local solar wind flux (Figure~\ref{fig:c_lc}). The $\frac{3}{4}$ keV LTE rate, however, was not.

It is perhaps fortuitous that \rosat\ was launched during solar maximum, when the variation in the solar wind is at its greatest. The \rosat\ All-Sky Survey, done in the first six months of the mission was ideally scheduled to maximize the probability of detecting the variation in the SWCX emission. Incidentally, both \xmm\ and \chandra\ were also launched during solar maximum, which increased their chances to detect SWCX emission as well (See Figure~\ref{fig:sunspot}). 

\begin{figure*}
\center{\includegraphics[width=12.5cm,angle=0]{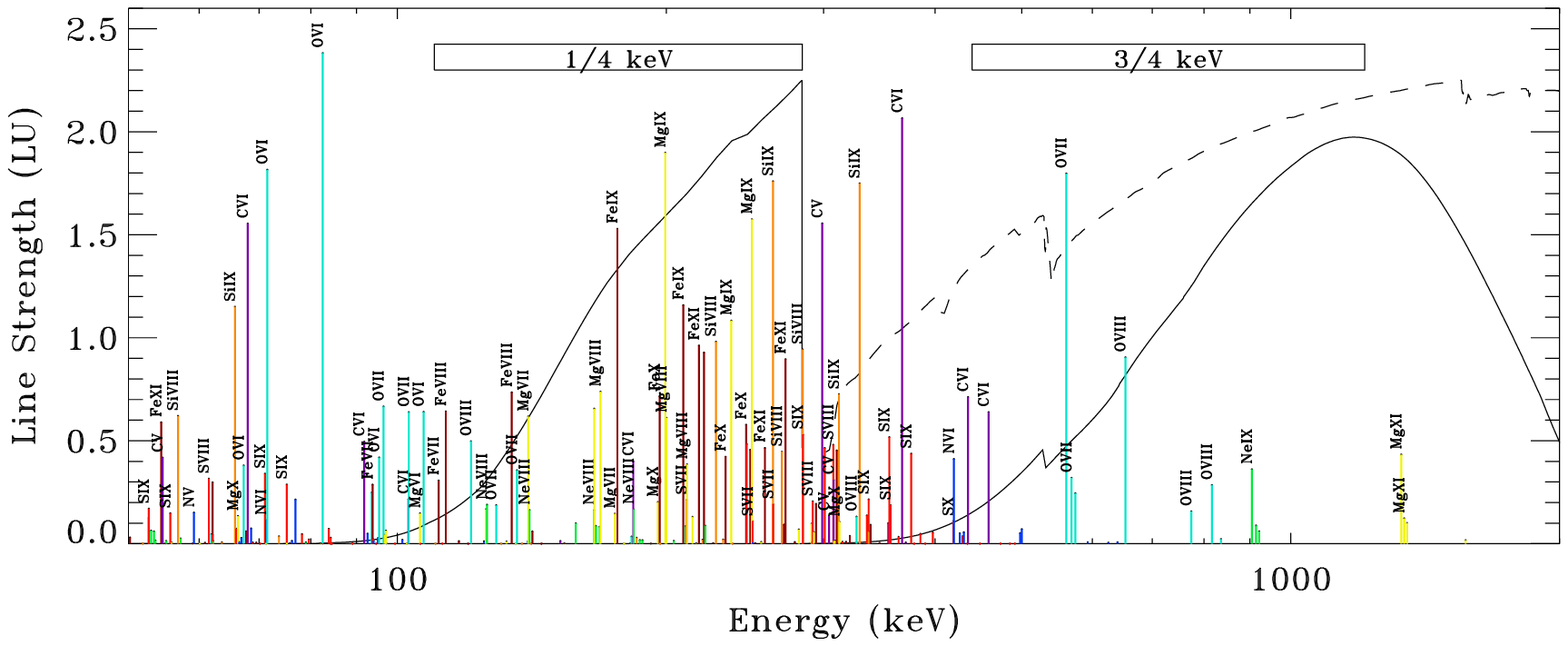}}
\vspace*{-0.5cm}
\caption{The spectrum of SWCX emission calculated for a slow solar wind from \citet{koutroumpa_etal_2009}. This spectrum was compiled from various theoretical calculations of cross sections and using carefully selected solar wind abundances. Without a doubt some of the higher $n$ lines have incorrect ratios and it misses some of the less important species. However, it demonstrates a number of important points about the SWCX emission. The bulk of the emission is in the $\frac{1}{4}$ keV band and is composed of many lines of many different species and ionization states. The $\frac{3}{4}$ keV band is dominated by just a few lines. Reduction of the ionization state of the solar wind would shift the location of the dominant lines within the $\frac{1}{4}$ keV band, but remove a significant amount of flux from the $\frac{3}{4}$ keV band.  The smooth curves show the response of \rosat\ (solid) and the \xmm\ MOS (dashed) after arbitrary normalizations.}
\label{fig:dimitra_spec}
\end{figure*}

Figure~\ref{fig:dimitra_spec} shows a SWCX spectrum. The density of lines below 0.4 keV is much higher than above that energy. The SWCX spectrum in the \rosat\ $\frac{1}{4}$ keV band is composed of many lines from many different species and ionization states. The SWCX spectrum in the \rosat\ $\frac{3}{4}$ keV band contains only a few lines. The strongest are those of O VII and O VIII\footnote{Since the study of SWCX falls at the intersection of multiple fields of study, it is very important to note confusing and conflicting conventions. For example, what an astrophysicist would refer to as an O VIII line (``oxygen eight''), a space physicist would refer to as O$^{+7}$ (``oxygen plus seven'' or sometimes just ``oxygen seven'') line. In this context, of course, the line is due to the recombination of O$^{+8}$ (``oxygen eight'') which can lead to a new level of confusion. This work will use the astrophysical convention for identifying lines, but will use the space physics convention when referring to the ions themselves.}, with occasional strong contributions from Ne IX and Mg XI. \citet{kuntz_etal_2015} attributed the strong correlation between the \rosat\ $\frac{1}{4}$ keV LTE and the solar wind (proton) flux to this plethora of different species. The argument is that {\it random} variation in abundances average out over multiple species. Further, the $\frac{1}{4}$ keV band contains multiple ionization states of the same species, so that a change in the mean ionization state changes which lines are strong, but does not change the overall emission as much. The $\frac{3}{4}$ keV band contains only a few lines, so a change in the ionization state, for example from O VII to O VI, would shift the emission out of the band, rather than redistributing it within the band. Thus \citet{cravens_etal_2001} were lucky that the band with the highest signal-to-noise ratio was also the band that averaged over many lines, otherwise the connection between LTEs and the solar wind flux might not have been convincing.
 
 \subsubsection{\chandra\ and \xmm\  \label{sec:hist_chan}}

Unfortunately, SWCX was discovered very late in the \rosat\ mission, so there was little opportunity to make follow-up observations. \xmm\ and \chandra , when launched, had a lower energy limit of $\sim$0.35 keV, and so miss the bulk of the SWCX emission. The first definitive detection of SWCX emission in the post-\rosat\ era was a detection of O VII, O VIII, and other lines towards the dark portion of the moon \citep{wargelin_etal_2004}. \chandra\ observed the moon through all or part of the magnetosheath, looking through the day side flanks of the magnetosheath. Calculated fluxes and observed fluxes agreed remarkably well. 

Shortly thereafter, \citet{snowden_etal_2004} discussed a series of four \xmm\ observations of the Hubble Deep Field North. Three of the observations, and the last quarter of the fourth, had identical spectra. The first part of the fourth showed strongly enhanced lines of Mg XI, Ne IX, O VIII, O VII, and C VI\footnote{These four observations were being used to test background subtraction software. The post-doc who was working on the problem assumed that there was something wrong his software for several weeks.}. \xmm\ had a special observing geometry during this observation; it was outside the bow shock, looking tangentially through the subsolar nose of the magnetosheath where the emission is expected to be the brightest. However, neither the period of the solar wind flux enhancement nor the period of enhanced O$^{+7}$/O$^{+6}$ matched the period of line enhancement as the line enhancement {\it preceded} the solar wind enhancements. Although the light curve was not consistent with a magnetospheric origin, it was not clear that it was consistent with a heliospheric origin due to the sharpness with which the enhancement ended. Since the heliospheric emission is integrated to the heliopause, one does not expect rapid changes in the heliospheric strength. \citet{collier_etal_2005} showed, from a correlation analysis using the \ace\ (Advanced Composition Explorer) and \wind\ data, that the enhancement in the solar wind was in the form of a tilted front, which would have moved into the line of sight before reaching \ace , making the observation consistent with a heliospheric origin. This event was further modeled as a heliospheric event by \citet{koutroumpa_etal_2006}.

Since the SWCX emission observed by \rosat\ and by \chandra\ were clearly magnetospheric, the \xmm\  observation seemed rather anomalous. The next several studies assumed that the bulk of SWCX events would be magnetospheric. This was in part due to the eye-catching modeling of the magnetospheric emission done by \citet{robertson_etal_2006} and the general impression that heliospheric emission varied on much longer time scales. 

\citet{kuntz_snowden_2008} pursued the strategy begun by \citet{snowden_etal_2004}, looking for targets with multiple observations. They found six sets besides the Hubble Deep Field North and found several SWCX events. Several lines of sight that seemed as if they should have passed through the nose of the magnetosheath and produced strong SWCX events, showed no sign of such, while other lines of sight that passed through the flanks showed very strong events. Further, there seemed to be little correlation with the solar wind flux. The former problem was attributed to an inadequate magnetosheath model. The authors had used a static model of the magnetosheath \citep{spreiter_etal_1966} scaled to the current solar wind flux using the relation given in that work:
\begin{equation}
R_{MP} = H^{1/3}(2\pi \rho_{\infty}v_{\infty}^2)^{-1/6}
\label{eqn:spreiter_mp}
\end{equation}
and
\begin{equation}
\Delta R_{MP} = 1.1 R_{MP} \frac{(\gamma+1)M_{\infty}^2}{(\gamma-1)M_{\infty}^2 +2}
\label{eqn:spreiter_delt}
\end{equation}
where $R_{MP}$ is the magnetopause stand-off distance, $\Delta R_{MP}$ is the distance between the magnetopause and the bow shock, both in units of R$_E$\footnote{For magnetospheric studies, the convenient length unit is the terrestrial radius, R$_E$. Although that quantity is rather ambiguous, it is generally taken to be 6371 km. For scale, note that the Moon's orbit has a semi-major axis of 60.4 R$_E$, while a geosynchronous orbit has a radius of 6.6 R$_E$.}, $H=0.312$ Gauss, and $\rho$, $v$, and $M$ are the solar wind mass density, velocity, and Mach number before it interacts with the Earth. It was assumed that the calculated magnetopause stand-off was incorrect and that the narrow \xmm\ beam had missed the magnetosheath. The latter problem was attributed to structures in the solar wind that did not pass near enough to the Earth to be noted by \ace\ but were still quite local. 

\citet{carter_etal_2011} and \citet{carter_sembay_2008} searched the \xmm\ archive for SWCX events by plotting the 0.5-0.7 keV flux (the SWCX band for \xmm ) against the 2.5-5.0 keV flux (a non-SWCX band) for 1 ks bins for each observation. Because the ``soft proton flares'' can produce variation in the SWCX band, comparison of the two bands allows the identification of SWCX-only variations. Although such analysis detects SWCX only in longer observations, they found 103 events. They determined that 60\% of SWCX events occurred when \xmm\ was on the day side of the Earth and, of those, most occurred when \xmm\ was near apogee. However, they found only a weak correlation between solar wind flux and SWCX emission. Comparison to simple models of the magnetosheath using scaled \citet{spreiter_etal_1966} models and \ace\ measured solar wind fluxes showed rather poor agreement. In retrospect, their results suggest that heliospheric events play a significant role.

\citet{henley_shelton_2010} and \citet{henley_shelton_2012} used the \xmm\ archive to measure O VII and O VIII over the entire sky. They found 69 (later 217) sets of multiple observations of (nearly) the same target. For each set they compared the relative solar wind fluxes with the relative X-ray line fluxes and found only a poor correlation. Thus, they concluded, the observed SWCX emission must be due to both magnetospheric and heliospheric components. 

It was expected that using better models for the magnetosheath would produce better results. \citet{wargelin_etal_2014} used BATS-R-US MHD models of the magnetosheath to model SWCX emission for a dozen strong solar wind events when \chandra\ could be reasonably expected to observe the resulting SWCX emission from the magnetosphere. That is, the events were chosen to reduce the magnetospheric/heliospheric ambiguity. Although the modeling was primarily for lines with well measured cross sections and concurrent solar wind abundances, the overall agreement was mediocre. Some events were well modeled, others were not. The disagreements sometimes appeared in light curve shape, and sometimes in emission strength. The authors noted uncertainty in the location of the magnetopause as a potential cause of the disagreements. This note has become a common refrain in more recent works. 

Since heliospheric variations are slower, they require longer baselines for modeling. Repeated observations of blank fields are not common, and even fewer have the observing geometries or cadence required for such a study. There have been a few such studies whose results are beginning to be published. 

\subsubsection{\suzaku\ }

\suzaku , being in low Earth orbit (LEO) like \rosat , has a different view of the magnetospheric emission. Roughly 3\% of \suzaku\ observations have detectable SWCX emission \citep{ishi_etal_2017}. Some events, such as those studied by \citet{ezoe_etal_2010} or \citet{ezoe_etal_2011} occurred in the flanks of the magnetosheath. Analyses generally assumed a magnetospheric origin and, given the observing geometry, special effort was made to demonstrate that the excess emission was indeed charge exchange, due to correlation with and lag from solar wind features, rather than scattered solar X-rays. Calculations using extremely simple models and a variety of cross sections usually found larger than expected emission strengths. Revisiting these observations with updated models and cross sections would be fruitful.

\suzaku's\ potential strength, however, is its observation of the cusps, the regions near the poles of the Earth's magnetic field where solar wind plasma can travel deep into the Earth's atmosphere (see  Figure~\ref{fig:ms_demo}). The cusps are impossible to observe with high Earth orbit (HEO) missions due to Earth avoidance angles. \citet{fujimoto_etal_2007} is the only published cusp observation, though given the \suzaku\ observing geometry, there should be more such observations. Time variations in observations through the cusp were attributed to changes in the distance from the last closed field line. Due to the observing geometry, this is equivalent to the change in the path length through the cusp. The cusp geometry is complex and there are currently no good models for the X-ray emission in the cusps. Thus the \citet{fujimoto_etal_2007} observation revealed the potential for such observations.

\suzaku's\ spectral resolution has allowed diagnosis of the elemental composition in Interplanetary Coronal Mass Ejections (ICME). Following the work of \citet{carter_etal_2010} with \xmm , \citet{ezoe_etal_2011} analyzed an ICME that fortuitously passed through the \suzaku\ FOV. Although the analysis primarily confirmed through X-ray means what had been known through {\it in situ} measurements, their work demonstrates the utility of X-ray observations for remote sensing of the conditions in the solar wind.

\subsubsection{Curbing the Chaos}

Were it possible to model SWCX emission for an arbitrary time and look direction, the history of SWCX research would not be over. Remote sensing of the solar wind and direct imaging of the magnetosheath are exciting possibilities for space physicists \citep{sibeck_ssr_2018, walsh_etal_2016}. As we will see in the following sections, modeling is still far from successful. However, continued interest in SWCX from space physicists, though even for their own ends, is key to solving the problem. Our uncooperative foreground emission is their signal, while our treasured cosmic X-ray background is their unfortunate background. For those of us who now have commitments in both worlds, the future is exciting. Now let us survey the boundaries between the known, the incompletely understood, and the completely unknown.
%\clearpage

\section{Introduction to the Problem \label{sec:intro}}

Charge-exchange occurs when an ion interacts with a neutral atom (or molecule) and an electron is transferred from the neutral atom to the ion. That is:
\begin{equation}
A^{q+}+B \rightarrow A^{*+q-1}+B^{+1}
\end{equation}
and then
\begin{equation}
A^{*+q-1} \rightarrow A^{+q-1} + \nu
\end{equation}
After the electron transfer, the ion formerly of charge q is in a highly excited state before experiencing a radiative decay. Thus, charge exchange is similar to a purely recombining plasma, but without an easily characterized distribution of electron energies. It is true that multiple-electron transfer can occur, but the cross section for that transfer is usually lower than for a single electron transfer \citep{krasnopolsky_etal_2004}. However, the probability of multi-electron transfer may not be negligible at low collision energies. Conversely, the principal neutral target for magnetospheric SWCX is hydrogen while the principal neutral target for SWCX from the inner heliosphere is helium, so while multiple electron transfer is important for cometary or planetary study, where the neutral targets have multiple electrons, it is not important for the SWCX contaminating astrophysical observations. Similarly, ion-ion interactions can occur, but again cross sections are low and, due to the low density of ions in the solar wind, the probability of an ion-ion interaction is low.

For a line of sight, the observed flux in a transition $j$ due to charge exchange between a neutral of species $k$ and a solar wind ion of species $s$ and charge state $q$ is given by
\begin{equation}
F_j = \int_{0}^{\infty} n_k n_{sq} v_{rel}~ \sigma_{sqk}(v_{rel})b_{sqj}~d\Omega dl/4\pi, \label{eqn:first}
\end{equation}
where the integral is along the line of sight from the observer.  The $n_k$ and $n_{sq}$ are the densities of the neutral targets and solar wind ions respectively. The value of $v_{rel}$, the relative velocity of the ion and the neutral, is given by
\begin{equation}
v_{rel}\sim(v_{r}^2+v_{t}^2)^{\frac{1}{2}}
\end{equation}
where the $v_{r}$ is the bulk velocity of the ions with respect to the neutrals, while $v_{t}$ is the thermal velocity of the ions, generally $\sqrt{(3kT/m_p)}$. In the free-flowing solar wind the ions have the same thermal velocities as the protons and $v_{t}\sim 0.1 v_{r}$, but in the magnetosheath where the solar wind enters a classic shock, the $v_{t}$ is larger than the $v_{r}$. The thermal and bulk velocities of the neutrals is usually negligible in comparison. 
% Do the ACE plot here?
The $\sigma_{sqk}(v_{rel})$ and $b_{sqj}$ contain the atomic data; the cross section for the interaction (as a function of $v_{rel}$) and the branching ratio, which is the probability of the emission of a photon in transition $j$ once the charge exchange has occurred. The remainder of the terms in Equation~\ref{eqn:first} are the geometric factors from the integral. In this form, $F_j$ would have units of photons cm$^{-2}$ s$^{-1}$ sr$^{-1}$. 

It might seem otiose to note that
\begin{equation}
n_{sq} = n_p \frac{n_s}{n_p} \frac{n_q}{n_s}
\end{equation}
but this substitution allows a useful simplification.

For a given bandpass
%\begin{equation}
%F= \sum_k \sum_s \sum_q \sum_j  \int_{0}^{\infty} n_k n_{sq} v_{rel} \sigma_{sqk}(v_{rel}) b_{sqj}~d\Omega dl/4\pi
%\end{equation}
\begin{equation}
F= \sum_k \sum_s \sum_q \sum_j  \int_{0}^{\infty} n_k n_p v_{rel}~\frac{n_s}{n_p} \frac{n_q}{n_s}\sigma_{sqk}(v_{rel}) b_{sqj}~d\Omega dl/4\pi
\end{equation}
where the sum is over all of the transitions falling within the band, and thus over all the appropriate ion species and charge states. 

In many cases this expression can be simplified. There are often only one or two types of neutral targets. In the case of magnetospheric emission, $n_{sq}/n_p$ is likely to be constant over the relevant pathlength. The dependence of $\sigma_{sqk}$ on $v_{rel}$ may be small. In such a case
\begin{equation}
F_k = Q_k \varsigma_k
\end{equation}
where
\begin{equation}
Q_k \equiv \int_{0}^{\infty} n_k n_p v_{rel} ~d\Omega dl/4\pi
\end{equation}
and
\begin{equation}
\varsigma_k\equiv \sum_s \sum_q \sum_j \left[ \frac{n_s}{n_p} \frac{n_q}{n_s}\langle\sigma_{sqk}\rangle b_{sqj} \right]
\end{equation}
This formulation has the advantage that it isolates what are thought to be well understood quantities in $Q$ from the poorly known quantities in $\varsigma$. This formulation is also useful in an observational context; under certain conditions, with a sufficiently broad bandpass, a time-averaged $\varsigma_k$ can be directly measured (as will be described in \S\ref{sec:bb_prod}).
 
In the heliophysics context it is often preferred to deal with the energy flux instead:
\begin{equation}
\mathcal{F}_j =h \int_{0}^{\infty} n_k n_{sq} v_{rel}~ \sigma_{sqk}(v_{rel})b_{sqj}\nu~d\Omega dl/4\pi, 
\end{equation}
which leads to 
\begin{equation}
\mathcal{F}_k = Q_k \alpha_k
\end{equation}
where
\begin{equation}
\alpha_k\equiv  h \sum_s \sum_q \sum_j \left[ \frac{n_s}{n_p} \frac{n_q}{n_s}\langle\sigma_{sqk}\rangle b_{sqj} \nu \right]
\end{equation}
however, there is some variation in the energy units used, so $\alpha$ is often ambiguous.

Although the lack of measured cross sections plagues the modeling of both the magnetospheric and heliospheric emission, the two emission regimes pose rather different modeling problems. The following two subsections sketch the issues for these two regimes so that the more detailed discussion of our understanding of the solar wind ions and their neutral targets that follows can be more readily placed in context.

\subsection{Magnetospheric Issues}

\begin{figure}
\center{\includegraphics[width=8.5cm,angle=0]{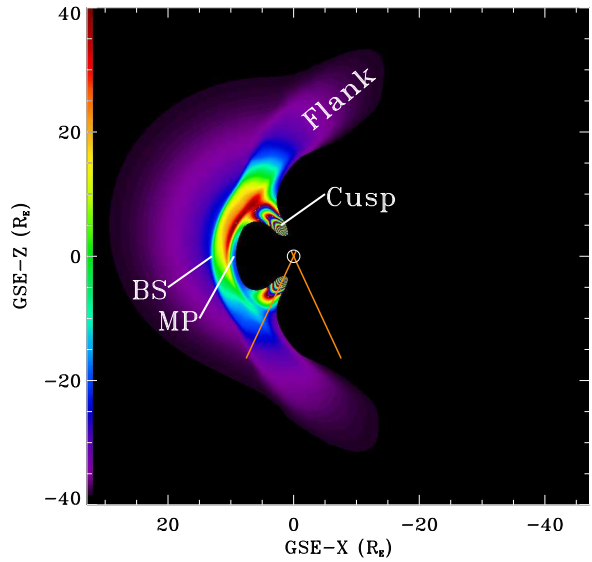}}
\caption{The X-ray emissivity in the near-Earth environment on the XZ$_{GSE}$ plane, which is defined by the Sun (to the left), the Earth, and the ecliptic north pole. (This is the midnight-noon plane, for those who are more used to thinking in terms of local time.) This plot was made for the Earth's equinox for the median solar wind pressure. The Earth is the white circle. The locations of the magnetopause (MP) and the bow shock (BS) at the nose of the magnetosheath are marked, as is the northern cusp. The color scale wraps multiple times in the cusp. Strong emission is expected in the cusp, but a MHD model has difficulties modelling the important kinematic effects in this region. The projection onto the XZ$_{GSE}$ plane of the most extreme \xmm\ orbit is shown, as well as the orbit six months later (red lines). The \chandra\ orbit is similar but with its apogee to the North. The \rosat\ and \suzaku\ orbits would be difficult to distinguish from the Earth symbol in this plot. This simulation was created from a BATS-R-US model run by the NASA CCMC.}
\label{fig:ms_demo}
\end{figure} 

The magnetospheric regime may be more readily modeled than the heliospheric regime. The solar wind is almost continuously measured at L1 ($\sim$235 R$_E$ upstream of the Earth), so we know the proton density, speed, and temperature in the solar wind, as well as the strength and direction of the interplanetary magnetic field (IMF). The \ace\ satellite originally provided ion density information for a number of species, but now provides ion ratios for only a few species. The solar wind data are the necessary input for magneto-hydrodynamic (MHD) models of the magnetosheath, which determine the structure and location of the shock, and thus the $n_p$ and $v_{rel}$ as a function of location. The neutral density, $n_n$, is given by a combination of models and extrapolations. When expanded into spherical harmonics, the neutral density distribution has rather low frequency deviations from a spherical $r^{-3}$ relation. The ion distribution, however, is very strongly structured. 

Figure~\ref{fig:ms_demo} shows a XZ$_{GSE}$\footnote{The Geocentric Solar Ecliptic (GSE) coordinate system is a right-handed coordinate system centered on the Earth that has as its +X axis the vector from the Earth to the Sun and as its +Z axis the direction to the north ecliptic pole. The +Y axis lies in the plane of the ecliptic in the direction opposite to the Earth's motion. The GSE coordinate system is like the ecliptic coordinate system, except the zero point is the Sun rather than the vernal equinox. The Geocentric Solar Magnetospheric (GSM) coordinate system is a right-handed coordinate system entered on the Earth that has as its +X axis the vector from the Earth to the Sun. The +Z axis is the projection of the Earth's magnetic pole on a plane perpendicular to the +X axis. The X$_{GSE}$ and X$_{GSM}$ axes are the same, and the other two are rotated around that axis with respect to one another.} plane cut through the distribution of the relative emissivity ($n_n n_p v_{rel}$) for typical solar wind conditions. The dark region immediately sunward of the Earth is the magnetosphere, where the terrestrial magnetic field excludes the solar wind; the magnetopause is its outer boundary, at $\sim$10 R$_E$ on the x-axis. The bow shock is a sharp discontinuity at $\sim$13 R$_E$ on the x-axis. There is further emission sunward of the bow shock due to the free-flowing solar wind interacting with the outermost portions of the exosphere. The cusps are formed by magnetic field lines that are anchored in the Earth but have reconnected with IMF field lines that travel with the solar wind. Here the solar wind can plunge deep into the atmosphere. Aurorae and direct observation with {\it in situ} instruments demonstrate that the cusps do contain high densities of solar wind particles \citep{walsh_cusp_2016} but, since the kinetic physics of the cusp are not described by hydrodynamics, the MHD results for the low-altitude cusp may not be accurate.

The \xmm\ orbit is marked in Figure~\ref{fig:ms_demo}, while the \rosat\ and \suzaku\ orbits are indistinguishable from the Earth. It is clear that the amount of magnetospheric SWCX emission seen will depend very sensitively on the location of the spacecraft and the look direction. Most X-ray satellites are constrained to observe roughly perpendicularly to the Earth-Sun line. For low Earth orbit missions (\rosat\ and \suzaku ), the line of sight usually passes through the relatively low emissivity flanks of the magnetosheath. Occasionally, the line of sight can pass through the cusp of the magnetosheath, which is expected to be very bright. For high Earth orbit missions (\xmm\ and \chandra ), the bulk of the lines of sight pass through the flanks, but some observations pass through the nose of the magnetosheath which is, of course, very bright.

As the solar wind pressure (roughly $nv^2$) increases, the magnetopause moves closer to the Earth and the densest part of the shock moves further into the Earth's exosphere where the neutral density is higher, thus increasing the emissivity. Since the solar wind pressure is variable on many time scales, the emission seen on a particular line of sight can depend very sensitively on the solar wind pressure. Thus, while the inputs to this system are relatively well characterized, whether or not the very narrow FOV of an X-ray satellite is correctly predicted to pass through a strongly emitting region will depend upon the accuracy of the MHD models (as discussed in \S\ref{sec:ms_mod}). 

\subsection{Heliospheric Issues}

The heliospheric emission occurs all along the line of sight from the satellite to the heliopause (at 100-200 au). Integration on such a long line of sight tends to smooth over the variation in the solar wind characteristics. Conversely, since the solar wind density decreases as $r^{-2}$, the bulk of the emission is nearer rather than farther, so one is still sensitive to variations in the solar wind. Indeed, we will see that that sensitivity depends upon the look direction; looking perpendicular to a solar wind ``front'' produces far less variation than looking tangentially to that front. Since the solar wind is strongly magnetized and its speed is variable, it produces a complex structure of shocks, reverse shocks, and other discontinuities. MHD models describing this structure do exist, but have somewhat limited application for this problem. 

\begin{table}
\caption{Characteristic Time Scales}
\label{tab:times}
\begin{tabular}[]{lr}
\hline
Source & Time Scale \\
\hline
solar rotation & 24.47 days (sidereal) \\
solar rotation & 26.24 days (synodic) \\
solar sunspot cycle & 131$\pm$14 months \\ %\citep{hathaway_wilson_2004}
solar magnetic cycle & $\sim$22  years \\
\hline
Assuming SW speed of 435 kms\\
\hline
correlation length (125-400 R$_E$) & 30-100 minutes \\
from the nose of the bow shock to the Earth ($\sim$13 R$_E$) & 3.2 minutes \\
from L1 (i.e. ACE) to bow shock & 53-61 minutes \\
$1\arcdeg$ across LOS at 1 au & 1.8 hours \\
from Sun to heliopause ($\sim$120 au) & $\sim$480 days \\
\hline
Assuming ISM neutral speed of 21 km s$^{-1}$\\
\hline
from heliopause to Sun & 28 years\\
\hline
\end{tabular}
\end{table}

One of the difficulties with modeling the heliospheric emission is the range of operative time scales, some of which are listed in Table~\ref{tab:times}. These time scales range from that of the turbulence in the solar wind, through the quasi-periodicity of the solar wind (due to solar rotation), to the length of time it takes to reach the heliopause. There is also temporal variation in the neutral density. The inflowing neutral ISM moves from the heliopause to the Sun in approximately two solar cycles. Those neutral atoms are partly ionized by photoionization, electron impact ionization, and charge exchange, the rates of which depend upon the solar cycle. 

One of the most significant problems is that, with few exceptions, we do not routinely monitor the solar wind anywhere except near the Earth. Therefore while we know the {\it average} conditions of the solar wind at high solar latitudes (which are generally high ecliptic latitudes), we have no information on the solar wind conditions through which an observation was made. Thus, no matter the quality of the models, we simply do not have the data required to accurately model the emission over much of the sky.

%\clearpage

\section{The Solar Wind \label{sec:sw}}

This section reviews the phenomenological aspects of the solar wind required for understanding SWCX. It is important to remember that the solar wind was proposed in 1951, and its existence confirmed by Mariner 2 in 1962, meaning that we have had fewer than three solar magnetic cycles to study it. Further, while the solar wind has been studied with near-Earth spacecraft since 1963, we have been able to study its three dimensional structure with only a single spacecraft (\ulys) for a single solar cycle. Thus, while a broad picture of the solar wind exists, many details need to be added, and some of those details are important for SWCX. A useful compendium of the properties of the solar wind is provided in Table~\ref{tab:solarwind}.

\begin{table}
\caption{Solar Wind Properties}
\label{tab:solarwind}
\begin{tabular}[]{lrrrrr}
\hline
Property$^a$ & Mean & Mode & 10\% & Median & 90\%\\
\hline
Density (cm$^{-3}$) & 6.57 & 3.00 & 2.23 & 5.06 & 12.7\\
Speed (km s$^{-1}$) & 434 & 375 & 323 & 411 & 588\\
Flux ($10^8$ cm$^{-2}$ s$^{-1}$) & 2.66 & 1.60 & 1.06 & 2.12 & 4.79\\
%Temperature K & 1.03e5 & 2.00e4 & 2.22e4 & 7.38e4 & 2.22e5\\ 
Temperature (log K)  & 5.0 & 4.3 & 4.34 & 4.86 & 5.34\\
Pressure (nPa) & 2.26 & 1.25 & 0.890 & 1.82 & 3.98\\
$B$ (nT) & 5.52 & 4.0 & 2.6 & 4.8 & 9.2\\
%$B_z$ (nT) & 0.0 & 0.0 & -3.1 & 0.0 & 3.1\\
$|B_z|$ (nT) & 2.03 & 0.5 & 0.2 & 1.4 & 4.5 \\
$\phi_B$ ($\arcdeg$ ecliptic) & 135 & 135 & 78 & 134 & 191 \\
$\phi_V$ ($\arcdeg$ ecliptic)$^b$ & -0.270 & -0.750 & -3.78 & -0.436 & 3.52 \\
\hline
Derived Quantities \\
\hline
$R_{MP}$ (R$_E$)$^c$ & 10.3 & 10.2 & 8.99 & 10.2 & 11.5\\
$\Delta$$^d$ & 2.94 & 2.93 & 2.58 & 2.93 & 3.29 \\
\hline
Abundances$^e$  \\
\hline
He/O & 89.1 & 83.6 & 47.8 & 86.0 & 125. \\
C/O & 0.614 & 0.617 & 0.474 & 0.624 & 0.729 \\
$\left<Q_{C}\right>$ & 5.18 & 5.28 & 4.82 & 5.20 & 5.50 \\
$\left<Q_{O}\right>$ &6.15 & 6.03 & 6.00 & 6.10 & 6.34 \\
Ne/O & 0.134 &0.108 & 0.0876 & 0.122 & 0.188 \\
Mg/O & 0.160 & 0.127 & 0.0991 & 0.145 & 0.233 \\
$\left<Q_{Mg}\right>$ & 8.79 & 8.86 & 8.19 & 8.83 & 9.31 \\
Si/O & 0.170 & 0.145 & 0.116 & 0.160 & 0.235 \\
$\left<Q_{Si}\right>$ & 8.89 & 8.68 & 8.14 & 8.81 & 9.74 \\
Fe/O & 0.163 & 0.10 & 0.0849 & 0.140 & 0.263 \\
$\left<Q_{Fe}\right>$ & 10.0 & 9.60 & 9.09 & 9.78 & 11.0 \\
C$^{+6}$/C$^{+5}$ & 0.965 & 0.700 & 0.222 & 0.793 & 1.83 \\
O$^{+7}$/O$^{+6}$ & 0.205 & 0.0400 & 0.0282 & 0.131 & 0.433 \\
\hline
\end{tabular}

$^a$ {These parameters are derived from the OMNI 5-minute database sampled between 1981 and 2016 which was obtained from ftp://spdf.gsfc.nasa.gov/pub/data/omni/high\_res\_omni/.}\\
$^b$ {The upwind direction.}\\
$^c$ {The magnetopause standoff distance, the distance between the center of the Earth and the magnetopause along the Earth-Sun line, as calculated from Equation~\ref{eqn:spreiter_mp}.}\\
$^d$ {The thickness of the magnetosheath along the Earth-Sun line as calculated from Equation~\ref{eqn:spreiter_delt}.}\\
$^e${These parameters are derived from the \ace\ SWICS 1.1 database sampled between 1998 and 2012 which was obtained from ftp://mussel.srl.caltech.edu/pub/ace/level2/ssv4/. No selection on solar wind type was applied.}
\end{table}

\subsection{Phenomenology}

\begin{figure}
\center{\includegraphics[width=8cm,angle=0]{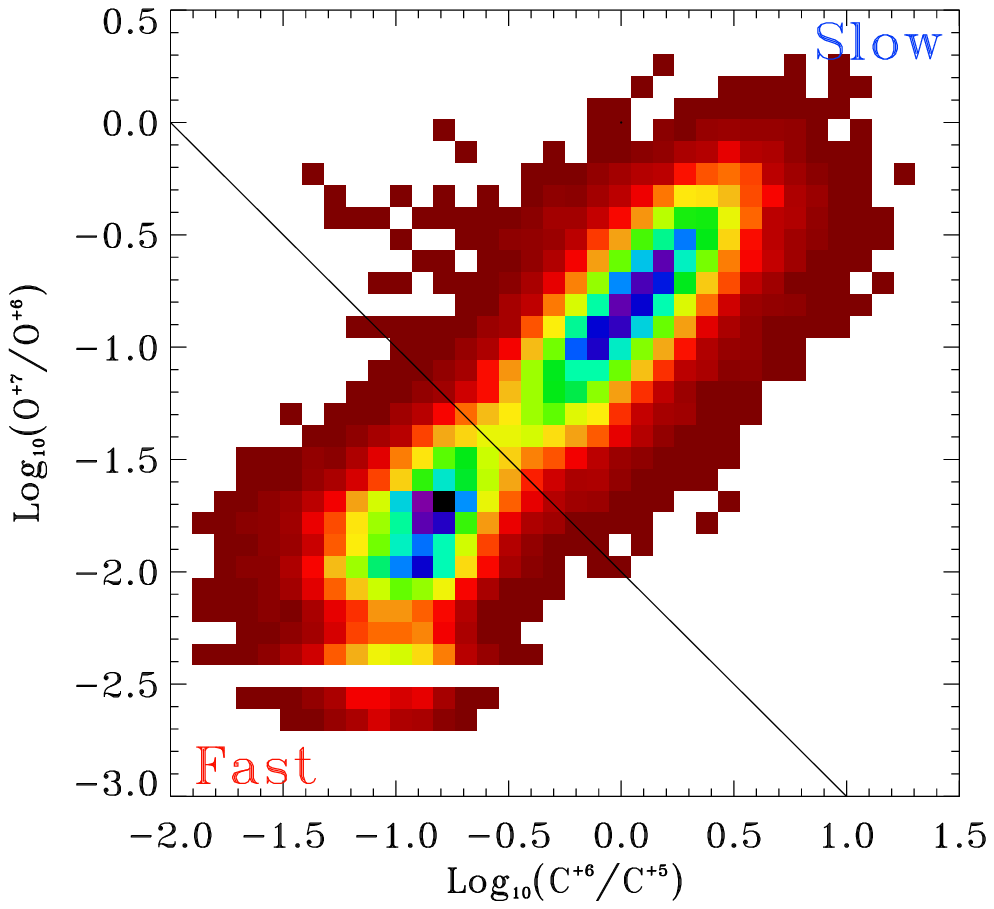}}
\vspace*{-0.75cm}
\center{\includegraphics[width=8cm,angle=0]{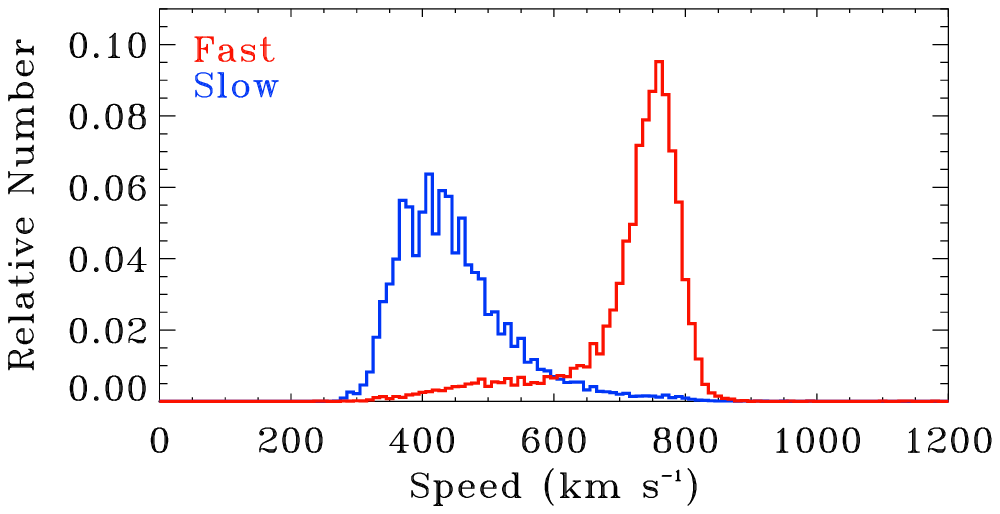}}
\vspace*{-0.75cm}
\center{\includegraphics[width=8cm,angle=0]{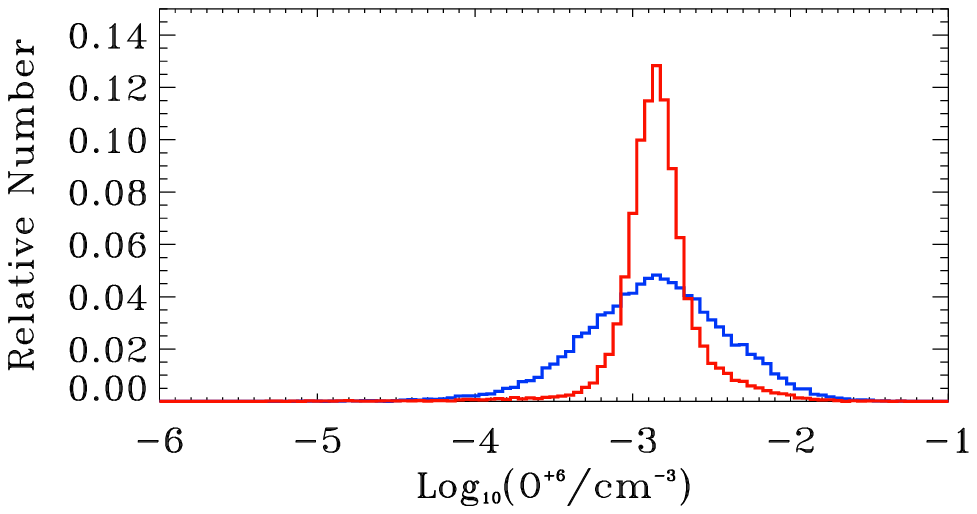}}
\vspace*{-0.75cm}
\center{\includegraphics[width=8cm,angle=0]{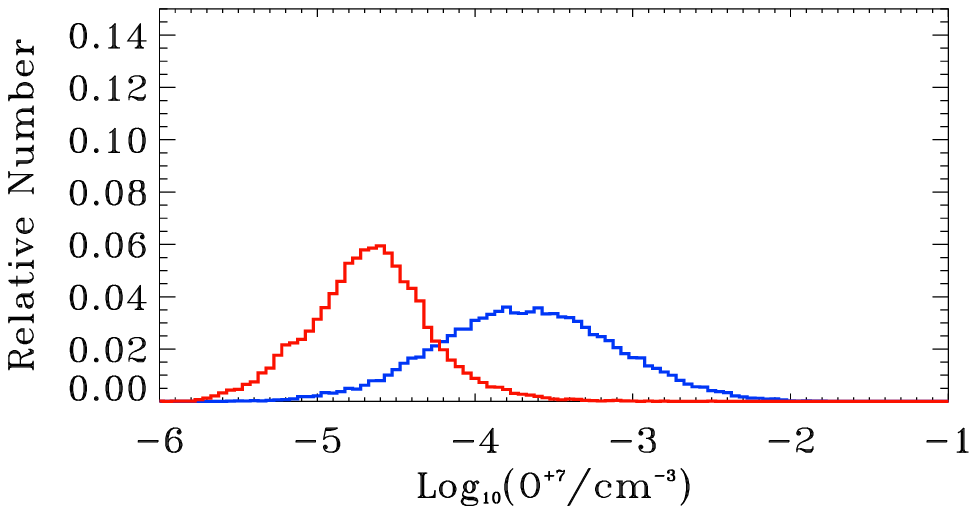}}
\caption{\ulys\ SWICS-SWIMS data demonstrating the differences and similarities of the fast and slow solar wind. ICMEs (defined by the mean iron ionization state being 12 or greater) have not been included.  {\it Top: } O$^{+7}$/O$^{+6}$ {\it versus} C$^{+6}$/C$^{+5}$, showing the bimodal distribution. The straight line is O$^{+7}$/O$^{+6}\times$C$^{+6}$/C$^{+5}=0.01$, which is the standard division between fast and slow. {\it Upper Middle: } The distribution of velocity for both fast (red) and slow (blue) solar winds. {\it Lower Middle: } Distribution of the O$^{+6}$ density for both fast and slow solar winds. {\it Bottom: } Distribution of the O$^{+7}$ density for both fast and slow solar winds. These figures are drawn from the \ulys\ final archive.}
\label{fig:vonsteiger}
\end{figure}

Besides coronal mass ejections (CME, sometimes noted as ICME when no longer close to the Sun), there are two types of solar wind, colloquially called ``fast'' and ``slow'' which, surprisingly enough, are not well distinguished by their speed. Instead, they are generally distinguished by ion ratios such as O$^{+7}$/O$^{+6}$, C$^{+6}$/C$^{+5}$, or some combination of the two \citep{vonsteiger_etal_2000, zhao_etal_2014}. The two ratios are strongly correlated and the distribution of each ratio is strongly bimodal \citep[][and see Figure~\ref{fig:vonsteiger}]{vonsteiger_etal_2010}. Low values of C$^{+6}$/C$^{+5}$ or O$^{+7}$/O$^{+6}$ correspond to higher solar wind speeds and {\it vice versa}. However, the distribution of velocities of the ``fast'' solar wind shows significant overlap with the distribution of velocities for the ``slow'' solar wind. Many of the observed properties of the solar wind (abundances, ionization balance, etc.) show bimodal distributions and/or are correlated with the ``fast''/``slow'' categorization. In general, the ``slow'' solar wind has, on average, higher charge states while the ``fast'' solar wind has lower charge states. However, it should be noted that the distribution of parameters such as density, abundance, or ion ratios is generally much narrower for the ``fast'' solar wind than it is for the ``slow'' solar wind.

The ``fast'' solar wind originates in regions occupied by coronal holes (regions of the sun where the magnetic field lines are open). The ``slow'' solar wind is related to coronal streamers (closed magnetic loops extending to several solar radii), but the ``slow'' solar wind would seem to originate from a broad range of different types or sizes of surface structures \citep{zurbuchen_etal_2002}. The greater uniformity of the source of the ``fast'' solar wind may be the cause for its narrower distribution of properties, but a connection between the distribution of the properties of the ``slow'' solar wind and the structure of the source regions remains ambiguous.

%Besides the ``fast'' {\it versus} ``slow'' division, the actual speed of the solar wind is an important parameter as the kinetic temperature of the solar wind is correlated with the speed and the density is anticorrelated with speed. 

\subsubsection{Structure \& Mechanics}

\begin{figure*}
\center{\includegraphics[width=6.5cm,angle=0]{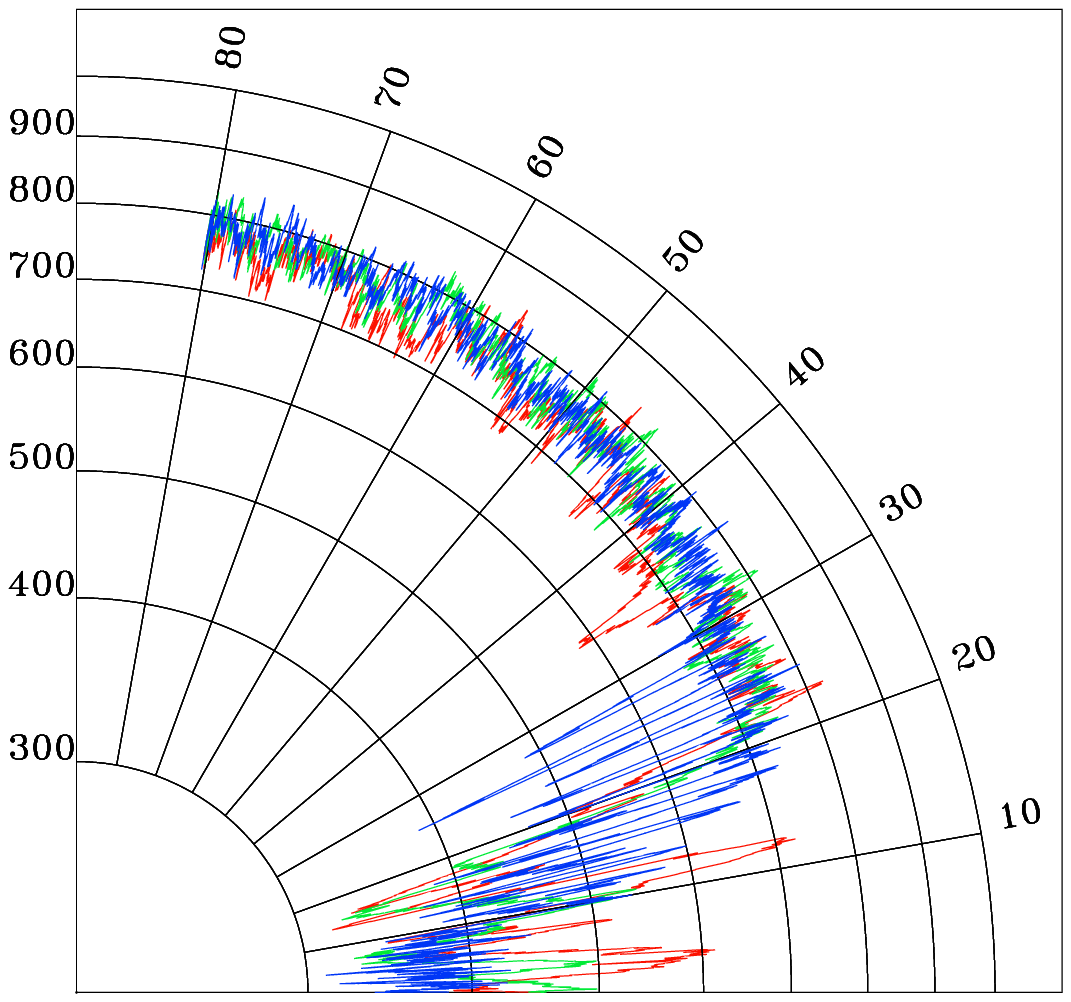}\hspace*{-1.0cm}
\includegraphics[width=6.5cm,angle=0]{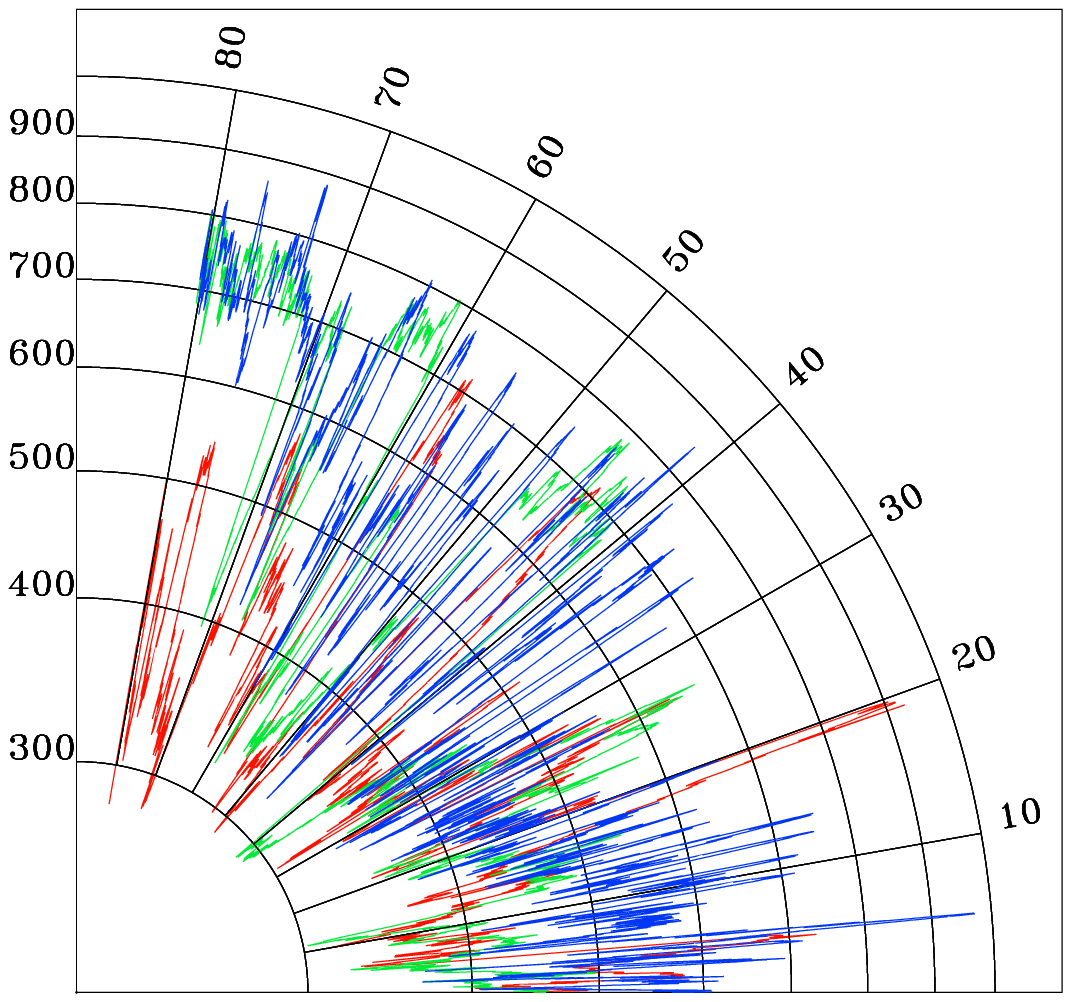}}
\caption{The \ulys\ view of the structure of the solar wind. The radial distance is the solar wind speed marked in km s$^{-1}$ while the solar latitude is marked in degrees. {\it Left: } The solar wind proton speed as a function of solar latitude for solar minimum. The data extend over three quarter-orbits (indicated by red, green, and blue curves) from day 256 of 1994 to day 349 of 1997. {\it Right: } The solar wind proton speed as a function of solar latitude for solar maximum. The data extend over three quarter-orbits from day 329 of 2000 to day 53 of 2004. The \ulys\ SWOOPS data were obtained from NSSDC.}
\label{fig:ulysses}
\end{figure*}

Our knowledge of the three-dimensional structure of the solar wind is due to \ulys , which had a 6.2 year polar orbit around the Sun from 1990 (solar minimum) through 2009 (the following solar minimum). \ulys\ showed that at solar minimum, solar latitudes $\gtrsim20\arcdeg$ contain a uniform fast solar wind associated with coronal holes while lower latitudes are dominated by the very irregular slow solar wind associated with coronal streamers. It should be noted that the latitude at which the fast solar wind was detected by \ulys\ ($\sim$1.3 au $<r<\sim$5.4 au) is not the latitude from which it left the Sun; the fast solar wind appears to be over-pressured with respect to the slow equatorial flow \citep{gosling_etal_1995}. 

As the Sun approaches maximum, the large polar coronal holes break up, irregular small coronal holes appear at all latitudes, and coronal streamers move to high latitudes. At solar maximum, the solar wind is very irregular with a mix of both types at all latitudes. This difference is well characterized by the solar wind velocity as a function of solar latitude, as seen in Figure~\ref{fig:ulysses} \citep{mccomas_etal_2008, mccomas_etal_2000}. 

The measured direction of the solar wind is narrowly peaked (within $\sim\pm3\arcdeg$) around purely radial. Since the Sun is rotating, in the length of time a parcel of solar wind moves outward $r=v_rt$, the Sun will have rotated an angle $\theta=-\Omega t$ so that $\theta=-\Omega r/v_r$, and the path describing the location of successive parcels of solar wind launched from the same location is an Archimedean spiral known as a Parker spiral from its original description by \citet{parker_1958}. As the solar magnetic field is dragged outwards from the location from which the solar wind is launched, while remaining anchored at that location, the same equation describes the magnetic field lines. Thus, the interplanetary magnetic field is predominately radial near the Sun.

\begin{figure*}
\center{\includegraphics[width=7.0cm,angle=0]{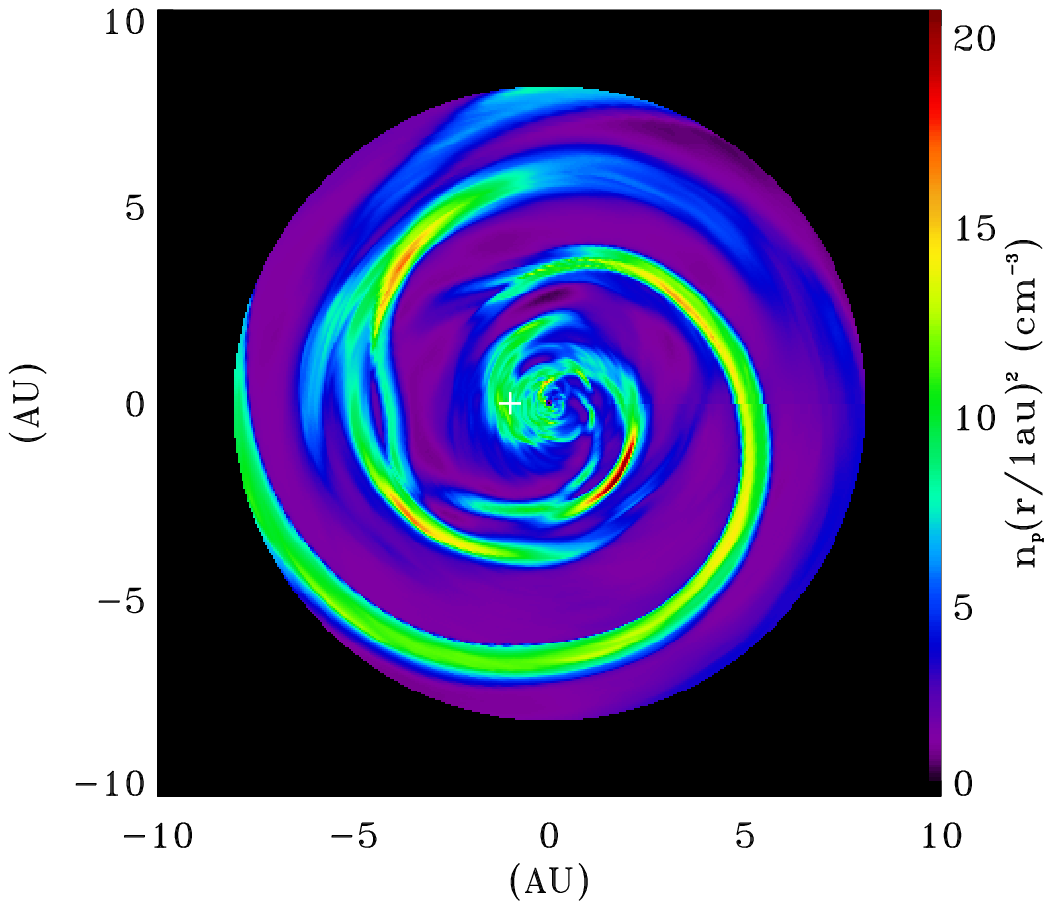}\hspace*{-1.0cm}
\includegraphics[width=7.0cm,angle=0]{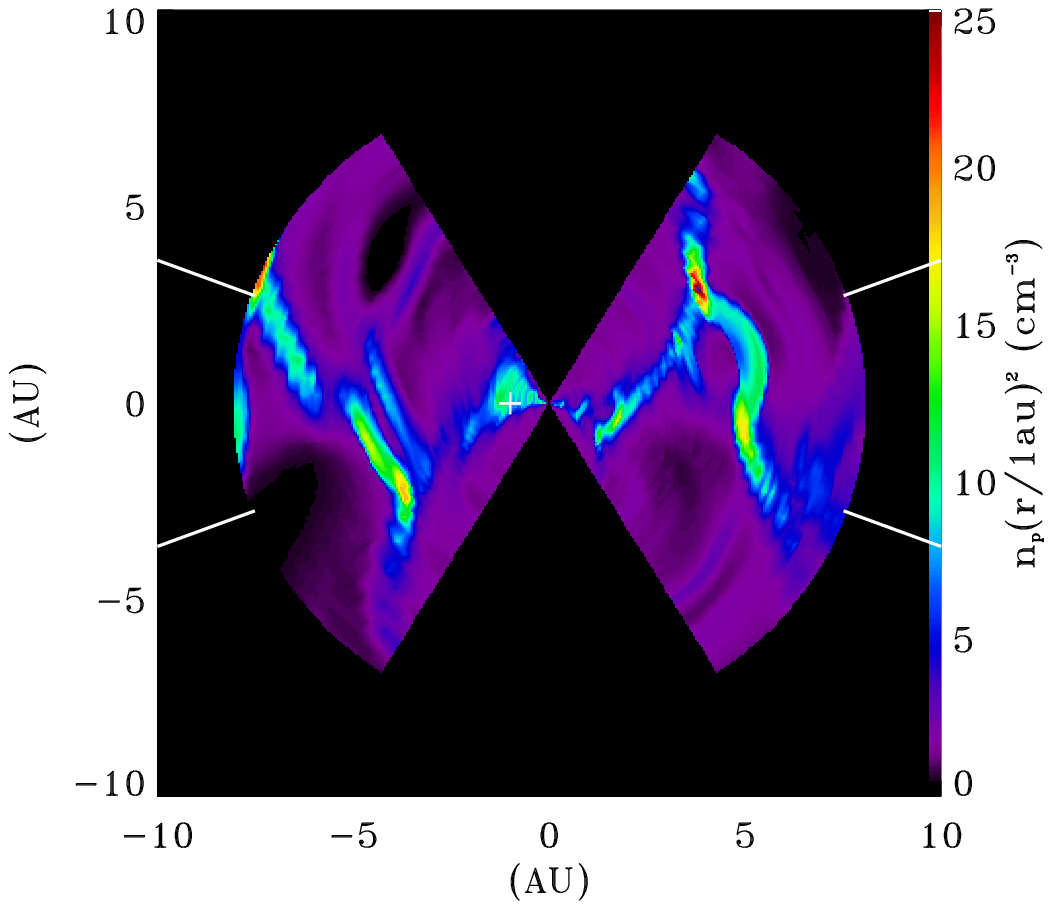}}
\caption{An ENLIL simulation of the solar wind density during solar minimum. The overall $1/r^2$ density dependence has been removed to accentuate the small-scale structure. The Sun is at the center and the location of the Earth at 1 au has been marked. {\it Left: } In the solar rotational plane. {\it Right: } A cut through the solar rotational plane. The white lines mark the $\pm20\arcdeg$ region that is dominated by the equatorial flow. This ENLIL run was produced by the CCMC for the author.}
\label{fig:enlil}
\end{figure*}

Since the average solar wind speed is 435 km s$^{-1}$ and solar rotation rate is $\Omega=4.41\times10^{-7}$ Hz, the Parker spiral forms an angle of $\sim43\arcdeg$ from the radial in the plane of the solar rotation at 1 au. Thus, if $\lambda_{as}$ is the ecliptic longitude of the anti-sun, lines of sight along longitudes $\lambda_{as}-43\arcdeg$ and $\lambda_{as}+133\arcdeg$ tend to be tangent to the Parker spiral. Figure~\ref{fig:enlil} shows an ENLIL MHD simulation of the solar wind in the plane of the solar rotation\footnote{The solar rotation axis is inclined to the ecliptic pole by $7.25\arcdeg$. The line of nodes is at an ecliptic longitude of $76\arcdeg$. Thus, while ecliptic latitudes are similar to solar latitudes, they are not the same. This difference can be very important when assessing whether a particular line of sight is within the solar equatorial flow or not.}; the Parker spiral clearly dominates other structures.

Of course the Parker spiral, as described here, is a gross oversimplification. The solar dipole is not aligned with the rotational axis (and the relative tilt is roughly a function of the solar cycle), the rotation rate of the Sun varies with solar latitude, and adjacent locations may launch the solar wind at very different speeds. Since the pitch angle of the interplanetary magnetic field (IMF) depends upon the speed of the solar wind, the variation in solar wind speed leads to the development (and continual evolution) of pairs of shocks in the solar wind. These shocks define ``corotating interaction regions'' (CIR). The CIR are not, in themselves, particularly important to the discussion of SWCX. However they do demonstrate that it is no trivial task to reconstruct the solar wind density and speed, even just for the equatorial flow, from only the data taken at L1. 

\subsubsection{Abundances and Ionization Structure\label{sec:conun}}

The abundances in the solar wind are not identical to those observed spectroscopically in the photosphere. As early as \citet{hovestadt_etal_1973} it was realized that, in the solar wind, elements with low first ionization potentials (FIP, $<10$ eV), such as Mg, Si, and Fe, were enhanced with respect to high FIP elements (FIP $>10$ eV) such as O, N, Ar, and Ne, compared to the optically determined abundances. This is typically expressed as the FIP fraction
\begin{equation}
(X/O)_{solar~wind}/(X/O)_{photospheric}
\end{equation}
where all the abundances are referenced to that of oxygen. In the ICMEs, the FIP fraction ranges from 3 to 5 for low FIP elements \citep{zurbuchen_etal_2016}. In the slow solar wind the FIP fraction is $\sim$3 for low FIP elements \citep{vonsteiger_etal_2000} while for the fast solar wind the FIP fraction is lower. \citet{vonsteiger_etal_2000} gives a value of 1.8-1.9, while \citet{vonsteiger_zurbuchen_2016}, citing the same paper, gives a value of 1.0-1.5. (The source of this discrepancy is not clear but is likely due to a change in the accepted photospheric abundances, particularly that of O.) The FIP fractions for elements with intermediate FIP, $\sim$10 eV, such as S and C, usually have abundance enhancements that are lower than those of the low-FIP elements. Various mechanisms have been proposed to explain the FIP effect \citep[see][for a review]{laming_2015}, but none is entirely satisfactory. The abundance of helium in both the slow and fast solar wind is reduced by a factor of 2-4 from the photospheric abundance. This depletion is thought to be due to the difficulty of accelerating helium through interactions with protons \citep{geiss_1982}. Thus, in general, the mechanisms producing the abundances in the solar wind are not well understood, though we have reasonably good abundance measurements. 

%As previously noted, for a given type of solar wind, there is a range of abundance values, though measurements show smaller abundance variation in the fast solar wind than in the slow solar wind \citep{vonsteiger_etal_2000}. However, the abundance/ionization differences between fast and slow solar wind are observed in the X-ray; \citet{bodewits_etal_2004} used comets as probes of the solar wind, using the X-ray and UV emission to diagnose the solar wind abundance/ionization through the charge exchange emission. Thus, the abundance/ionization structure differences are important for modeling SWCX.

The ionization structure of the solar wind is set by its passage through the first several solar radii. In these regions the electron density and the electron temperature vary strongly. When the time scale for transforming the ions in state $q$ to state $q+1$ or {\it vice versa} is longer than the time required for the ions to travel through an electron density scale height, the ion abundance is said to be frozen in, and $q^{i+1}/q^i$ reflects the electron temperature at the freezing radius. Thus ion abundance ratios for successive states, $q^{i+1}/q^i$ {\it versus} $q^{i+2}/q^{i+1}$, will reflect different freeze-in temperatures, and the freeze-in temperatures for a single element can show a wide range of values. Further, the ionization structure in a particular packet of solar wind will reflect the coronal structure of the particular region from which the solar wind was launched.

\begin{figure}
\center{\includegraphics[width=8cm,angle=0]{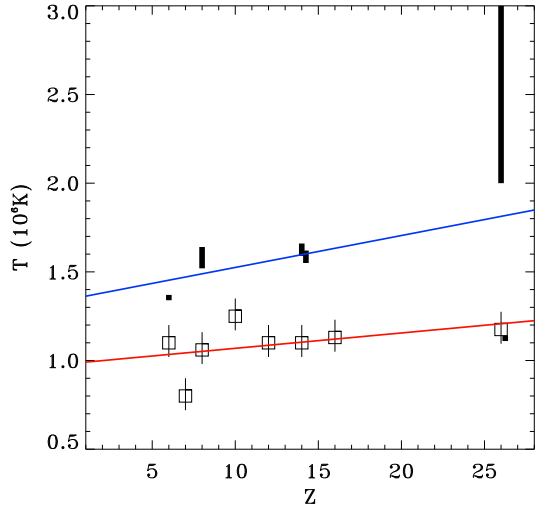}}
\caption{Freeze-in temperatures for the solar wind as a function of element. Open boxes with error bars are shown for the fast solar wind \citep[data from][]{gloeckler_geiss_2007}.  For the slow solar wind the range of freeze-in temperatures are shown by thick bars \citep[data from][]{vonsteiger_etal_2000}. For Si, the lower charge states (+7 through +10) showed much lower freeze-in temperatures than the highest charge state (+12), so two values are shown for Si. Similarly for Fe, the lower charge states (+6 through +13) are shown separately from +16.}
\label{fig:freeze-in}
\end{figure}

Although the abundances and ionization states of the solar wind are set within the first 3 to 4 R$_{\sun}$, the minor ions continue to interact with the protons and He$^{+2}$ that dominate the solar wind (i.e., Coulomb collisions) as well as with Alfv{\'e}n waves. It is thought that the increasing importance of wave-particle interactions at greater distances from the Sun explains the observed changes in the kinetic properties of the ions with distance, even though the details of the wave-particle interactions are not well understood \citep{vonsteiger_zurbuchen_2006}. However, the effects are sufficiently subtle that, with our current state of knowledge, they can be ignored. At 1 au, the radial velocities of different ions differ by a few tens of km s$^{-1}$, with greater discrepancies at greater velocities \citep{hefti_etal_1998} while velocity differences are negligible by 5 au \citep{vonsteiger_zurbuchen_2006}. Similarly, while at 5 au, the thermal velocities of different species are nearly the same (the temperatures being proportional to the mass), at 1 au there is an additional low temperature component for which the temperatures are equal. However, for most SWCX purposes, the approximations that all species have the same bulk velocity and that the temperature scales with mass are sufficient.

\begin{figure}
\center{\includegraphics[width=8cm,angle=0]{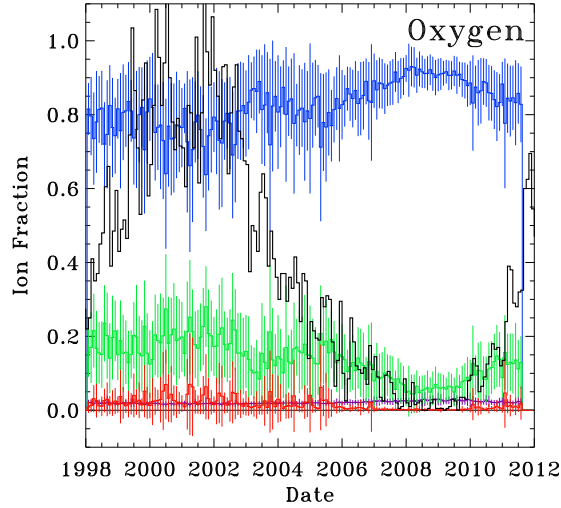}}
\caption{The ion fractions for oxygen as a function of date. Each data point is the mean for a month, extracted from the ``2 hour'' \ace\ data. The {\it black} histogram is the normalized sunspot number. {\it Purple} is the O$^{5+}$ fraction (almost hidden by the red line), {\it blue} the O$^{6+}$ fraction, {\it green} the O$^{7+}$ fraction, and {\it red} the O$^{+8}$ fraction. Also shown as error bars are the dispersion for each month-long interval.}
\label{fig:demo_oxy}
\end{figure}

In our introduction to the solar wind, the properties of the solar wind were set in the context of the fast-slow dichotomy. However, the fast-slow dichotomy is not the entire story. Figure~\ref{fig:demo_oxy} shows the oxygen ionization states over the interval in which \ace\ was producing good measurements. The dominant state is O$^{+6}$, the subdominant is O$^{+7}$, and both O$^{+8}$ and O$^{+5}$ are trace states. We first see that there is a strong trend with solar cycle, but that there is a large dispersion within each month-long interval. As expected from the discussion of the slow solar wind, even during solar minimum, there is a large dispersion in the measured values of individual ionization state. However, the trend is not what one might expect; during solar maximum, when we expect a mixture of fast and slow solar wind and thus, on average, a lower O$^{+7}$/O${^+6}$ ratio, we actually see a higher O$^{+7}$/O${^+6}$ ratio. Comparison of this ratio from the SWICS instrument on \ulys\ and its near duplicate, the SWICS instrument on \ace , for periods when both were sampling the equatorial flow are consistent, so this effect is not a cross-calibration issue. Therefore, the solar cycle plays an important role in the ion ratios.

In summary, solar wind abundances and ionization structure show clear trends, but it is difficult (if not impossible) to reconstruct either from just the measured solar wind speed. {\it In general,} the fast solar wind has lower ionization temperatures, and thus lower abundances of the high ionization species that tend to dominate the X-ray emission. Since the higher ionization species tend to produce lines at higher energies (on average) than lower ionization species, the fast solar wind should produce a softer X-ray spectrum. However, such a statement ignores many complications, such as FIP fractions or solar cycle effects. Further,  the slow solar wind has a very broad distribution of properties, so even if the solar wind at a particular phase of the solar cycle is ``slow'', the properties required for modeling the SWCX emission would still be poorly defined. 

Although it is difficult to extrapolate ion ratios from solar wind speed, the opposite is not true as the abundance/ionization differences between fast and slow solar wind are observed in the X-ray band. \citet{bodewits_etal_2004} used comets as probes of the solar wind, using the X-ray and UV emission to diagnose the solar wind abundance/ionization through the charge exchange emission.

\subsection{Data}

\subsubsection{Sampling}

Most solar wind missions have sampled the solar wind from the slow equatorial flow; only \ulys\ has taken a significant sample of solar wind outside the equatorial flow. There is currently no monitoring of the solar wind outside the equatorial flow. Within the equatorial flow, there are several solar wind monitors at L1 and, at times, \stereo\ measurements at 1 au from the sun and at some distance ahead and behind the Earth. At times there have also been solar wind monitors on spacecraft at Mars, Venus, Jupiter, and Saturn. Thus, there is a limited amount of data from which to reconstruct the solar wind along the line of sight. 

\subsubsection{Characteristic Scale-lengths \label{sec:sw_scale}}

Understanding the usefulness of solar wind data obtained from L1 requires knowledge of the characteristic spatial scale of variation in the solar wind. A spacecraft at L1, \ace\ for example, executes a complex orbit around L1 ($\sim$235 R$_E$ upstream of the Earth) with a $\Delta$(x$_{GSE}$,Y$_{GSE}$,Z$_{GSE}$) of ($\sim$20,$\sim$42,$\sim$25) R$_E$. As the solar wind is also not strictly radial as seen in the Earth's inertial frame, a packet of solar wind passing \ace\ may not strike the Earth's magnetosheath at all. Thus it is important to understand the correlation lengths of the solar wind.  

Cursory inspection of solar wind data from any L1 spacecraft reveals variations on many time scales, so there is clearly a size {\it spectrum}. From the Parker spiral geometry one might expect that the characteristic size(s) depend upon the direction (radial, azimuthal, or polar) in which they are measured. Although solar wind parameters are correlated with one another, different solar wind parameters have different characteristic scales. Finally, given these observations, there is the issue of {\it defining} a characteristic scale length. Several different methods have been used and, while they are not equivalent, they yield similar results. Perhaps the easiest to understand is that used by \citet{richardson_paularena_2001}; the characteristic size is the scale at which the correlation coefficient falls by 0.1 from its maximum.

The characteristic scale sizes for a particular parameter of the solar wind are determined by measuring the correlation between time series for that parameter as measured by two or more spacecraft at different locations, correcting for the expected time of flight between spacecraft (the advection time) and the motion of the Earth/spacecraft in that interval. The measured correlation coefficient increases with the length of the time series used; six-hour time series typically produce maximum correlation coefficients of 0.7. This is not surprising; as noted by \citet{matsui_etal_2002}, high frequency variations are likely due to turbulence which would not be correlated. Turbulence is thought to dominate the spectrum at time scales smaller than $\sim$2 hours. The balance between turbulence and coherent variation may also explain the fact that the correlation coefficient increases as the variance of the density increases.

%The correlation coefficient is larger for B_i than for V_i or n.

The characteristic radial scale length (i.e., in the $-$X$_{GSE}$ direction) is 200-300 \er\ or longer \citep{matsui_etal_2002, richardson_paularena_2001}. The characteristic transverse scale length (i.e., perpendicular to the X$_{GSE}$ direction) has been reported to be $\sim$45 \er\  for components of $\vec{B}$ \citep{collier_etal_2000,richardson_paularena_2001,matsui_etal_2002,collier_etal_1998}, $67\pm6$ for $|V|$, and $119\pm21$ for $n$ \citep{richardson_paularena_2001} though \citet{matsui_etal_2002} found a weaker dependence of the correlation coefficient on spacecraft separation and thus, possibly, a longer scale length. 

\begin{figure}
\center{\includegraphics[width=8cm,angle=0]{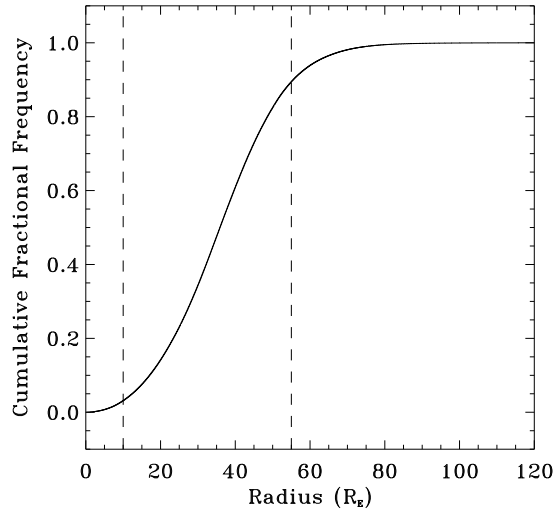}}
\caption{The cumulative fraction of the time that a packet of solar wind measured by \ace\ passes within a given radius of the the center of the Earth. The vertical lines mark the approximate radius of the magnetopause, and the magnetopause distance plus a solar wind correlation length. The relations for other L1 based spacecraft would be similar.}
\label{fig:ace_hit}
\end{figure}

Thus we may return to the question of what fraction of the time does the solar wind, as measured by an L1 satellite, such as \ace , actually represent the solar wind striking the Earth's magnetosheath. Figure~\ref{fig:ace_hit} shows the cumulative fraction of time that a packet of solar wind measured by \ace\ passes the Earth at a particular radius from Earth's center. Roughly 3\% of measured solar wind packets will pass within 10 R$_E$ of the Earth, which is roughly the radius of the magnetopause. This would suggest that the \ace\ measurements should have little correlation with the behaviour of the magnetosheath. Conversely, taking the correlation length of $\sim$45 R$_E$, nearly 90\% of the solar wind measurements fall within a correlation length of the magnetopause, suggesting that \ace\ does a passable job of measuring the upstream solar wind that will strike the magnetopause. However, lack of correlation between individual structures in the solar wind and the magnetospheric response to the solar wind should not be surprising.

\subsubsection{Time Scales}

The characteristic radial correlation length is 200-300 \er\ or longer which, for a mean solar wind speed of 435 km s$^{-1}$, corresponds to 2-3 ks, which is much shorter than a typical X-ray observation. However, the correlation length is not necessarily the scale length of interest for the problem(s) at hand. The solar wind varies on many time scales, from the rotational period of the Sun, to the size of an active region, to turbulence. Integrating the heliospheric SWCX emission from the Earth to the heliopause will tend to suppress the shorter times scales. Conversely, the magnetospheric emission responds almost immediately to variation in the solar wind, so the short times scales are important. Discussion of the relevant solar wind time scales is delayed to \S\ref{sec:play}.

\subsubsection{Abundances}

\begin{figure}
\center{\includegraphics[width=8cm,angle=0]{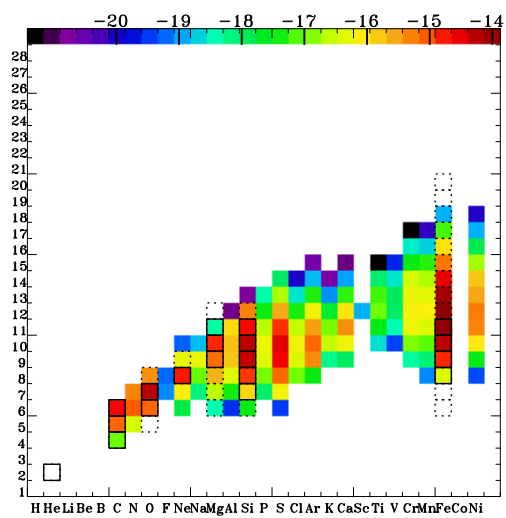}}
\caption{A comparison of the elements/ionization states that are important for X-ray spectroscopy and the elements/ionization states measured by \ace . The logarithmic {\it color scale} shows the relative strength of the X-ray emission expected from the solar wind. The {\it solid black borders} denote the elements/ionization states that are measured in standard \ace\ data products and the {\it dashed black borders} denote the elements/ionization states for where there is sporadic or low signal-to-noise coverage by \ace . The color scale was calculated using APEC to determine the emissivity of an element assuming an equilibrium temperature equivalent to the freeze in temperature and an \citet{anders_grevesse_1989} abundance pattern.It is clear that some of the more important X-ray producing species are not well measured by the \ace\ data.}
\label{fig:apec_ace_rosat}
\end{figure}

\ulys\ had a polar orbit around the Sun and the \ace\ is in an orbit about L1. Both missions had instruments named SWICS to measure the abundances of the ``minor species'', that is, anything but hydrogen and helium. The SWICS instruments measure energy per charge, time of flight, and total energy of each particle, and those quantities are converted to mass and mass per charge. The uncertainties in these quantities are sufficiently large that the error ellipses for successive ionization states or successive elements have significant overlaps \citep[see, for example Figure A1 of ][]{vonsteiger_etal_2000}. The standard data products result from the equivalent of a two dimensional deconvolution of the M {\it versus} M/q data; an atomic number/ionization state combination with low abundance that falls close to an atomic number/ionization state combination with a high abundance will thus have higher backgrounds and will be more uncertain. However, the less abundant species do not necessarily have less impact on the X-ray spectrum.
%low abundance elements/ionization states near high abundance elements/ionization states will thus have higher backgrounds and will be more uncertain. However, the less abundant species do not necessarily have less impact on the X-ray spectrum.

In order to understand this issue, we have calculated the relative contribution of each of the minor species to a typical X-ray bandpass using approximate freeze-in temperatures for the slow solar wind. As noted above, the freeze-in temperature for the slow solar wind is poorly characterized and varies with $q$. However, from Figure~\ref{fig:freeze-in} it is clear that the freeze-in temperatures for the slow solar wind are roughly 1.6 MK. Given this temperature, APEC \citep[Astrophysical Plasma Emission Code,][]{smith_etal_2014} was used to determine the relative emissivity for the elements/ionization states found in the 0.1-2.0 keV band (the colored boxes in Figure~\ref{fig:apec_ace_rosat}). The elements/ionization states well measured in the standard \ace\ data (before 23 August 2011)\footnote{After 23 August 2011, only a much more restricted set of elements/ionization states are available and then only as abundance ratios.} are marked with black borders, while those elements/ionization states that are either poorly measured or only sporadically well measured are marked with dashed borders. As can be seen from the figure, many important species, such as O VIII, the higher ionization states of Fe, or any state of S, are not well measured by \ace . 

One further technical issue concerning abundances should be noted. The instruments which measure ion abundances (i.e., SWICS on \ace ) measure He$^{+2}$, but not protons. Other instruments (i.e., SWEPAM on \ace ) measure He$^{+2}$ and protons. However, the He$^{+2}$ measurements are not necessarily consistent between instruments \citep[see][for further discussion]{koutroumpa_etal_2019} which introduces significant uncertainties.

\subsection{Validating the Models \label{sec:sw_mod}}

The most common code used to model the solar wind in the heliosphere is ENLIL\footnote{ENLIL is not an acronym, it is the name of the Mesopotamian god of, among other things, the winds.} \citep{odstrcil_2003} which is available for public use through the Community Coordinated Modeling Center (https://ccmc.gsfc.nasa.gov/). ENLIL is a 3D MHD numerical model capable of modeling the solar wind to $\sim$10 au for solar latitudes to $\pm60\arcdeg$. The inner boundary conditions at 21.5 R$_{\sun}$ are set by a model of the corona, typically the Wang-Sheeley-Arge (WAS) model though others are available. The input for the coronal model is coronographic observations of the Sun by \soho\ and \stereo.

The extent to which ENLIL models accurately describe the solar wind in the inner heliosphere is a matter of active study. The bulk of studies attempting to verify ENLIL performance concentrate on CMEs, which are a special case, as the properties of the CME must be inserted at the inner boundary. Thus, CME modeling is dependent on both the robustness of the MHD code and the uncertainty of the input parameters. Prediction of CME passages have a $\sim$50\% success rate, that is, about half of CMEs predicted to hit the Earth or a particular spacecraft actually do. The mean error in the arrival time is $\sim$10 hours \citep{wold_etal_2018}.

Of greater interest are studies of CIRs since these more adequately reflect the large scale structure of the solar wind. However, studies comparing ENLIL predictions and measured CIRs are more scattered and have less statistical weight. Although ENLIL appears to reproduce the structure of CIR quite well, the predictions of CIR passage have errors of $\sim$2 days at 1 au and $\sim$4 days at 5.4 au \citep{jian_etal_2011, prise_etal_2015, lee_etal_2009}. It should be noted these studies primarily measure the solar wind within the equatorial flow.

Thus, while ENLIL can provide a reasonable facsimile of the solar wind structure, the timing uncertainties make it an unreliable predictor of the solar wind along a particular line of sight at a particular time. Any calculation made with ENLIL should check the variation over a several day interval to ensure that it has not been affected by a mis-timed feature. Such timing uncertainties have been an important factor in understanding SWCX emission from other planets such as Mars \citep{koutroumpa_etal_2012} or Jupiter \citep{kimura_etal_2016}. 

%\clearpage

\section{Solar Wind in the Magnetosheath \label{sec:sw_ms} }

\begin{figure}
\center{\includegraphics[width=8cm,angle=0]{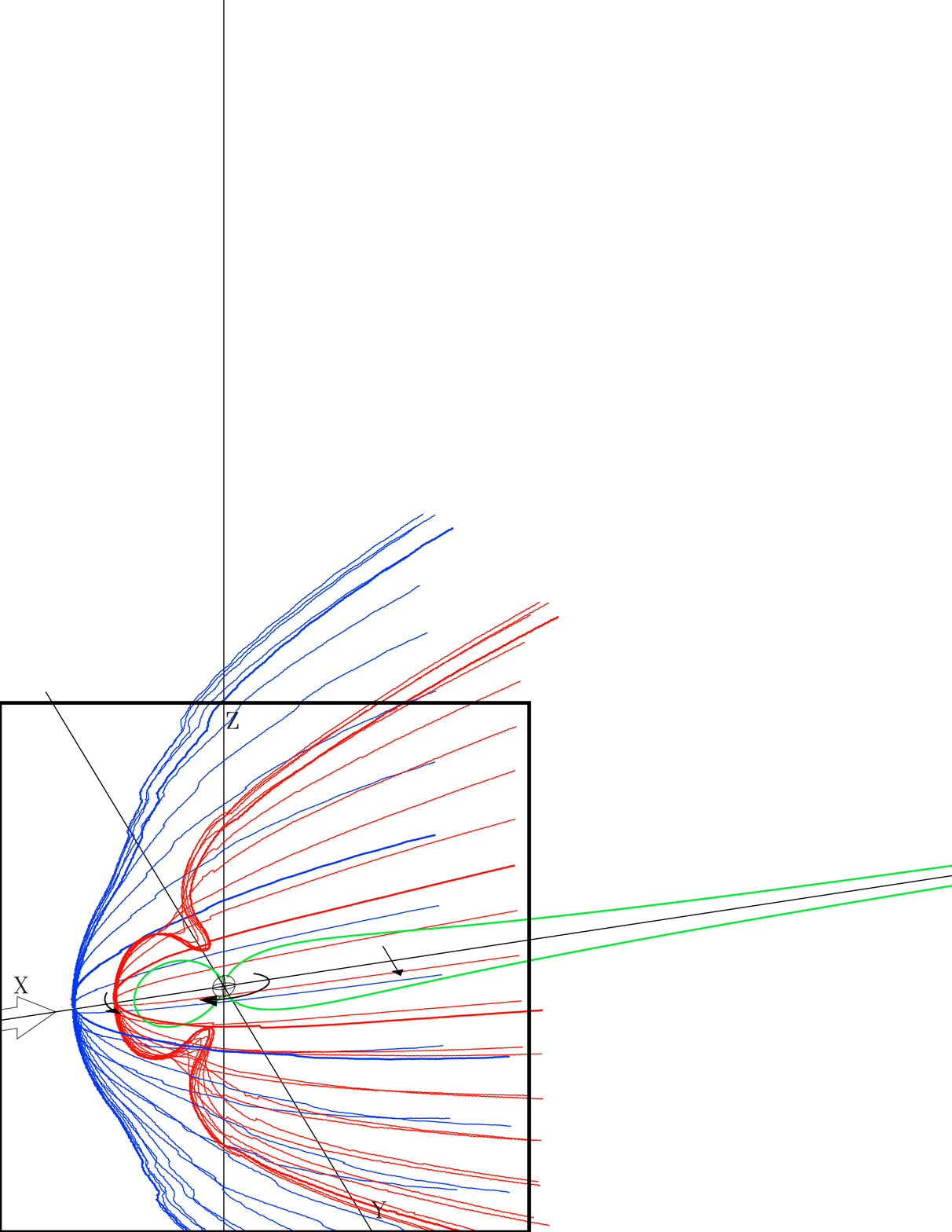}}
\caption{The magnetosheath and the near-Earth environment. The {\it blue} grid shows the location of the bow shock while the {\it red} grid shows the location of the magnetopause with a stand-off distance of 10 R$_E$. The grid lines in the XY and XZ planes are drawn more thickly. Grid lines in the near upper quadrant have not been drawn. The {\it green} line shows a closed field line in the XZ plane; the line passes through (8,0,0) R$_E$. The large arrow shows the direction of the solar wind. The smaller arrows show the direction of magnetopause current (just outside the magnetopause), the ring current (around the Earth) and the cross-tail current. The irregularities in the magnetopause and bow shock are due to limited resolution in the outer part of the MHD simulation from which the values were taken, which is the same used in Figure~\ref{fig:ms_cut}.}
\label{fig:textbook}
\end{figure}

The previous section discussed the solar wind in the context of the distribution of the ion populations producing SWCX emission throughout the heliosphere. The problem for the magnetosheath is more restricted and tractable. There are a number of solar wind monitors at L1 that provide some information about solar wind conditions in the near-Earth environment. However, the physics of the solar wind/terrestrial magnetic field interaction provides another level of complexity and uncertainty. MHD models of the magnetosphere/sheath are the workhorses for simulating the SWCX emission from the near-Earth environment, but have some poorly-recognized limitations. Before addressing the issues of MHD models of the magnetosheath it is useful to make a {\it brief} excursus on the physics that the MHD models may or may not capture. We can then understand the uncertainty in our knowledge of the size and shape of the magnetosheath at any given time.

\subsection{The Physical System \label{sec:mhd}}

The standard textbook depiction\footnote{Much of this summary is drawn from \citet{cravens_book} and \citet{kivelson_russell}. Together, these sources provide a good introduction to the magnetosphere.} of a {\it simplified} magnetosphere (Figure~\ref{fig:textbook}) is still a confusing system with multiple different plasmas and multiple current systems. 

The magnetized solar wind, whose field orientation is usually along the Parker spiral, approaches the Earth roughly along the X$_{GSE}$ axis. Between the bow shock and the magnetopause, the dynamic pressure of the solar wind is converted to a thermal and magnetic pressure, and the balance of that pressure with the magnetic and (generally much lower) thermal pressure of the magnetosphere sets, to first order, the distance from the Earth to the magnetopause (the magnetopause stand-off distance). When the hot protons and electrons in the magnetosheath impinge upon the terrestrial magnetic field they create a current perpendicular to the terrestrial magnetic field, due to the Lorentz force, known as the magnetopause current. The effect of this eastward current is to increase the magnetic field inside the magnetopause, thus increasing the stand-off distance.

A second important plasma is the {\it ring current} which is confined by the terrestrial dipole field. It is the low energy ($kT_i\sim10$'s of keV) ion equivalent of the trapped electron radiation belts, also known as the van Allen belts. This plasma is subject to magnetic gradient and curvature drifts, which produce a westward current. This produces a magnetic field that has the opposite sense of the Earth's dipole at the Earth's surface, but enhances field strengths at the magnetopause. The ring current plasma is injected from the magnetotail during magnetic storms, so the current density is strongly time-variable. Therefore, its contribution to determining the stand-off distance will be similarly variable.

The magnetotail, the anti-sunward lobes of the Earth's magnetic field, stretches well beyond $100$ R$_E$ behind the Earth. This stretching causes anti-aligned magnetic fields to lie in close proximity on opposite sides of the equatorial plane. The plasma sheet in the equatorial plane contains a hot plasma ($kT_i\sim5$ keV) and the magnetic configuration produces a {\it cross-tail current}\footnote{For those who would note that a current cannot be sustained without a loop, the cross-tail current is connected to the tail current which flows across the surface of the lobes of the magnetotail back to the other side of the plasma sheet. The above description is not an exhaustive description of all of the plasmas and current systems, just a description of those most salient for the issue of MHD modeling.} in the equatorial plane perpendicular the X$_{GSE}$ axis. Magnetic reconnection across the plasma sheet is the process that injects plasma into the ring current during magnetic storms (due to a strong solar wind pressure impulse followed by a prolonged interval of southward IMF) and substorms (due to energy release in the magnetotail).

The tailward reconnection is ultimately the result of magnetic reconnection that occurs on the surface of the magnetosphere. The interplanetary magnetic field (IMF) is generally not aligned with the field in the outer parts of the terrestrial dipole. As the solar wind sweeps past the Earth, the IMF becomes draped over the magnetosphere. For southward and ecliptic IMF orientations, magnetic reconnection of the IMF and the terrestrial field occurs on (primarily) the sunward side of this interface. The sweep of the solar wind past the Earth pulls these newly reconnected field lines down the magnetotail. Since the magnetic field lines cannot accumulate in the tail indefinitely, magnetic reconnection occurs in the mid-plane of the magnetotail where the field lines are anti-aligned. Exactly how and where reconnection occurs on the day side is an outstanding problem in space physics which can be addressed using charge-exchange emission from the magnetosheath \citep{sibeck_ssr_2018}. Reconnection modifies the outer magnetospheric magnetic field strength and thus the pressure balance with the magnetosheath, causing changes in the stand-off distance until pressure balance is restored.

Neither reconnection nor the injection of plasma into the ring current  (nor ``shadowing losses'' from the ring current where it runs into the magnetopause) are MHD processes. Thus, while MHD models can produce the distribution and characteristics of the plasma in the magnetosheath, they rely on other models of the ring current (among other things) to produce the inner boundary condition. Different MHD codes have different methods of introducing the effects of reconnection and particle kinetics.

%[Birkelund currents?
%[Convection?

\subsection{MHD Codes and Validation \label{sec:ms_mod}}

There are multiple MHD models of the magnetosphere, the most popular of which are BATS-R-US \citep{toth_etal_2005}, Gumics \citep{janhunen_etal_2012}, LFM \citep{lfm_2004}, and OpenGGCM \citep{raeder_etal_2001}. There are differences in the implementation of the MHD equations (gridding and refinement), differences in the treatment of boundary conditions, and differences in the ring current model to which they are coupled. The NASA Community Coordinated Modeling Center (CCMC) holds copies of all of the cited codes so one can request runs for a given set of solar wind input conditions\footnote{https://ccmc.gsfc.nasa.gov/requests/requests.php}. Thus, these are the most commonly used codes for simulating the magnetospheric SWCX emission.

\begin{figure}
\center{\includegraphics[width=6.75cm,angle=0]{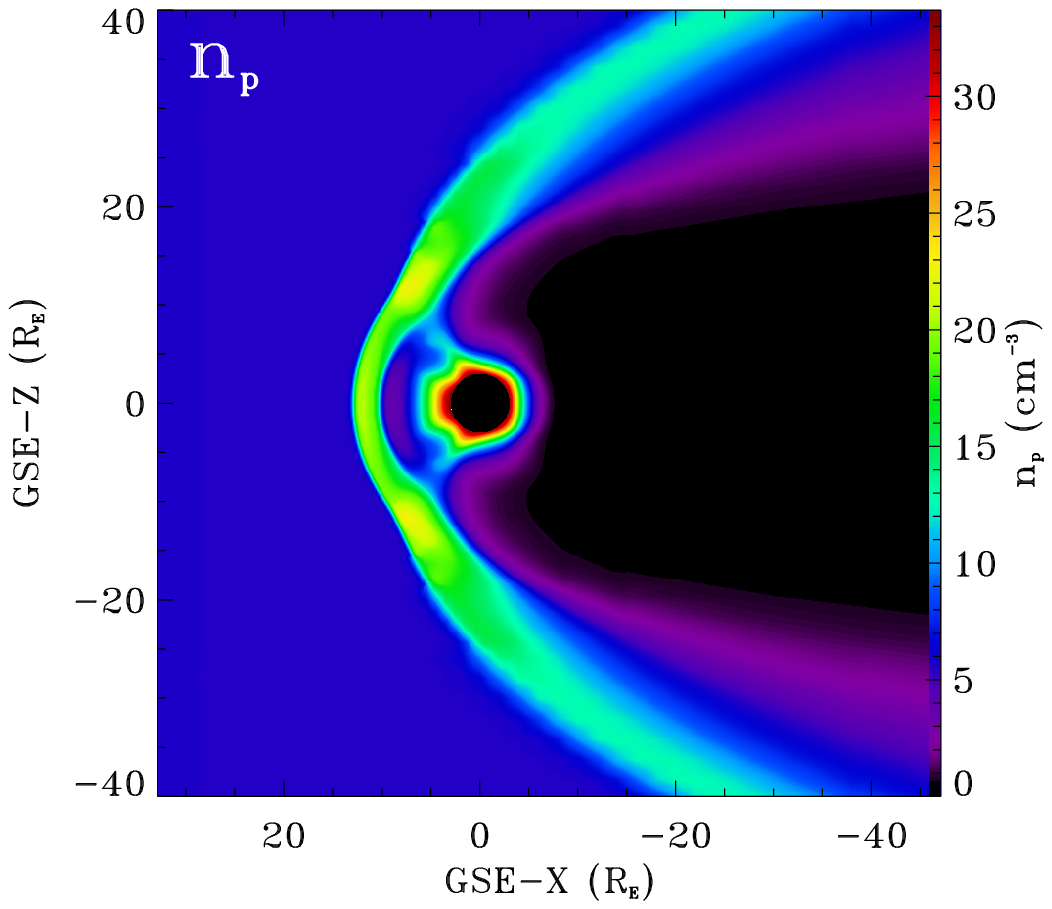}\hspace*{-0.75cm}
\includegraphics[width=6.75cm,angle=0]{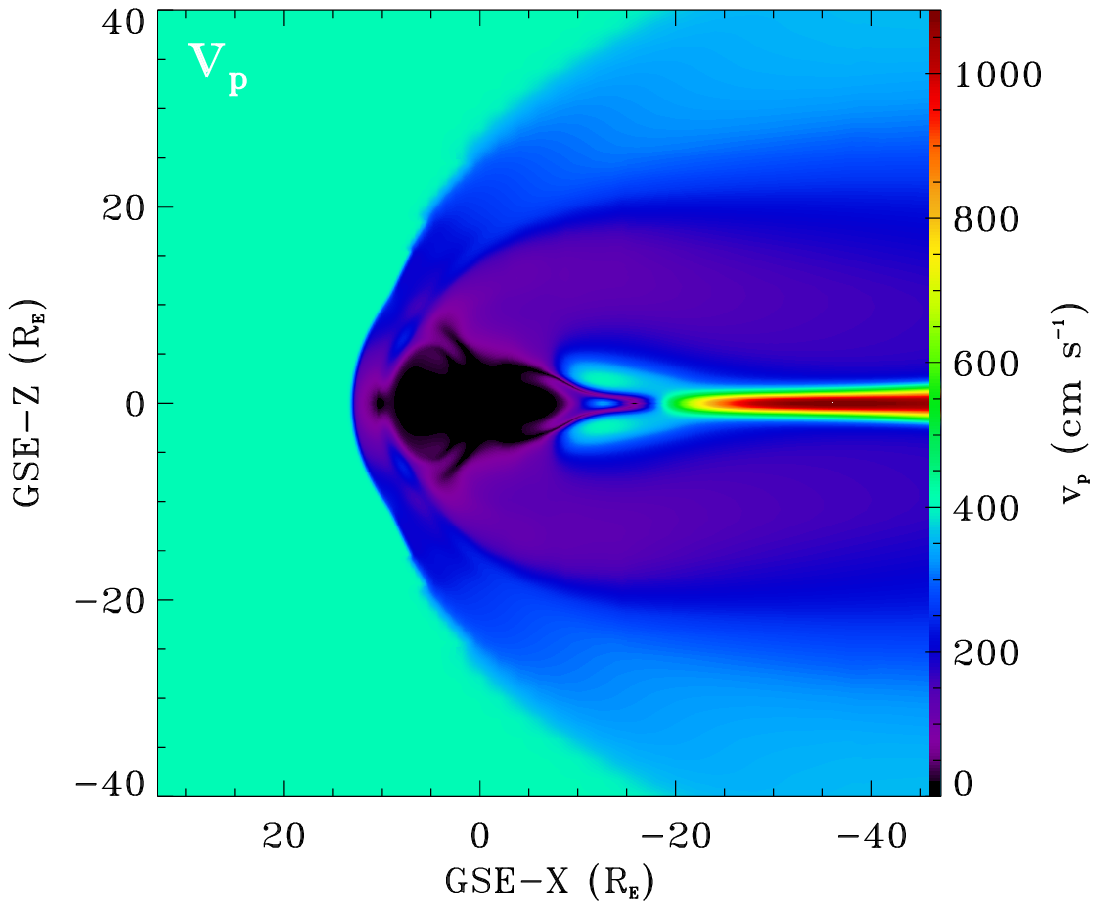}}
\vspace{-0.5cm}
\center{\includegraphics[width=6.75cm,angle=0]{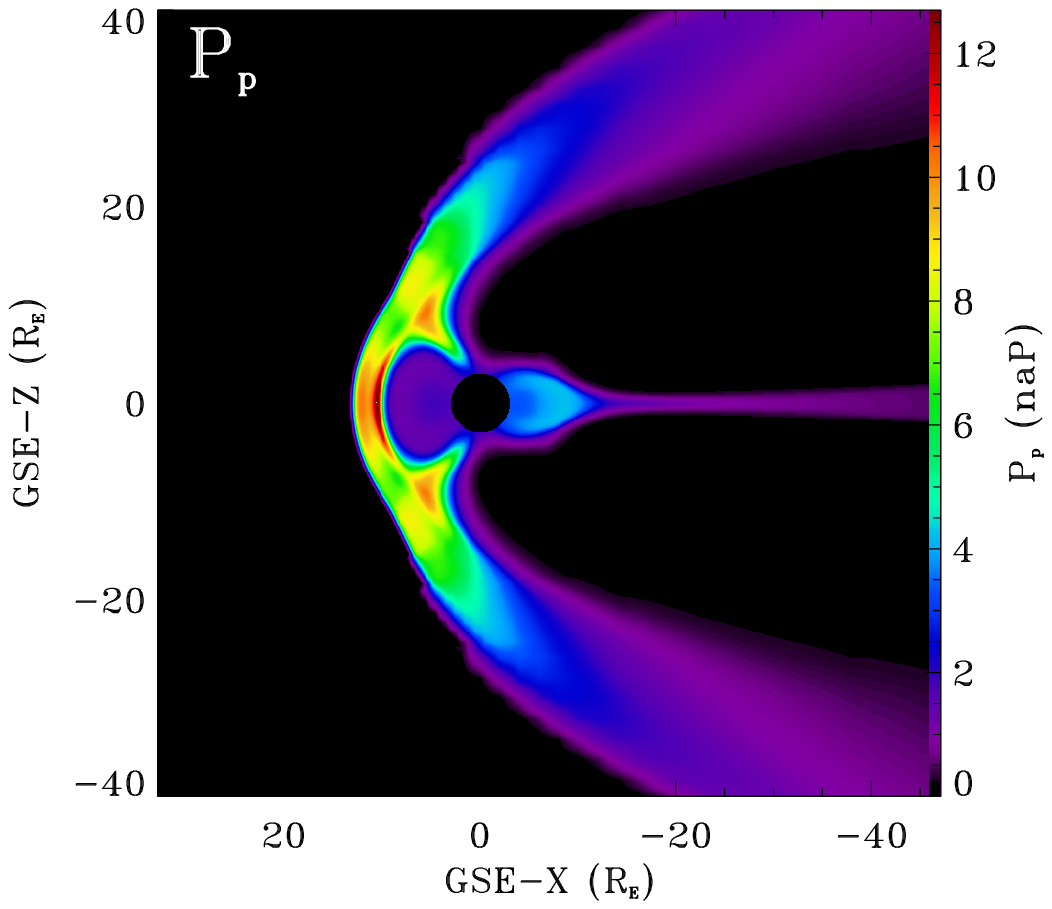}}
\caption{XZ$_{GSE}$ slices through an MHD model for a typical solar wind and IMF:  $n_p=5.5$ cm$^{-3}$, $v_{\infty}=405$ km s$^{-1}$, $B_z=5$ nT. The dynamical solar wind pressure is $0.909\times10^{16}$ g cm s$^{-2}$. The simulation was done for the equinox. {\it Top Left: } Proton density {\it Top Right: } Proton bulk velocity {\it Bottom: } Proton thermal pressure. Note that in all cases the solar wind protons are not distinguished from plasmasphere protons. The region with $R<3$ R$_E$ has been removed as the MHD models are not valid there. These were extracted from a BATS-R-US model run by the NASA CCMC.}
\label{fig:ms_cut}
\end{figure}

All of these codes have the same issue: the code tracks protons, but does not distinguish between solar wind protons and protons originating in the plasmasphere, a region with $R\lesssim4$ R$_E$ containing cold ($kT\sim1$ eV) plasma. In practice, these codes can allow plasmaspheric plasma to creep out to the magnetopause, which can be seen in Figure~\ref{fig:ms_cut} (n$_p$), filling the region between the plasmasphere, the cusps, and the magnetopause. Since this plasma, whose ultimate origin is the ionosphere, does not contain the ions producing the charge-exchange emission, one cannot simply take the proton density from a model and multiply it by $n_{ion}/n_p$ to determine the density of high charge state ions. Instead, one must first remove the plasmaspheric protons, typically by making the assumption that the solar wind plasma does not enter the dayside magnetosphere.Since the gyroradius for a solar wind proton entering the magnetosphere is $\sim$0.02 R$_E$, direct penetration of the magnetopause by the solar wind occurs on scales much smaller than the typical model grid size ($\sim$0.1 R$_E$). However, dayside reconnection allows solar wind ions to enter the magnetosphere, forming a ``boundary layer'' the thickness of which thickness of which increases with distance from the reconnection site and may range from 0.1 to 1.0 R$_E$ \citep[see][and references therein]{tkachenko_etal_2008}. This point is raised because this process of removing the plasmaspheric protons can artificially sharpen boundaries or leave significant artifacts if not implemented correctly. Multi-fluid codes, where the solar wind protons are tracked separately from the plasmasphere protons exist, but are not yet publicly available.

Unfortunately, what these codes do not share is uniformly consistent results. Only recently have there been significant efforts to compare MHD models with one another and simultaneously to measurements. The most salient part of the models is the magnetopause standoff distance. For the same solar wind inputs, different codes predict different stand-off distances at the $\pm1$ R$_E$ level \citep{colladovega_etal_2015}. Some test cases show little consistency in either location or temporal trends while other test cases, run in the same way, show reasonable agreement. It is not yet clear what solar wind conditions are more or less problematic. 

Comparison to {\it in situ} measurements is not trivial; keep in mind the uncertainty in the input solar wind parameters discussed in \S\ref{sec:sw}. There are a limited number of spacecraft measuring the location of the magnetopause on the order of once or twice per orbit, so there are a limited number of data points for a given MHD run. \citet{garcia_hughes_2007} detailed a number of further technical issues and compared LFM predictions of magnetopause locations to a variety of empirical models. The empirical models were, in turn, based on databases of {\it in situ} measurements of magnetopause crossings by spacecraft. They found that the LFM model consistently placed the magnetopause Earthward by 0.5-1.0 R$_E$ at local noon, and 1.-2.0 R$_E$ Earthward at the terminator. They attributed this discrepancy to an insufficient ring current model. \citet{colladovega_etal_2015} found that the times with the best agreement between model and measurement seem to be periods of relatively constant solar wind conditions. It is not yet clear when models best characterize reality \citep{colladovega_etal_2018}.

\subsection{The Cusps}

The cusps clearly show bright SWCX emission \citep{fujimoto_etal_2007}. Quantifying that emission through modeling is difficult. The ion density in the cusps is set more by kinetic processes than by the fluid assumptions in MHD. Because of this discrepancy, MHD models do a poor job of simulating both the size of the emitting region and the emission strength. \citet{walsh_etal_2016} found that the density of the cusp in MHD simulations is usually less than half that of the cusp density as measured by the {\it Polar} mission. The observed cusp has an opening angle of $\sim4\arcdeg$, which roughly half as wide (in latitude) as the cusp produced by MHD simulations. MHD simulations do usually produce the large-scale motion of the cusp resulting from changes in the driving solar wind magnetic field vector \citep{zhang_etal_2013}. On the whole, MHD models of the cusp regions are useful guides to where SWCX emission will be strong, but cannot be used to accurately quantify that emission.

Because the cusp regions are magnetically connected to the magnetopause where magnetic reconnection occurs, they are of great interest to space physicists. {\it CuPID}, a cubesat-scale mission to launch in 2019, will study the X-ray emission in the cusps and characterize the angular size of the cusps and the emission strength in the $\sim\frac{1}{4}$ keV band as a function of solar wind conditions.

\section{The Neutrals in the Heliosphere \label{sec:neut_hs}}

\begin{table}
\caption{ISM Flow Parameters$^a$}
\label{tab:ism}
\begin{tabular}[]{lcc}
\hline
Parameter & H & He \\
\hline
Upwind $(\lambda,\beta)$ & $(252.3\arcdeg,8.5\arcdeg)$ & $(254.7\arcdeg,5.3\arcdeg)$ \\
Upwind $(\ell,b)$ & $(5.1\arcdeg,19.6\arcdeg)$ & $(3.4\arcdeg,15.9\arcdeg)$ \\
$n$ & 0.1 cm$^{-3}$ & 0.015 cm$^{-3}$\\
$v$ & 21 km s$^{-1}$ & 26.2 km s$^{-1}$ \\
$T$ & 13000 K & 6300 K \\
\hline
\end{tabular}

$^a$ {Values taken from \citet{koutroumpa_etal_2006} and the sources therein.}
\end{table}

\subsection{The Model \label{sec:enlil}}

\begin{figure*}
\center{\includegraphics[width=6.75cm,angle=0]{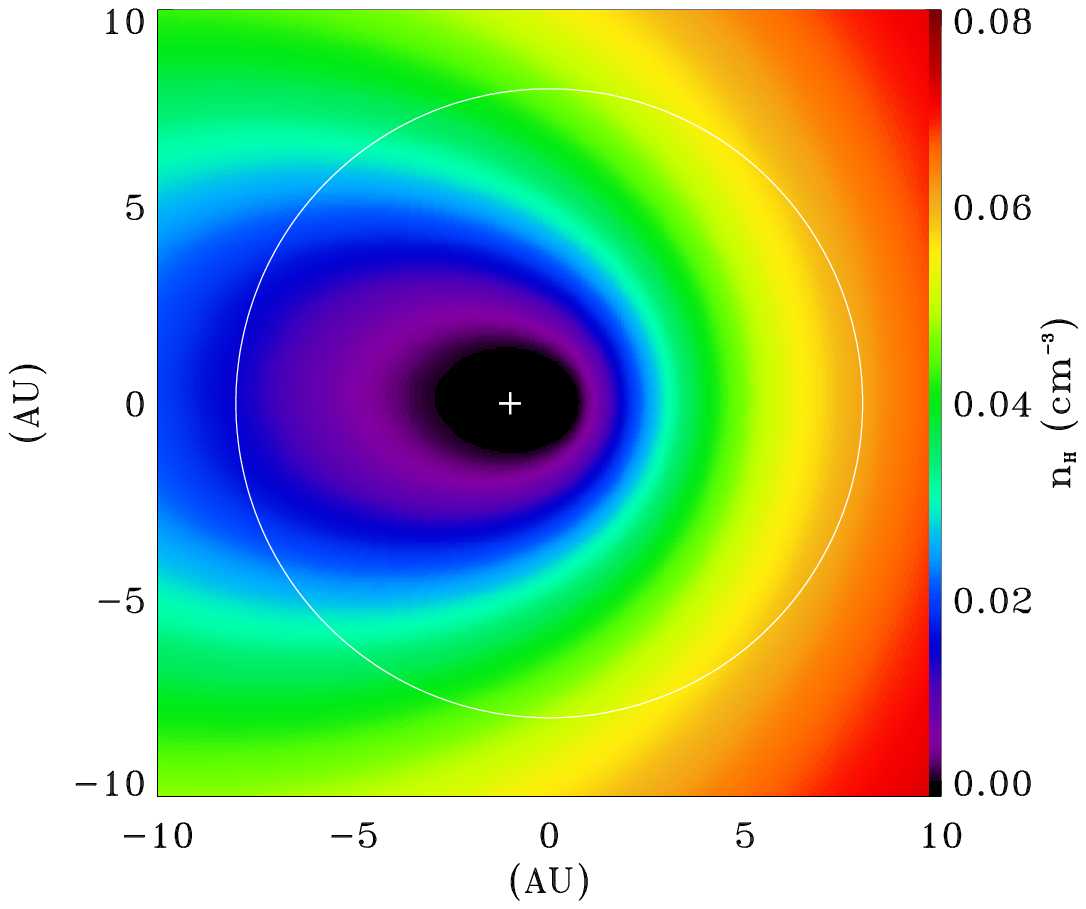}\hspace*{-0.75cm}
\includegraphics[width=6.75cm,angle=0]{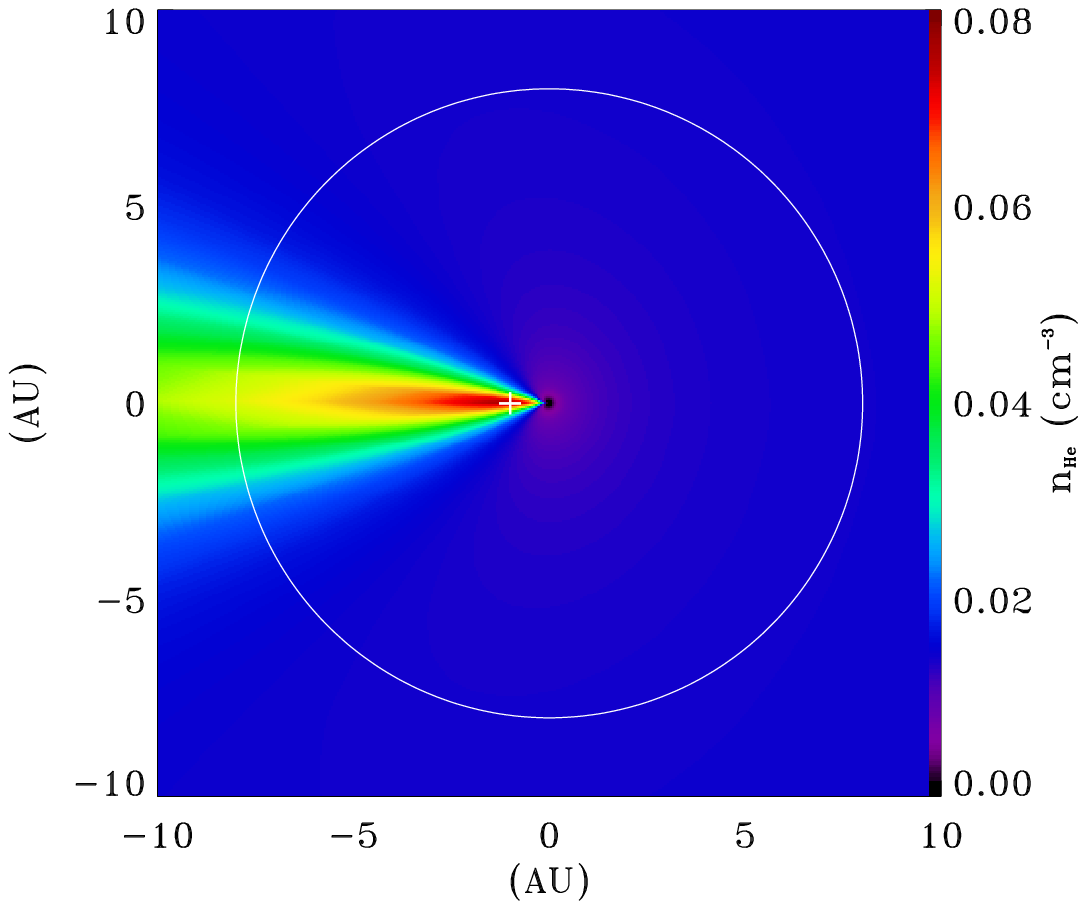}}
\center{\includegraphics[width=6.75cm,angle=0]{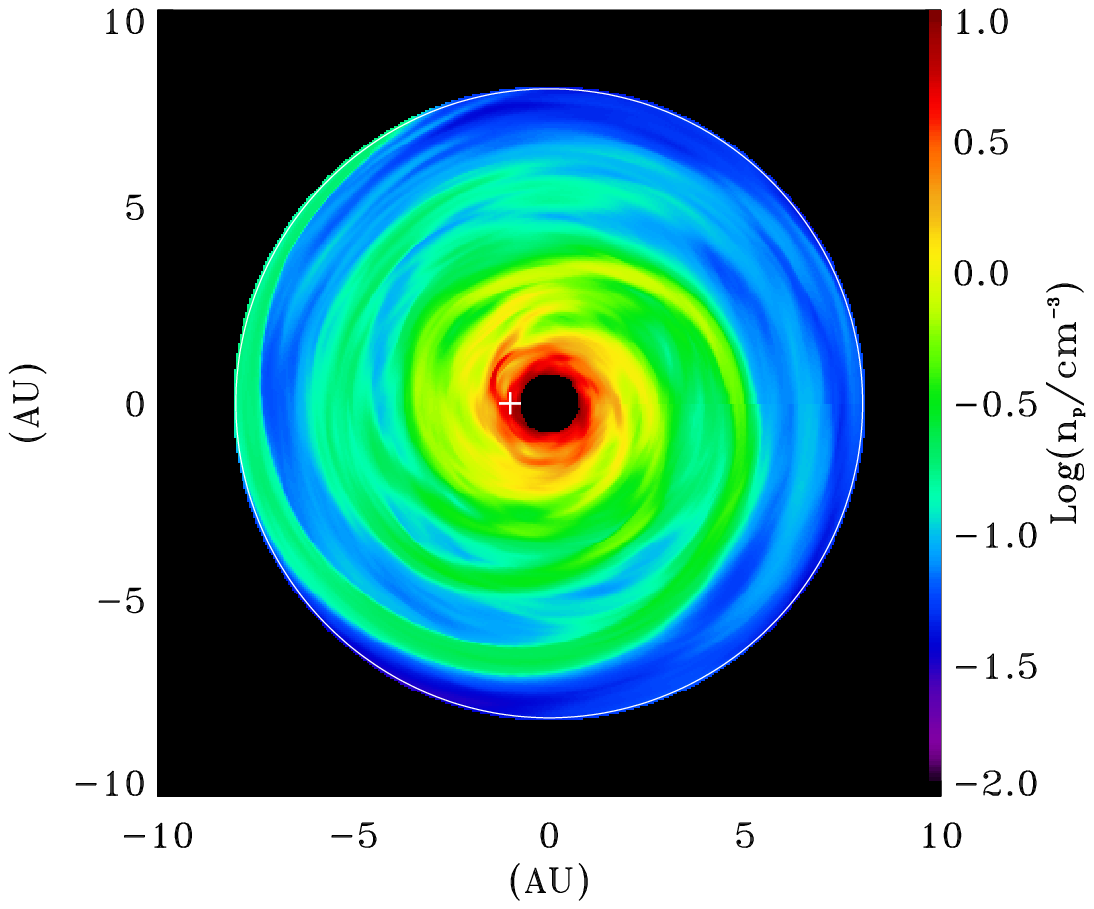}\hspace*{-0.75cm}
\includegraphics[width=6.75cm,angle=0]{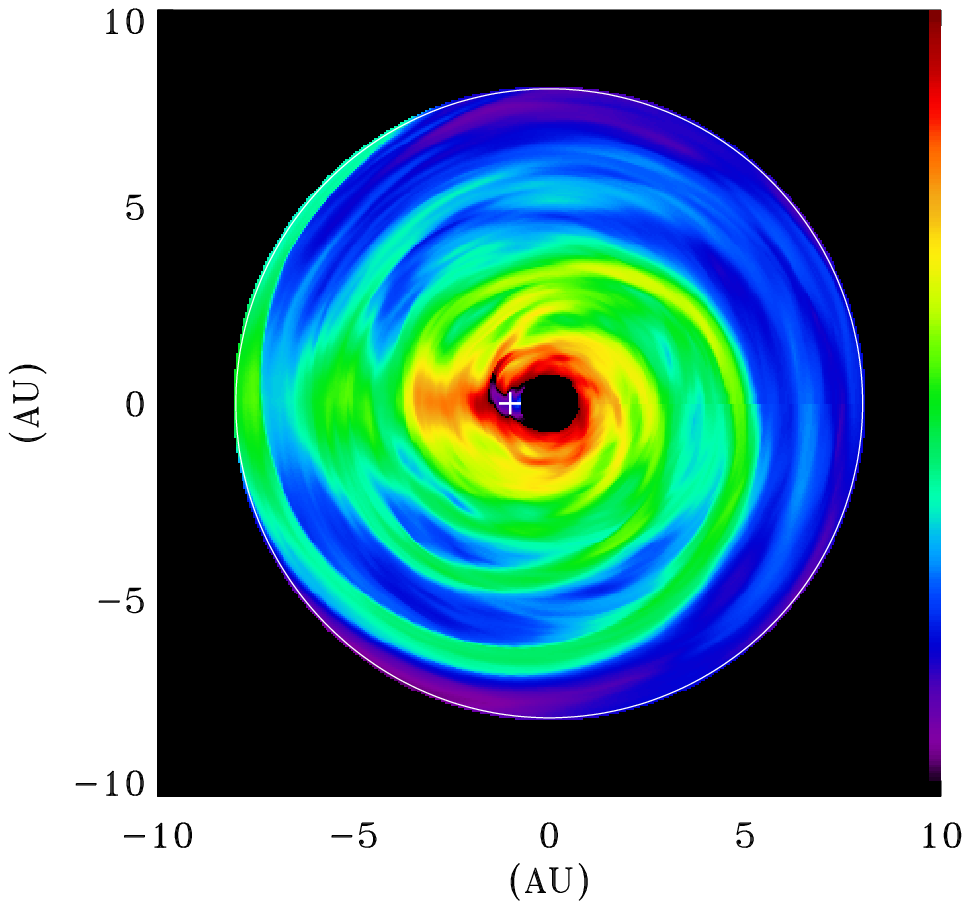}}
\caption{The distribution of the neutral and proton densities in the solar rotational plane. Each plot is centered on the Sun and the location of the Earth on 7 December is marked. The upwind direction is to the right. {\it Upper Left: } the hydrogen distribution calculated from a hot Fahr-type model by D. Koutroumpa for solar minimum. Note the deficit of neutral hydrogen in the inner solar system. {\it Upper Right:} The helium distribution calculated from a hot Fahr-type model by D. Koutroumpa for solar minimum. Note the helium focussing cone. {\it Lower Left:} The solar wind proton distribution from the ENLIL model. The center 0.75 au radius region has been masked out for display purposes. The model extends to 8.08 au, which is marked by the white circle in all panels. The Parker spiral is readily visible. {\it Lower Right: } The X-ray emissivity calculated from $n_p v_{rel} (\sigma_H n_H + \sigma_{He} n_{He}$).The coordinate system is Heliospheric Numerical Model system, where the sun is at the center and the Earth is fixed at 180$\arcdeg$.}
\label{fig:fahr_demo}
\end{figure*}

The Sun moves with respect to the ISM. As a result, there is a heliopause and a bow shock in the direction of the solar motion with respect to the ISM. As with the solar wind interacting with the Earth's magnetic field, the ionized fraction of the ISM does not penetrate the heliopause, while the neutral component flows through\footnote{The direction of the Sun's motion with respect to the local standard of rest is towards $(\ell,b)=(47\fdg8,23\fdg7)$ while the upwind direction is $(\ell,b)=(5\fdg6,19\fdg57)$ though the neutrals are not entirely undeviated. Since the upwind direction is so close to the Galactic center, X-ray studies of charge exchange emission from the nose of the heliosphere are infeasible. \citet{frisch_1996} has drawn attention to this peculiarly infelicitous alignment which prevents the X-ray study of the nose of the heliosheath.}. Modeling the distribution of neutral ISM particles as they flow through the solar system has a long history starting with \citet{fahr_1968}, with further developments by \citet{fahr_1971}, \citet{lallement_etal_1985}, and many others. These ``hot'' models are based on a distribution of particle orbits modified by various loss processes. More recently, gasdynamic models have been used, which are claimed to provide greater accuracy, though at the cost of greater complexity  \citep{katushkina_izmodenov_2010, katushkina_etal_2013}. The traditional ``hot'' models are described below because they have been the workhorse for SWCX calculations. The inputs to those models are summarized in Table~\ref{tab:ism}.

It is generally assumed that the neutral atoms, having crossed the heliopause, can be described as a uniform parallel flow with a particular temperature. In reality, the situation is likely more complex \citep{katushkina_izmodenov_2010}. Neutral hydrogen and helium atoms entering the heliosphere are generally modeled as having hyperbolic orbits under the influence of the solar gravity and radiation pressure. Since both of these forces are proportional to $r^{-2}$, one generally scales the gravitational force by $(1-\mu)$, where $\mu$ is defined as the ratio of the radiation pressure to the gravitation and varies from 0.9 to 1.6 \citep{woods_etal_2000} for hydrogen, and is insignificant for helium. When $\mu>1$, typical of solar maximum, the net force is repulsive and hydrogen is prevented from entering the inner solar system. Helium, on the other hand, experiences only an attractive force, causing it to be ``focussed'' in the down-wind direction. The distance from the Sun to the nose of the heliopause is roughly 122 au, though it clearly depends upon the solar wind pressure \citep{burlaga_ness_2014, washimi_etal_2017, gloeckler_fisk_2016}. Assuming a mean ISM speed of 28 km s$^{-1}$, it takes $\sim$28 years for an interstellar neutral to travel from the heliopause to the Sun, or roughly three solar cycles. 

Assuming a uniform density and a parallel flow of neutrals upstream of the heliosphere and an average value of $\mu$, one can derive a closed form solution for the density of particles flowing through any particular volume element \citep{fahr_1971,lallement_etal_1985}. This is the ``cold'' model. The ``hot'' model assumes that the input neutrals have a distribution in speed and direction; it is a weighted superposition of multiple ``cold'' models with different input speeds and upwind directions. Although early forms of the ``hot'' model assumed an isotropic distribution of the thermal component, anisotropy is readily implemented in this model. 

There are a number of loss processes that must be considered as well. The primary losses are due to photoionization and charge exchange with solar wind protons. Photoionization by the solar UV emission will be a function of $r^{-2}$. The losses due to charge exchange are also a function of $r^{-2}$, but since the charge exchange is a function of the solar wind flux, they will necessarily be a function of solar latitude and the phase of the solar cycle. During solar minimum the total ionization rate for hydrogen at the poles is 60\% of that observed at the equator \citep{quemerais_etal_2006}. The total ionization rate for helium is roughly an order of magnitude smaller than that of H. Thus, even when hydrogen experiences a net attractive force, it has a low density in the inner solar system due to ionization. A third loss mechanism is electron impact ionization which is latitude and solar cycle dependent \citep{mcmullin_etal_2004,rucinski_fahr_1989}. 

The resulting distribution of the density of neutral hydrogen and neutral helium in a plane containing the Sun and the upstream direction and close to the solar equator is shown in Figure~\ref{fig:fahr_demo}.

The hot Fahr model is conceptually very simple and has a limited set of input parameters ($n$, $v$, and $\vec T$ for each species) which have been fitted to match the observed data; the Ly$\alpha$ backscatter \citep[e.g.][and many others]{costa_etal_1999}, He I 584 \AA\ backscatter, neutral helium, and helium pickup ions \citep{mobius_etal_2004}. The complexity arises in modeling the sources of ionization, which has been done in part by modeling the asymmetries in the Ly$\alpha$ backscatter \citep[e.g.][]{bzowski_etal_2003}. Due to continued refinement by successive data, the hot Fahr model is quite robust. More recent MHD-based models of the neutral atom flow would appear to be consistent with the hot Fahr model \citep{katushkina_izmodenov_2010} given an non-Maxwellian input distribution. However, it should be noted that the loss processes depend upon the solar wind and thus upon, in particular, high solar latitude solar wind data. 

\subsection{The Heliopause}

The heliopause, like the magnetopause, is roughly parabolic, with a standoff distance set by the time-variable solar wind pressure and, potentially, the variation in the ISM pressure. Voyager 1 is thought to have crossed the heliopause at 122 au \citep{krimigis_etal_2013,burlaga_etal_2013,stone_etal_2013,webber_mcdonald_2013}\footnote{These papers would form an interesting study in the history and philosophy of science. None of them {\it actually} claims a heliopause crossing, just the abrupt crossing into a region that is clearly not the inner heliopause. However, later papers cite these in retrospect in terms of the heliopause crossing.} but there is still some disagreement \citep{gloeckler_fisk_2016}. Calculation shows that the bulk of the ions in the solar wind do not charge exchange by the time they reach the heliopause. Using the upwind neutral model from Koutroumpa for solar minimum and assuming cross sections between O$^{+7}$ and hydrogen of $5.5\times10^{-15}$, and helium of $1.8\times10^{-15}$, then roughly 80\% of O$^{+7}$ ions survive their trip to the heliopause . If all of the surviving ions recombined upon crossing the heliopause, the emission observed from Earth would be a fraction of a percent of the emission seen due to recombination occurring between the Earth and the heliopause. Thus the heliopause is not a significant source of emission.

%o7 x+y+z+w sigma=55e-16 for 400 m/s
%o8 40e-16

Although it is quite clear that the solar wind ions do not suddenly recombine when they hit the heliopause, our understanding of what happens is, perhaps, less clear now than it was a decade ago. The Voyager results suggest that the heliopause is a much more complicated structure than previously thought, as can be seen in the different approaches taken in the 2013 papers \citep{krimigis_etal_2013,burlaga_etal_2013,stone_etal_2013,webber_mcdonald_2013} on the Voyager 1 crossing event. It is generally assumed that the solar wind ends its radial flow at the termination shock and is diverted to flow away from the nose of the heliosheath behind the heliopause. The solar wind will, no doubt, continue to charge exchange during this later flow as well.

%\clearpage

\section{The Neutrals in the Magnetosphere \label{sec:neut_ms}}

\subsection{The Model}

\begin{figure}
\center{\includegraphics[width=6.4cm,angle=0]{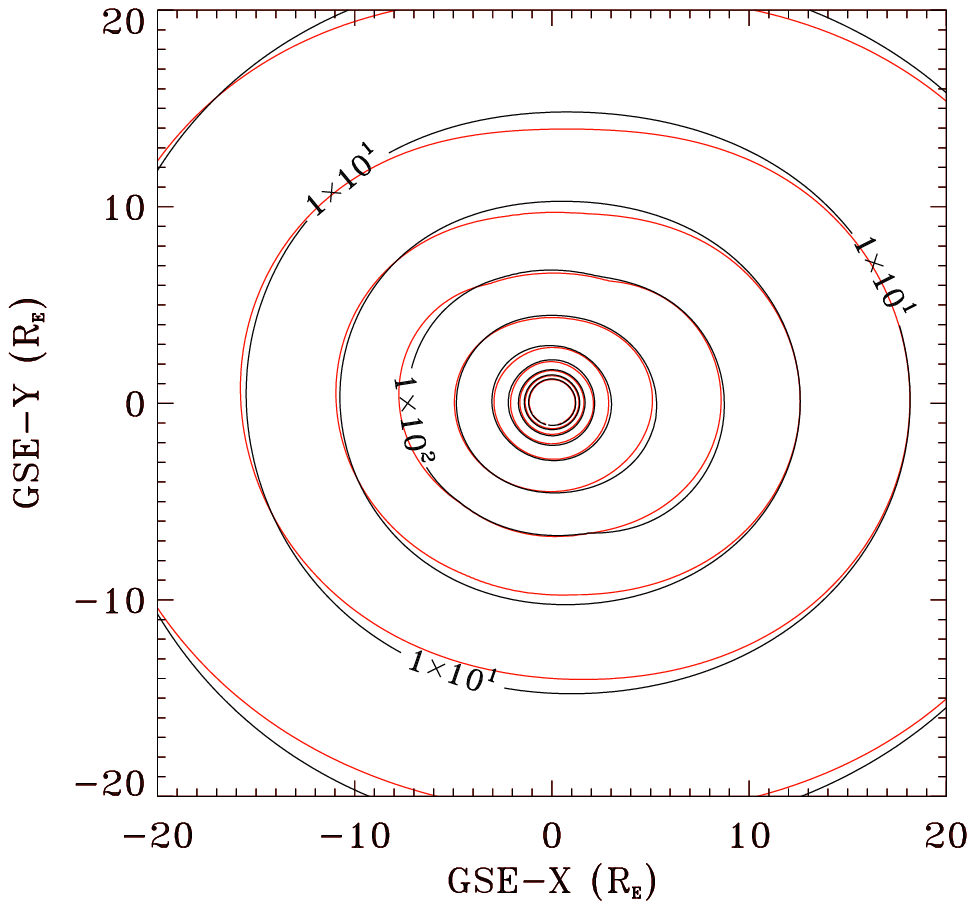}\hspace*{-0.75cm}
\includegraphics[width=6.4cm,angle=0]{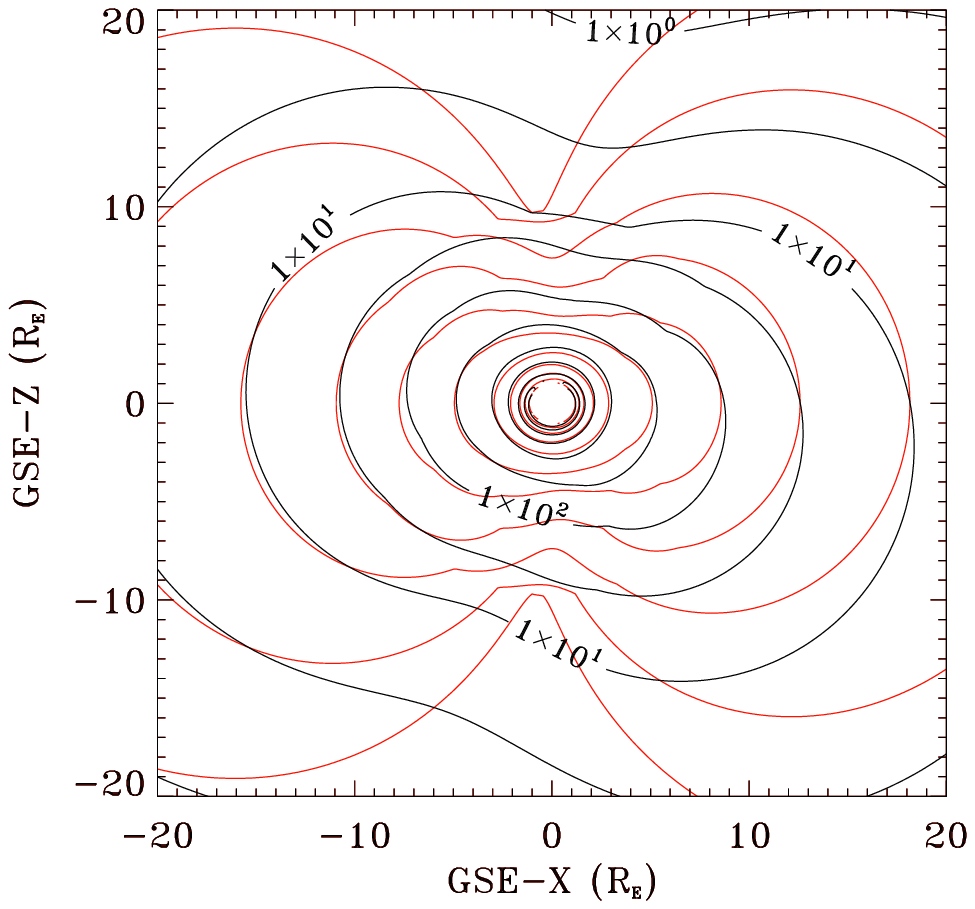}}
\caption{The exospheric hydrogen density from the Hodges model. {\it Left: } in the XY$_{GSE}$ plane. {\it Right: } in the XZ$_{GSE}$ plane. In both panels the {\it red} contours are for the equinox and the {\it black} contours are for the solstice.}
\label{fig:hodges_demo}
\end{figure}

The exosphere is the region in which the particle density is too low for the particles to act like a true gas. That is, the particle motion is not dominated by collisions but by ballistic motions. This is not to say that the region is collisionless, but that the particles move substantial distances between collisions. The Earth's exosphere is dominated by hydrogen. This hydrogen is subject to escape and is replenished from below. However, there are numerous source/loss mechanisms including chemical reactions, charge exchange, and ionization. The equilibria of these processes depend upon the solar ionizing flux as well as the total insolation upon the atmosphere. The primary information about the distribution of hydrogen comes, not from data, but from Monte Carlo simulations of the trajectories and interactions of individual atoms \citep{hodges_1994}.

The resulting model of the exosphere is a function of the insolation (as characterized by the 10.7 cm solar flux denoted `f10.7') and the terrestrial season. The Hodges model is specified in terms of a multipole expansion for radii from an altitude of 480 km to 10 R$_E$\footnote{Should any reader wish to implement this model, be aware that the plots in \citet{hodges_1994} were calculated with a higher order model than the model whose coefficients are published!}. The distribution is roughly spherical, with higher densities in the anti-solar direction and, to a lesser extent, the solar direction. The asymmetry is stronger at the solstices than at the equinoxes. The asymmetry decreases as f10.7 increases. Figure~\ref{fig:hodges_demo} displays XZ$_{GSE}$ and XY$_{GSE}$ plane cuts at the equinox and solstice. 

Data with which to confirm this model are scarce. \citet{hodges_1994} compared the model to a limited amount of Balmer-$\alpha$ data and found that the model matches reasonably well at low f10.7, but underestimates the density at high f10.7. More recent measurements of the hydrogen distribution using Lyman-$\alpha$ scattering by the exosphere shows the same asymmetries predicted by the Hodges model. \citet{bailey_gruntman_2011} also showed that during solar minimum the scattered solar Lyman-$\alpha$ measured by the TWINS mission is consistently lower than predicted by the Hodges model in the range 3 R$_E<R<8$ R$_E$, but only by a factor of $\sim$1.5. \citet{zoennchen_etal_2013} showed that during solar minimum agreement between TWINS measurements and the Hodges model was very good at the equinox, but poorer at the solstices, when the Hodges model was a factor of 1.8 too high. The same authors made a similar study at solar maximum \citep{zoennchen_etal_2015}. Although they did not make a comparison with the Hodges model, they do show that the measured density at solar maximum is a factor of two higher than at solar minimum. It is difficult to know precisely what the Hodges model should predict, but assuming an increase in f10.7 from solar minimum to solar maximum, one finds that the Hodges model predicts the density at solar maximum to be lower than at solar minimum by a factor of (very roughly) 0.7. Thus there are clear discrepancies between the Hodges model and measurements, but only by factors of two, at least for the seasons and insolations tested.

%\clearpage

\section{Atomic Data \label{sec:atom}}

\subsection{Theory}

\begin{figure*}
\center{\includegraphics[width=6.5cm,angle=0]{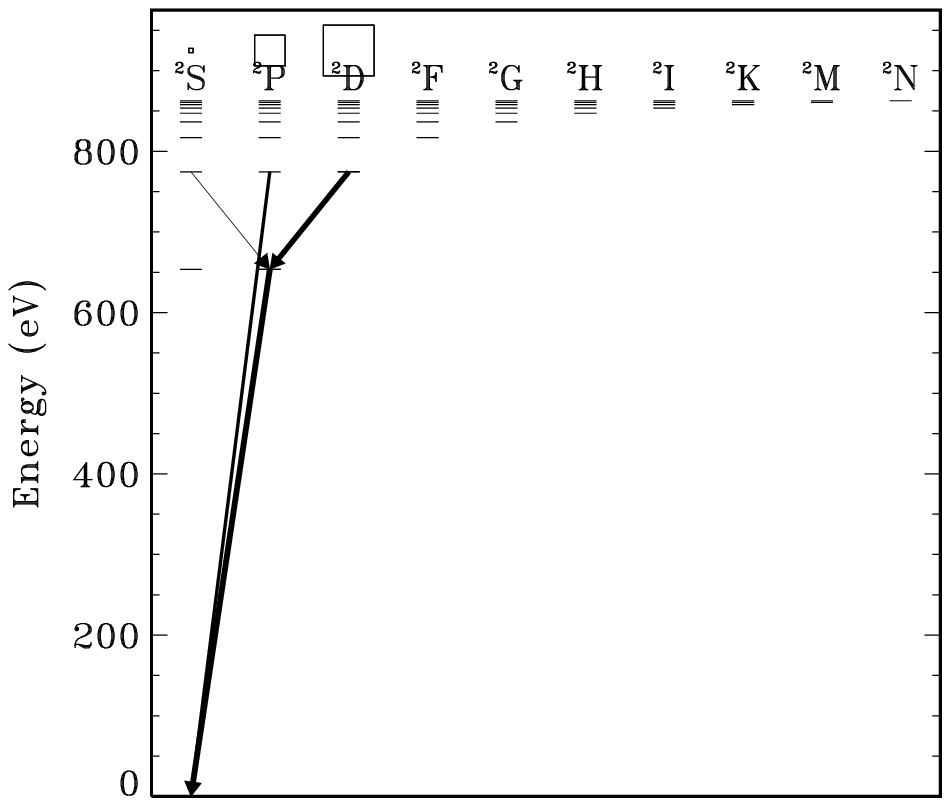}\hspace*{-0.75cm}
\includegraphics[width=6.5cm,angle=0]{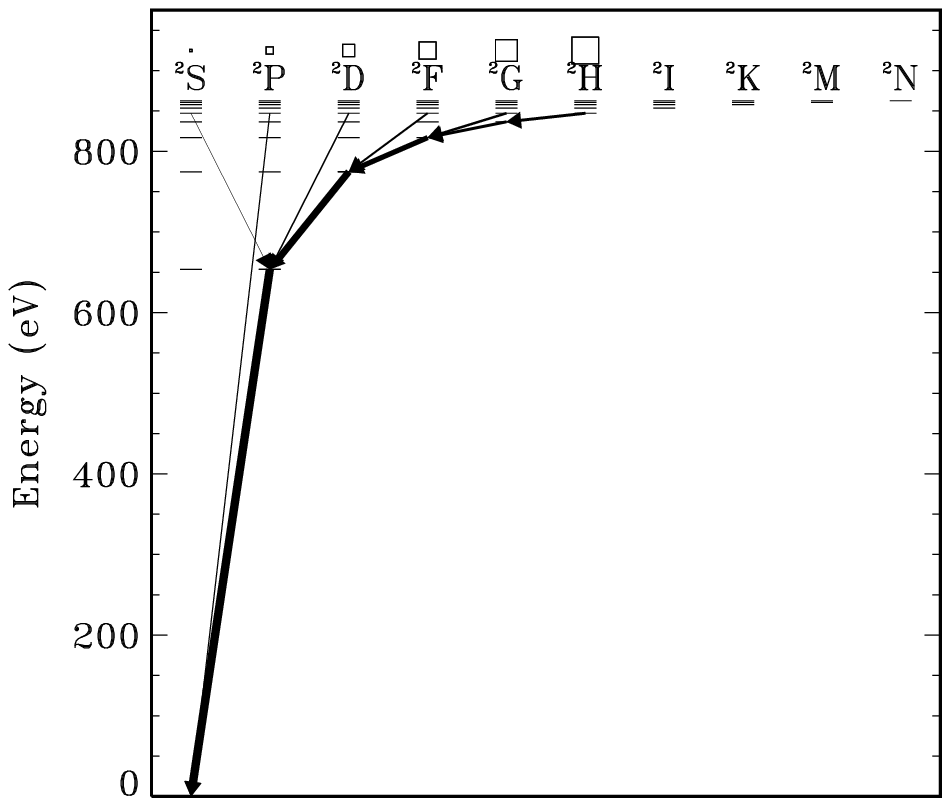}}
\vspace*{-0.5cm}
\caption{Grotrian diagrams for hydrogen-like oxygen. {\it Left:} schematic plot of transitions for low-v collisions where $n_i=3$ as calculated from Equation~\ref{eqn:jw}. For a real case there would be some small contribution from $n_i=4$ states. {\it Right:} schematic plot of transitions for high-v collisions. Here $n_i$ was set to 6. In both cases the sizes of the boxes shows the the relative $l$-distribution while thickness of the arrow indicates the approximate relative strength of the line.}
\label{fig:grot_h}
\end{figure*}

As suggested by \S\ref{sec:motiv}, the charge-exchange cross sections remain a significant source of uncertainty as many remain unmeasured and many calculations do not adequately reflect the peculiar behavior of real atoms. As there have been a number of extensive articles on the physics of charge exchange \citep{wargelin_beiersdorfer_brown_2008,krasnopolsky_etal_2004,betancourt_2017}, those results will be summarized here only so far as they are needed for the discussion of the current state of atomic data and the spectroscopic tools that use it.

First, we must distinguish between the total cross section (the probability that an electron is transferred in an interaction with a given collisional energy) and the $n,l$ resolved cross section (the probability that an electron is transferred to a particular $n,l$ state in an interaction with a given collisional energy).

The total cross sections are dependent upon the energy of the interaction, but that dependence can be relatively weak over an extended range of collision energy. For typical solar wind velocities (see Table~\ref{tab:solarwind}) the bulk velocities correspond to 0.5-1.8 keV/amu. The measured temperatures in the free-flowing solar wind corresponds to 0.0001-0.001 keV/amu (for oxygen) while temperatures within the magnetosheath can rise to 4-5$\times10^{6}$ K for typical solar wind values ($kT=0.34-0.43$ keV or 0.02-0.025 keV/amu for oxygen). Thus, for typical solar wind values, the collision energies fall within a fairly narrow range. However, it should be noted that the total cross section can be strongly dependent upon the neutral target \citep{beiersdorfer_etal_2003}. While this is vitally important for the study of solar wind charge exchange with comets and planetary atmospheres (other than the Earth's), it is not as important for the charge exchange that contaminates astrophysical observations, which is due primarily to solar wind interactions with hydrogen and helium.

Total cross sections are useful when determining the fraction of a solar wind ion that recombines over a given trajectory. However, for understanding the contamination of astrophysical observations, one needs to know individual line strengths, which requires knowing the probability of charge exchange into an initial state $n_i,l_i$, and then the selection rules and other atomic data required to determine the fraction of recombinations to $n_i,l_i$ that produce the line of interest. Much of the atomic data for higher $n$ values has been recently added to spectral models such as APEC in response to the need for charge-exchange spectra.

It is generally accepted that the principal quantum number $n_i$ to which there is the greatest probability of transfer is given by the approximation from \citet{janev_winter_1985}:
\begin{equation}
n_i\sim q\left(1+\frac{q-1}{\sqrt{2q}}\right)^{-1/2}\left(\frac{I_H}{I_n}\right)^{1/2}
\label{eqn:jw}
\end{equation}
where the distribution of $n_i$ is relatively narrow for the collision energies typical of SWCX. $I_H$ is the ionization potential of hydrogen and $I_n$ is the ionization potential of the neutral target, both in atomic units, and $q$ is the charge of the ion. This formula is probably good to $\pm1$, though larger discrepancies have been noted \citep{betancourt_2017}. The formula lacks an energy dependence of $n_i$, while measurement shows that $n_i$ does depend on energy, albeit weakly.

Unlike the total cross section or the initial principal quantum number, the initial distribution of $l_i$, the orbital angular momentum, depends sensitively upon the collision velocity. Here the collision velocity is usually compared to the classical velocity of a bound electron. At low collision velocities, low angular momentum values are favored while at higher collision velocities the higher angular momentum values are favored. The former may be represented by one of several different functions, while the latter is represented by the statistical weights \citep[see][for discussion]{smith_etal_2012,gu_etal_2016}. 

For bare nuclei becoming hydrogen-like ions through charge exchange, the distribution of input $l_i$ values has important consequences. If a low $l_i$ is favored, the radiative decay path is either directly from $n_i,l_i=1$ to $n=1,l=0$ producing a Lyman series photon, or transition from $n_i,l_i$ ($l_i$ being 0 or 2) to $n=2,l=1$, followed by the emission of a Lyman-$\alpha$ photon. If a high $l_i$ is favored, then the radiative decay path usually starts with a $n_i,l_i$ to $n=l_i,l=l_i-1$ transition followed by a Yrast cascade, where the bulk of the emission is due to $\Delta n=\Delta l=-1$ transitions leading to a Lyman-$\alpha$ photon. 

For a hydrogen-like ion becoming a helium-like ion through charge exchange, the difference between low-$l_i$ and high-$l_i$ favored initial states produces similar differences. However, here attention focuses on the ratio of entry into triplet states to singlet states. Electrons that begin in singlet states are likely to produce a K$\alpha$ line (the resonance or $w$ line). Electrons that begin in a triplet state cascade to either the $n=2,l=0$ state or a $n=2,l=1$ state, followed by a transition to the ground level singlet state producing a $z$ (forbidden), $y$, or $x$ (intercombination) line. For radiative recombination, entry into triplet states is preferred due to their larger statistical weight, decreasing the strength of the $w$ line compared to a plasma in collisional ionization equilibrium. 

\begin{figure}
\center{\includegraphics[width=6.5cm,angle=0]{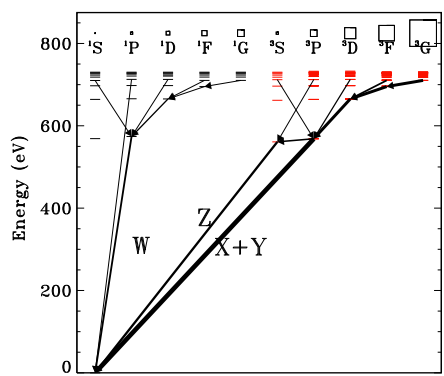}}
\vspace*{-0.5cm}
\caption{Grotrian diagram for helium-like oxygen with a schematic plot of transitions for high-v collisions. The thickness of the arrow indicates the approximate relative strength of the line. The triplet states are shown in red. Here we have assume a 3:1 ratio of triplet:singlet statistical weights. This is a vexed issue, see \citet{krasnopolsky_etal_2004} for further discussion.}
\label{fig:grot_he}
\end{figure}

\subsection{Tools}

The current choices for constructing SWCX spectra are the models produced at CfA \citep[available through APEC,][]{smith_etal_2012, smith_etal_2014} and those produced at SRON \citep[available through SPEX,][]{gu_etal_2016}. Both  groups assume that the total cross sections are relatively simple functions of the charge of the ion. The SRON method of calculating the total cross sections is slightly more sophisticated as the total cross section is taken to be a function of the collision energy as well. Both groups use the \citet{janev_winter_1985} function (Equation~\ref{eqn:jw}) to describe the distribution of the principal quantum number, $n_i$, of the initial state of the ion after charge exchange. Both groups implement a variety of {\it ad hoc} functions \citep[also from][]{janev_winter_1985} to describe the distribution of the initial electron orbital angular momentum ($l_i$). Here too the SRON  method is slightly more sophisticated: by comparing the simple analytic functions from \citet{janev_winter_1985} to {\it theoretical calculations} of $l$-distributions they determined that the best function for the $l$-distribution is a function of the collision energy. So the SRON implementation selects which $l$-distribution to use depending upon the ion velocity and the principal quantum number, while the CfA implementation applies the same user-selected $l$-distribution to all species. \citep[See the more detailed discussion of this issue in][]{betancourt_2017}.

Given what is known from laboratory measurements, production of the SWCX spectrum due to solar wind of a particular speed and type is not a straight-forward process for either implementation.
%neither implementation produces a ``spectrum'' of the SWCX emission due to solar wind of a particular type or speed in a straight-forward manner. 
The CfA implementation provides line ratios for a given species/velocity, given a choice of $l$-distribution, but that $l$-distribution will not be applicable to all the other species. The SRON implementation, while more sophisticated, has the same problem. By forcing the user to choose the $l$-distribution, the CfA implementation does remind the user of its limitations, while the SRON implementation allows its sophistication to obscure its limitations. Neither implementation handles element-dependent and state-dependent\footnote{As shown in Figure~\ref{fig:freeze-in}, the ionization temperatures calculated from the ratio of the number of ions in state $n+1$ to the number in state $n$ are a function of $n$. Thus using a single ionization temperature for an element could be problematic if there a number of different ionization states represented within the bandpass of interest.} ionization temperatures, which are essential for modeling the SWCX spectrum. For a given bandpass, most elements have strong lines from only a few ($<3$) ionization states. Thus, one can build up a relatively good approximation of the spectrum on a species-by-species basis, using a separate normalization and ionization temperature for each element. Some of these issues are discussed in \citet{henley_shelton_2015}.

\subsection{Laboratory Astrophysics}

\begin{figure}
\center{\includegraphics[width=8.25cm,angle=0]{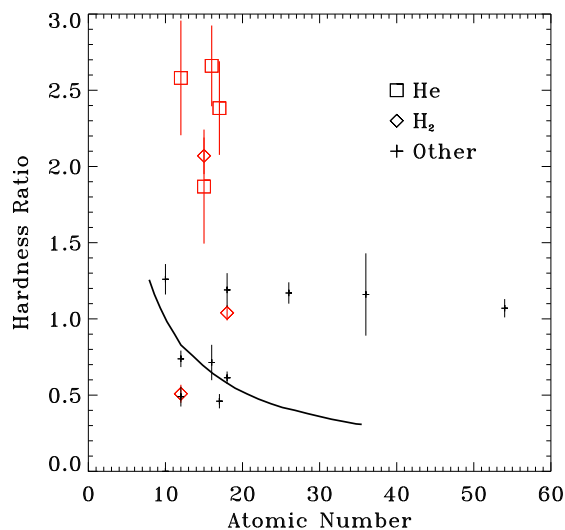}}
\caption{The hardness ratio ${\mathcal H}$ as a function of atomic number measured for hydrogen-like ions in experiments with collision energies $\lesssim25$ ev amu$^{-1}$. The measurements made with helium and H$_2$ as the neutral target are shown by boxes and diamonds respectively. The other points were measured using a variety of targets. The curve is the CTMC calculation for interactions with neutral hydrogen from \citet{wargelin_etal_2005}. This plot and the plotted data are adapted from \citet{betancourt_etal_2014}. Apparent are the differences between measurements (using H$_2$ and helium) and theory (calculated for atomic hydrogen). Also apparent are the differences between elements when when interacting with the same neutral target species.}
\label{fig:betan}
\end{figure}

Any model for charge exchange emission will be based on a combination of data, theory, and scaling.  How problematic is this? The most simple theoretical calculations that are still capable of providing $n_i,l_i$-resolved cross sections (such as classical trajectory Monte Carlo or CTMC calculations) show relatively good agreement with experimental values at high collision energies \citep[$\lesssim5$keV][though the agreement is not always great]{ali_etal_2010}, but poor agreement at the low collision energies typical of SWCX \citep[i.e.,][]{beiersdorfer_etal_2000}. A typical measure with which to compare theory and measurement is the hardness ratio,
\begin{equation} 
{\mathcal H} = \frac{\sum_{n=3}^{\infty}F_{n\rightarrow 1}}{F_{2\rightarrow 1}}
\end{equation}  
which is the ratio of the sum of the high-order Lyman series lines to Lyman-$\alpha$. \citet{beiersdorfer_etal_2000} showed that at low collision energies ($E<15$ eV/amu), whereas CTMC predicts ${\mathcal H}<1$ and $\mathcal H$ to decline with Z, measured values for hydrogen-like Ne, Ar, Kr, Xe, Au, and U were all somewhat greater than unity. \citet{leutenegger_etal_2010} measured $\mathcal H$ simultaneously for hydrogen-like Ar and P and measured ${\mathcal H}_{Ar}=1.04\pm0.05$ and ${\mathcal H}_P=2.07\pm0.12$ which demonstrates that similar elements have very different $l$-distributions even given the same neutral target under the same conditions. \citet{betancourt_etal_2014} measured $\mathcal H$ for hydrogen-like Mg, S, Cl, and Ar and found values of $\sim$2 for several species interacting with helium. They found the same species to have ${\mathcal H}$ lower than expected from CTMC predictions when interacting with H$_2$.  \citet{otranto_etal_2006} showed that the measured $\mathcal H$ for hydrogen-like O was lower than CTMC predictions, but here they were using H$_2$O as the neutral target. Even after the application of the following caveats, it is clear that CTMC does not reflect the real physics and one cannot simply scale by Z \citep[as demonstrated by][]{leutenegger_etal_2010} for which there are now multiple examples \citep{betancourt_etal_2014}. These results are summarized in Figure~\ref{fig:betan}.

One possible explanation for the discrepancy is that the neutral targets are usually not H, which is the target for which CTMC calculations are done. In most cases, the neutral target was chosen to have a first ionization potential similar to that of H, and a much larger second ionization potential to minimize multi-electron capture. However, even minimized, multi-electron capture is a significant possibility for many of the neutral targets used. This explanation is not likely to explain the bulk of the discrepancies; the \citet{leutenegger_etal_2010} measurements, made with H$_2$, showed some of the strongest discrepancies between similar elements under conditions most closely matching those required for CTMC calculations. That detailed comparisons of measured spectra to predictions using various $l$-weighting schemes show significant discrepancies should not be surprising since these {\it ad hoc} schemes do not include quantum mechanical or quasi-molecular effects.

More sophisticated and computationally complex methods exist for calculating the $n,l$ resolved cross sections, but they have not yet been deployed in any significant manner for this problem. \citet{cumbee_etal_2018} make an attempt in this direction using a variety of computation methods (QMOCC, AOCC, MCLZ, and CTMC) for various species and collisional energies. However, such calculations need to be checked against reality. Laboratory measurements continue, and the use of atomic hydrogen as the neutral target in a well characterized way is becoming more possible. From the results thus far, however, it is clear that a broad range of species will need to be measured since scaling from one species to another is not always successful.

\subsection{Broad Band Production Factors \label{sec:bb_prod}}

\begin{figure}
\center{\includegraphics[width=6.25cm,angle=0]{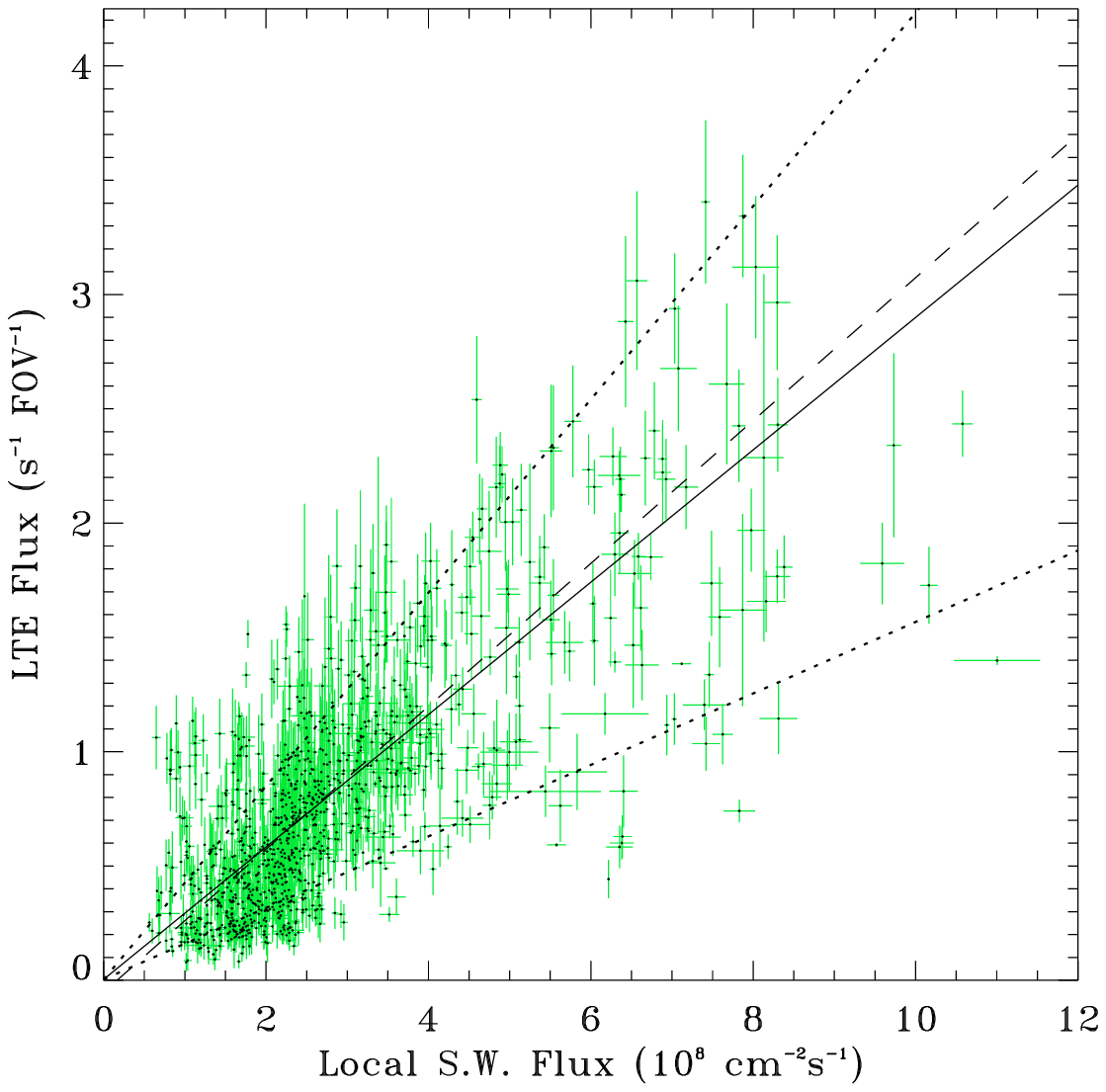}\hspace*{-0.5cm}
\includegraphics[width=6.25cm,angle=0]{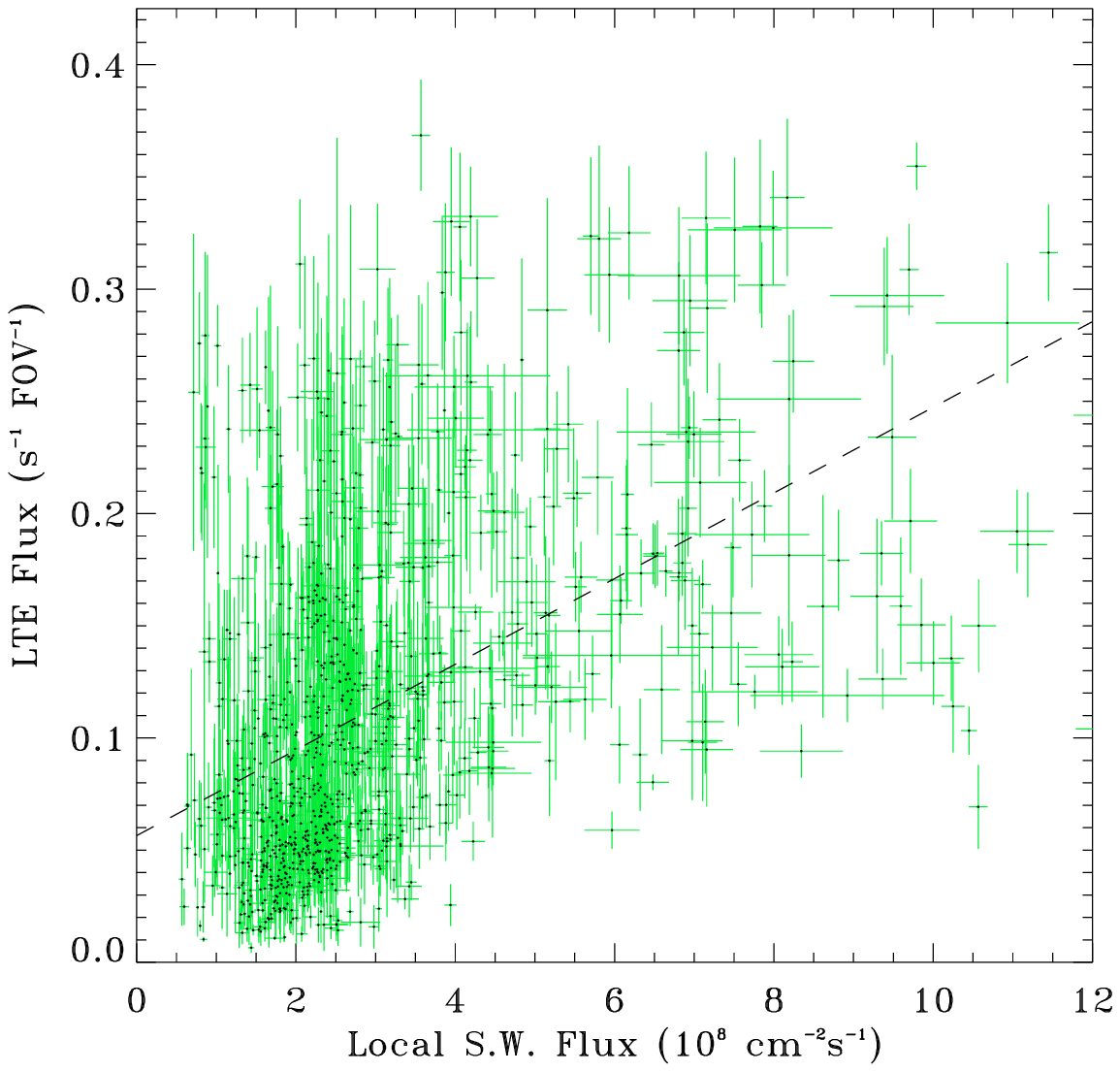}}
\caption{The correlation between \rosat\ Long Term Enhancement rates and the solar wind proton flux. {\it Left: } for the $\frac{1}{4}$ keV band, {\it Right: } for the $\frac{3}{4}$ keV  band. \citet{kuntz_etal_2015} demonstrated that the scatter in the $\frac{1}{4}$ keV LTE-$nv$ relation (dotted lines) was consistent with the scatter in the $Q$-$nv$ relation derived from MHD models. The scatter in the $\frac{3}{4}$ keV-$nv$ relation is due to the $\frac{3}{4}$ keV LTE flux being dominated by a few lines for which the ion densities were not measurable at the time.}
\label{fig:lte_nv}
\end{figure}

As noted in \S\ref{sec:contro}, understanding the amount of SWCX emission in the \rosat\ All-Sky Survey is central to resolving the issue of the Local Hot Bubble. Given the very broad bands or the \rosat\ PSPC, calculating the SWCX emission requires summing over many species and over an even greater number of lines. Two groups made such calculations and found production factors ($\varsigma_{R12}$) for the \rosat\ $\frac{1}{4}$ keV band that were within a factor of 2.5 of each other.

However, Snowden realized that one could measure $\varsigma_{R12}$ directly from the \rosat\ data. In his construction of the \rosat\ diffuse emission maps, he had removed, to the extent possible, the emission due to the Long Term Enhancements (LTEs) by comparing successive observations of the same location. Using only a portion of the available data, \citet{cravens_etal_2001} had shown that the LTE rates were well correlated with the solar wind proton flux. \citet{kuntz_etal_2015} used the correlation of the LTE rate and the solar wind flux together with the ``correlation'' between the $Q$ for \rosat\ lines of sight and the solar wind flux to derive $\varsigma_{R12}$. They also attempted to do the same for the $\frac{3}{4}$ keV band, but found that, since that band is dominated by the strong O VII and O VIII lines, and thus the O$^{+7}$ and O$^{+8}$ densities, neither of which were measured at the time, the correlation between the $\frac{3}{4}$ keV LTE rate and the solar wind proton flux was insufficient. Evidently, the fact that the $\frac{1}{4}$ keV band is composed of many lines from many species allows a much tighter correlation.

The observationally derived $\varsigma_{R12}=(3.86\pm0.20)\times10^{-20}$ count deg$^{-2}$ cm$^4$ is a production factor for the interaction of the solar wind with hydrogen in the Earth's exosphere. It is a factor of 1.8 higher than the theoretically derived value from \citet{koutroumpa_lbb} and a factor of 4.5 higher than the theoretically derived value from \citet{robertson_lbb}. Both of the calculated production factors were for the slow solar wind (i.e., higher abundances for the higher ionization states that produce the bulk of the X-rays, however, see the discussion in \S\ref{sec:conun}) while the measurements were taken during solar maximum, when the solar wind is a mixture of fast and slow. \citet{galeazzi_etal_2014} used the refurbished Wisconsin All-Sky Survey \citep{wass} proportional counters on a sounding rocket to measure the SWCX emission due to the helium focussing cone. By comparing that measurement with the \rosat\ measure of the same location on the sky, when it was not superposed upon the helium focussing cone, they were able to derive the $\varsigma_{R12}=(4.66\pm0.68)\times10^{-21}$ for helium. The large difference between $\varsigma_H$ and $\varsigma_{He}$, a factor of 8, is larger than expected from simple scaling arguments, a factor of 2 to 3, which may not be completely unreasonable, considering the above discussion.

%\clearpage

\section{Playing Before Working \label{sec:play}}

One hears the term ``toy model'' applied to simplified or over-simplified models of complex systems. It is often unclear whether ``toy'' is being used in the pejorative sense of ``useless'', or merely to imply that these models are useful for ``playing around'' to see which parameters are important. Since a ``real'' model of the SWCX emission would require models (and data) which may not yet exist as well as substantial computational effort, simplified models are useful for understanding various aspects of SWCX emission.

\subsection{A Simple Model of the Heliospheric Emission}

\begin{figure}
\center{\includegraphics[width=8cm,angle=0]{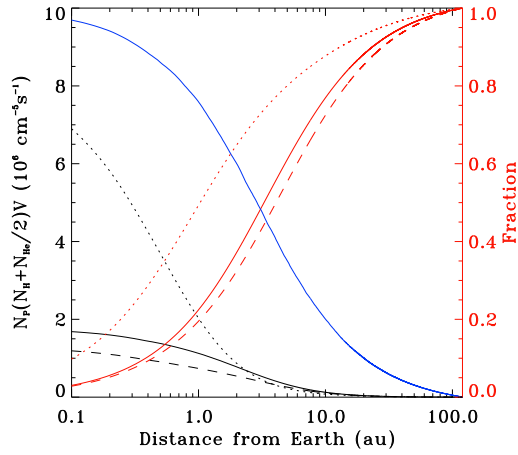}}
\caption{ The quantity $n_p n_n v_{rel}$, which is proportional to the emissivity, as a function of distance from the Earth for three directions, upwind ({\it black solid}), cross-wind ({\it black dashed}) and downwind ({\it black dotted}). In each case the emissivity was calculated from the Earth, radially outward from the Sun. The solar wind was assumed to have uniform $r^{-2}$ distribution with a density of 5 cm$^{-3}$ at 1 au from the Sun and with a constant speed of 435 km s$^{-1}$. The neutral density model is that calculated by D. Koutroumpa for solar minimum. The downwind value is very different from other directions due to the helium focussing cone; 20 degrees off of the focussing cone the curves are more similar to the cross-wind direction. The {\it red} lines show the cumulative fraction of SWCX emission reached as a function of distance for each direction. The {\it blue} line shows the temporal variance from the emission beyond $r$ compared to the variance observed from Earth (to be read on the right-hand scale). }
\label{fig:helio_emis}
\end{figure}

The model of the distribution of the neutral material alone provides useful information about the expected behavior of the heliospheric SWCX emission. We use here the results of a neutral model developed by Dimitra Koutroumpa who, very kindly, loaned us the output for solar minimum. If we assume that the solar wind proton density has a uniform $r^{-2}$ dependence and that the charge exchange with helium has an emissivity that is roughly half that of charge exchange with hydrogen (as one expects from theory), then one can build the simple model of the total emission shown in Figure~\ref{fig:helio_emis}. 

Each of the models shown assumes that one is observing from Earth in the anti-sunward direction. For most of the year (excluding December when one is looking down the helium focussing cone), $\sim$25\% of the SWCX emission comes from the nearest 1 au, $\sim$50\% arises from the nearest 3.3 au, and $\sim$75\% arises in the nearest 10.5 au. Thus, the bulk of the SWCX emission is relatively close. The total $Q=\int n_p n_n v_{rel} dl$ looking towards the anti-sun for this simple model is $9.5\times10^{19}$ cm$^{-4}$ s$^{-1}$ in the upwind direction, $7.3\times10^{19}$ cm$^{-4}$ s$^{-1}$ in the cross-wind direction, and substantially higher in the helium focussing cone. It should be noted that these values can be only approximate; we have not, for example, taken into account the shape of the heliopause.

%Similarly the bulk of the temporal variation in the SWCX emission arises close to the Earth; the temporal r.m.s. variation in the SWCX emission from $r>1$ au is 75\% of the variation observed from the Earth, and the variation from $r>2.8$ au is only 50\% of that observed from the Earth. The values for the variation were determined from simulations using the above simple model with the real solar wind density history measured in 1 hour ($\sim0.01$ au) bins, again assuming an anti-sunward look direction. 

\begin{figure}
\center{\includegraphics[width=8cm,angle=0]{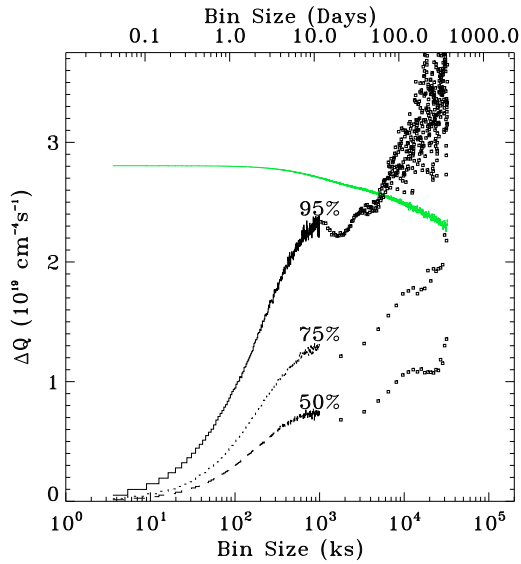}}
\caption{The variation in the SWCX emission as a function of the time bin size as calculated from the simple upwind model. Plotted in {\it black} is the upper limit of the value of $\Delta Q$ between two successive $n$ ks time bins at a given percentile. Presumably the dip at $\sim$20 days is due to the solar rotation period. Plotted in {\it green} is the RMS variation of $Q$ at each binning. The data were extracted from the OMNIWeb 1 hour data.}
\label{fig:delta_q}
\end{figure}

We can use this simple model to determine the typical variation in the heliospheric emission. We again use the upwind neutral distribution as, with the exception of the helium focussing cone, it provides an upper limit to the neutral density. We have used the OMNI database of solar wind proton density from 1995 to 2018, and have assumed a constant solar wind speed of 435 km s$^{-1}$. This model ignores the complexity of the evolution of corotating interaction regions and a number of other issues, but since we are looking radially outward, and the relative contribution declines rapidly with distance, we are probably justified in this simplification. We used the 1 hour resolution solar wind data, which provide a 0.01 au resolution at the mean solar wind velocity. For each time step we calculate $Q$ using the previous $1.2\times10^4$ hours of solar wind data. Once we have calculated $Q$ for each step ($\sim1.7\times10^5$ values), we bin the data by $n$ and calculate the $\Delta Q$ between each successive step. Figure~\ref{fig:delta_q} shows the 95th, 75th, and 50th percentile for the distribution of $\Delta Q$ for each bin size.

For a line of sight radially away from the Sun, the contribution to the SWCX emission of the last hour of solar wind is small compared to the integrated emission, and thus we expect low $\Delta Q$. At the 95th percentile $\Delta Q<0.5\times10^{18}$ cm$^{-4}$ s$^{-1}$. One expects $\Delta Q$ to increase with the bin size up to time scales comparable to the solar cycle. Indeed, given than no two solar cycles have the same strength, one would expect inter-cycle variation as well. The curve in Figure~\ref{fig:delta_q} should be at least representative of the variation at a given time scale for many lines of sight. Of course, this overly simple calculation is a lower limit to the variation that would be seen along the Parker spiral. Given that caveat, the 95th percentile $\Delta Q$ on month-scales is $2.3\times10^{19}$ cm$^{-4}$ s$^{-1}$, while for the 50th percentile it is $\sim7\times10^{18}$ cm$^{-4}$ s$^{-1}$.

\subsection{A Less Simple  Model of the Heliospheric Emission \label{sec:play_enlil}}

One of the key arguments of \citet{kuntz_etal_2015} was that the observed LTE/solar wind flux correlation was not due to heliospheric emission. They argued this point using an ENLIL simulation from a month during the RASS observations, which were made during solar maximum. (See their Figures 10 and 13.) They argued that the solar wind approaches from the direction of the Sun, but that the solar wind ``front'' is tilted by roughly $\pi/4$ due to the Parker spiral structure. If one is looking perpendicular to that front, the pathlength through any particular parcel of solar wind is quite short, so that variation due to that parcel is small compared to the emission from the entire line of sight. The blue curve in Figure~\ref{fig:helio_emis} is an adequate description of the contribution to the variation as a function of distance. If, however, one observes tangentially to the front, then one sees large variations as successive density peaks move past. One is looking through a long ($\sim$1 au) nearby pathlength of uniformly high (or low) emission, thus increasing the local contribution to the temporal variation. Thus, one expects to see the greatest variation in the SWCX emission to be {\it roughly} along the Parker spiral. Large variation implies that one should see the strongest and weakest heliospheric emission in the direction of the Parker spiral. One difficulty with the simulation used in \citet{kuntz_etal_2015} is that it covered only solar latitudes from $-30\arcdeg$ to $+30\arcdeg$ and gave little indication of what might be happening at higher ecliptic latitudes.

%The ENLIL simulations in \citet{kuntz_etal_2015} were made for solar maximum and show that the greatest variance in SWCX emission appears when looking along the Parker spiral. Those simulations also showed that the correlation between SWCX emission and solar wind flux measured at the Earth was typically low, $<0.2$, but increased to $\sim0.6$ in some regions towards the Parker spiral, as one might expect. \textcolor{blue}{One difficulty with the simulation used in \citet{kuntz_etal_2015} is that it covered only solar latitudes from $-30\arcdeg$ to $+30\arcdeg$.}

%In the ecliptic coordinate system where  $\lambda_{as}$ is the longitude of the anti-sun, the Parker spiral can be seen at $\sim\lambda_{as}-70\arcdeg$ (to the right) and $\sim\lambda_{as}+150\arcdeg$ (to the left). \textcolor{blue}{ Parker spiral does not }

%Those simulations also show that the correlation between SWCX emission and solar wind flux measured at the Earth was typically low, $<0.2$, but increased to $\sim0.6$ in some regions towards the Parker spiral, as one might expect. \textcolor{blue}{One difficulty with the simulation used in \citet{kuntz_etal_2015} is that it covered only solar latitudes from $-30\arcdeg$ to $+30\arcdeg$, and thus did not 

\begin{figure}
\center{\includegraphics[width=13.5cm,angle=0]{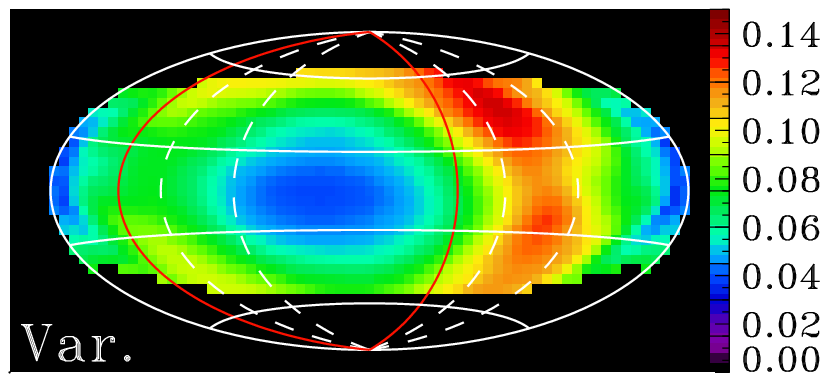}}
\vspace*{-3.0cm}
\center{\includegraphics[width=13.5cm,angle=0]{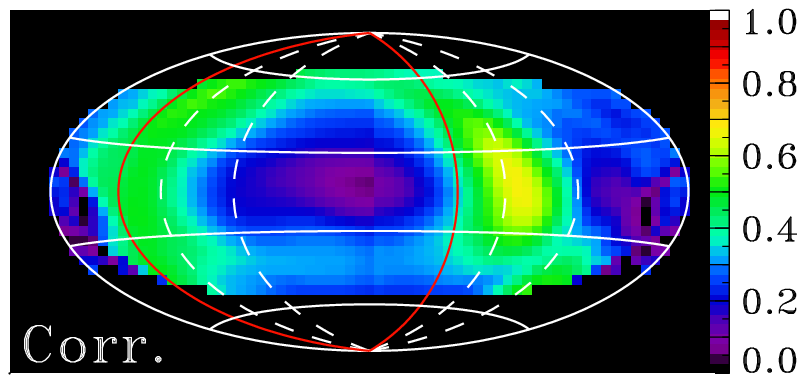}}
\vspace*{-1.5cm}
\caption{The SWCX emission variability (top) and correlation with the locally measured solar wind flux (bottom) as a function of location on the sky. In both cases the map is an Aitoff projection in ecliptic coordinates relative to the anti-sun (at the center), with negative values to the right. The Sun itself is at  the extreme left and right of the plot. Latitude lines are marked at $\pm20\arcdeg$ and $\pm60\arcdeg$. The location of the mean Parker spiral ($43\arcdeg$ from the propagation direction) is shown in red. The region typically observed by \xmm\ (within $20\arcdeg$ of the perpendicular to the Earth-Sun line) is shown by dashed white lines. The values were derived from ENLIL models of solar minimum. The shape of the map is set by the model volume, so the poles had pathlengths too short to be statistically useful. {\it Top: }The variability map is the temporal r.m.s. variation of the SWCX flux on time scales shorter than a week. The color bar runs from 0 to 0.15. {\it Bottom: } The absolute value of the correlation between the SWCX emission and the solar wind flux measured at the Earth.}
\label{fig:enlil_maps}
\end{figure}

The plane defined by the Parker spiral passing through the Earth contains the ecliptic poles, so one might expect the SWCX emission towards the ecliptic poles to have a high temporal variation. However, at least during solar minimum, the equatorial flow of the solar wind extends only $\sim$0.4 au along the line of sight, and the remainder is dominated by the more uniform polar flow, thus reducing the temporal variation. However, it is not clear which effect dominates. For this work, we requested an ENLIL simulation covering the solar rotational plane $\pm60\arcdeg$ (the largest opening angle possible for this model) and extending to $\sim$8 au (also the largest possible distance for this model) during a three-month period of solar minimum. A representative time step is shown in Figure~\ref{fig:enlil}, while maps of the temporal r.m.s. variation and correlation as a function of position are shown in Figure~\ref{fig:enlil_maps}. 

The maps of variation and correlation are constructed from the terrestrial viewpoint 1 au from the Sun and $7.25\arcdeg$ below the solar rotation plane. (The inclination of the solar equator to the ecliptic is $7.25\arcdeg$.) The bulk of the maps is calculated from lines of sight that extend to the edge of the ENLIL simulation $\sim$7 au from the Earth, but at latitudes $>\pm60$ the pathlengths drop rapidly to only 1.5 au. The effects of the helium focussing cone have been removed from these maps.

The map of the correlation between the SWCX emission and the solar wind measured at the Earth shows much of the expected structure. In the ecliptic coordinate system where  $\lambda_{as}$ is the longitude of the anti-sun, the Parker spiral can be seen at $\sim\lambda_{as}-70\arcdeg$ (to the right) and $\sim\lambda_{as}+130\arcdeg$ (to the left). Note that the Parker spiral does not appear at the calculated $-43\arcdeg$. This rotation (from $43\arcdeg$ to $75\arcdeg$) has two sources. First, due to the curvature of the Parker spiral, a line of sight  at negative longitudes intersects the spiral twice at more negative angles, so the emission from the Parker spiral at negative longitudes is smeared $10\arcdeg$ to $20\arcdeg$ to more negative longitudes, while emission from the Parker spiral at positive longitudes is smeared to more positive longitudes. Second, the solar wind velocity in the three months of solar minimum simulated here was lower than during the simulation during solar maximum used by \citet{kuntz_etal_2015}, which produces a $5\arcdeg$ to $15\arcdeg$ rotation from the position seen at solar minimum. Thus the Parker spiral as a whole moves to more negative longitudes for lower solar wind velocities. 
%This latter effect is relatively small; a $\pm40$ km s$^{-1}$ change from the mean produces a rotation of $\sim2.5\arcdeg$. 
In the north, where the Earth is further from the edge of the model, we see that the correlation due to the Parker spiral does indeed extend to higher latitudes.

The overall correlation strength seen here is similar to slightly lower than those seen in the \citet{kuntz_etal_2015} maps; the green regions are correlations of $>0.4$ while yellow are $>0.6$. The lack of a strong extension of the Parker spiral in the south is likely due to the Earth having been $7.25\arcdeg$ below the solar rotation plane in the construction of these maps.

The map of the fractional temporal variation is also similar to that of \citet{kuntz_etal_2015}, though also rotated. It should be noted that over the simulation period the solar wind varied on many time scales. Variations longer than a week were removed since a three month simulation does not sample such scales sufficiently. The lowest variation is perpendicular to the Parker spiral. If one disregards the regions with latitudes greater than $\pm60\arcdeg$ because their pathlengths are so short, a latitudinal variation is still seen. Reference to the right-hand panel of Figure~\ref{fig:enlil} suggests that the variation is due to a smaller number of strong structures at high latitudes.

Overall, the ENLIL simulations here and in \citet{kuntz_etal_2015} show that there is a stronger correlation of the SWCX emission with the local solar wind flux in the direction of the Parker spiral. The location of Parker spiral varies with the solar wind velocity. Thus, at times, the Parker spiral in the ecliptic plane is well away from the region typically observed by X-ray missions (between $70\arcdeg$ and $110\arcdeg$ from the Sun), while at times it will overlap, as shown in Figure~\ref{fig:enlil_maps}. The direction of the Parker spiral is also the direction in which one would expect the strongest temporal variation in the SWCX emission strength. The longitude range from the perpendicular to the Parker spiral to the anti-sun has the lowest temporal variation.

Thus the usefulness of the locally measured solar wind flux as an indication of the likelihood of heliospheric SWCX emission varies over the sky. In most directions the locally measured solar wind flux is poorly correlated with SWCX, as has been noted by \cite{henley_shelton_2010} but for slightly different reasons.

\subsection{Simple Use of a Complex Magnetospheric Model \label{sec:play_ms}}

\begin{figure}
\center{\includegraphics[width=4cm,angle=0]{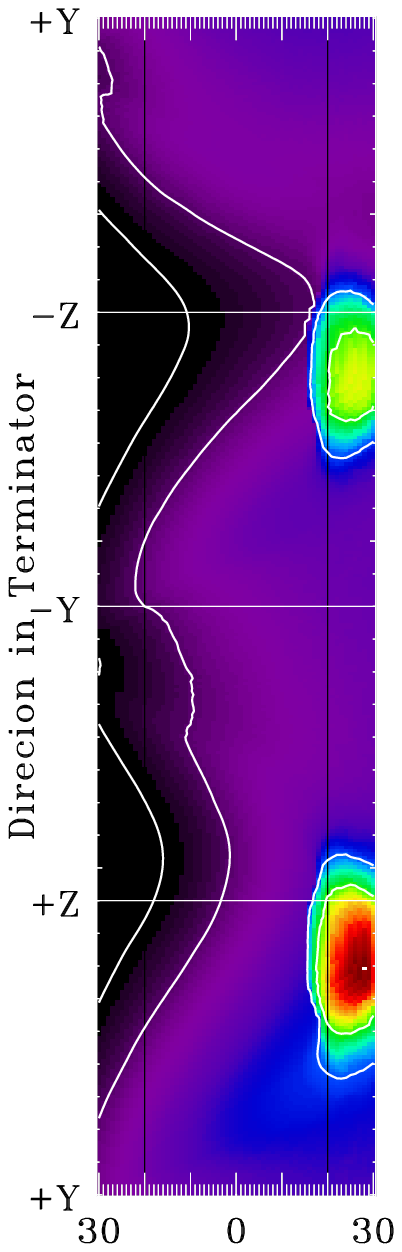}
\includegraphics[width=2.77cm,angle=0]{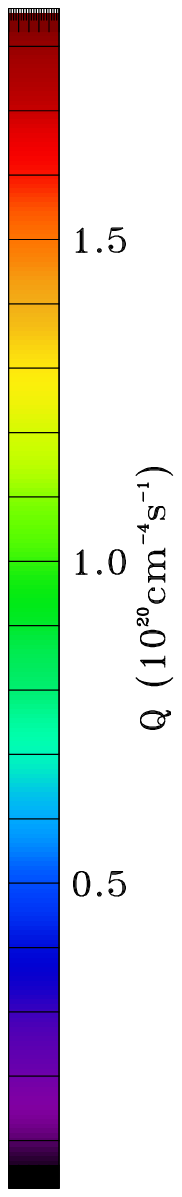}}
\caption{The value of $Q=\int n_p n_n v_{rel} dl$ due to the magnetosphere for lines of sight looking radially outward from the Earth through the flanks of the magnetosheath. The abscissa is the angle from the terminator, while the ordinate notes the direction in GSE coordinates. This simulation was done on the equinox for mean solar wind pressure. The bright polar cusps can be seen to the right.}
\label{fig:flank_band}
\end{figure}

The SWCX emissivity of the flanks of the magnetosheath as seen by a LEO mission (\rosat ) was studied at length in \citet{kuntz_etal_2015}. From a large collection of BATS-R-US simulations of the magnetosheath, they constructed a relation between the mean $Q$ measured through the flanks and the local solar wind flux. They considered only typical solar wind fluxes ($\lesssim6\times10^8$ cm$^{-2}$ s$^{-1}$) where the relation is quasi-linear. For the mean solar wind flux the typical $Q$ through the flanks is $\sim10^{19}$ cm$^{-4}$ s$^{-1}$. This is a bit of a fuzzy number as it depends upon how much one excludes as belonging to the cusps, which are inadequately defined in MHD models. $Q$ increases in the sunward direction. Asymmetries in both the ion distribution and the neutral distribution cause $Q$ to vary from $\sim 3\times10^{18}$ to $\sim5\times10^{19}$ cm$^{-4}$ s$^{-1}$ for the region between $70\arcdeg$ and $110\arcdeg$ of the Sun, excluding the cusps, given the same caveat.

%The cusps clearly show bright SWCX emission \citep{fujimoto_etal_2007}. Quantifying that emission is difficult. The ion density in the cusps is set more by kinetic processes than by MHD, so MHD models do a poor job of simulating both the size of the emitting region and the emission strength. Because the cusp regions are magnetically connected to the magnetopause where magnetic reconnection occurs, they are of great interest to space physicists. {\it CuPID}, a cubesat-scale mission to launch in 2019, will study the X-ray emission in the cusps and characterize the angular size of the cusps and the emission strength in the $\sim\frac{1}{4}$ keV band as a function of solar wind conditions.

The SWCX emissivity as seen by a HEO mission such as \xmm\ is more strongly variable with position. Most observations are still made through the flanks, but the spacecraft can be much further into the tail, where the emission is lower, or closer to the nose, where the emission is higher. For a few orbits per year the \xmm\ apogee is outside the nose of the magnetosheath, so observations made then can be entirely free of magnetospheric emission or can slice through the strong emission of the nose. At a mean solar wind pressure, the $Q$ through the nose (admittedly difficult for \xmm\ to observe) is $1.1\times10^{20}$ cm$^{-4}$ s$^{-1}$, seven times higher than the mean emission through the flank. As the solar wind pressure increases, the nose is pushed closer to the Earth, is more easily observed by \xmm , and the $Q$ increases dramatically. While the flank emission increases roughly linearly with $nv$, the nose emission increases roughly quadratically. Notably, the first SWCX observation with \xmm\ \citep{snowden_etal_2004} occurred when the solar wind pressure was strongly elevated.

\begin{figure}
\center{\includegraphics[width=9cm,angle=0]{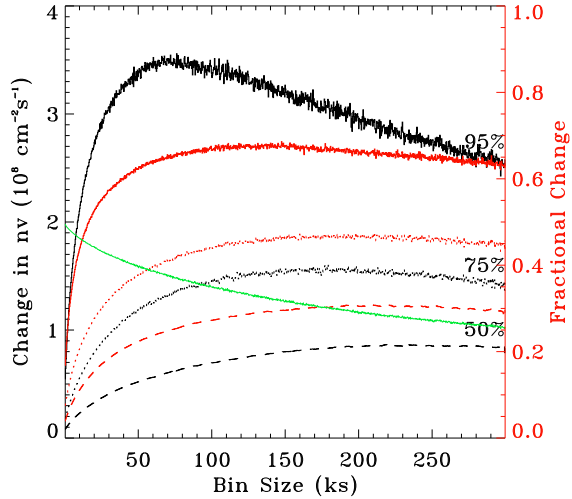}}
\caption{Absolute ({\it black}) and fractional ({\it red}) change of the solar wind flux between successive time bins as a function of the time bin size. The {\it green} line is the RMS variation of the solar wind flux at each time scale. The solar wind data are from the OMNIWeb database at a five minute resolution.}
\label{fig:sw_time_scale}
\end{figure}

The magnetosphere responds rapidly to changes in the solar wind, so the time scales of variation are mostly those of the solar wind. Of more importance to a given observation may be the changing observation geometry, particularly for observations made from HEO. Figure~\ref{fig:sw_time_scale} characterizes the temporal variation in the solar wind in a manner useful for the problem at hand. Shadowing observations  requires two observations (on- and off-cloud) of roughly the same length. Such observations are often scheduled consecutively to reduce the variation in the SWCX emission. The probability of the change in SWCX emission being greater than some $\Delta Q$ for a pair of observations of length $t$ can be calculated. Observing through the flanks of the magnetosheath, where the shape and thickness of the magnetosheath does not change as substantially with solar wind pressure \citep[though see][for a more accurate discussion]{petrinec_russell_1996}, we find that the change in the magnetospheric SWCX emission is proportional to the change in the solar wind flux, $nv$.

We extracted the solar wind proton density and speed from the OMNI database at a five-minute resolution. After masking out bad data intervals (as specified by OMNI), we binned the data by $n$ bins, calculated $\Delta nv$ for each pair of immediately adjacent bins, and accumulated the distributions. Figure~\ref{fig:sw_time_scale} shows the 95th, 75th, and 50th percentile of the distributions as a function of the size of the time bins. 

For a pair of typical X-ray observations, say 50 ks, the solar wind flux will change by $\sim0.5\times10^8$ cm$^{-2}$ s$^{-1}$ at 50th percentile. Perhaps more usefully, the solar wind flux will change by 22\%. Thus, 50\% of paired 50 ks observations would see at least a 22\% change in the magnetospheric SWCX emission. This is a useful figure of merit for evaluating shadowing observations, as well as for determining the variation over segments of longer exposures.

\subsection{Some Consequences}

As discussed above, there is sometimes an ambiguity as to the source of a particular SWCX enhancement; heliospheric {\it versus} magnetospheric. From the simple upwind model, the approximate $Q$ is $9.5\times10^{19}$ cm$^{-4}$ s$^{-1}$. Since the median $\Delta Q$ on month time scales is $<10^{19}$ cm$^{-4}$ s$^{-1}$, the bulk of the heliospheric emission does not vary. In most cases it would be measured as part of the Local Hot Bubble emission rather than as a separate component. The mean magnetospheric emission observed through the flanks is comparable to the variation in the heliospheric emission, so the 22\% variation of the magnetospheric emission is relatively negligible. 

However, most of the noticeable events are more likely to be the 95th percentile events at the relatively short time scale of a day or less. For the heliospheric emission, this is $\Delta Q = 2.5\times10^{18}$ cm$^{-4}$ s$^{-1}$ in the anti-sun direction, probably several times that level along the Parker spiral. For the magnetospheric SWCX, the 95th percentile change in the solar wind strength is 65\% which, for a nominal flank observation, is a $\Delta Q\sim 6.5\times10^{18}$ cm$^{-4}$ s$^{-1}$. Since some fraction of observations through the flank are in or near the Parker spiral, the heliospheric change and the magnetospheric changes are comparable. 

Thus, noticeable SWCX events will often have ambiguous origins. If there is no strong enhancement in the solar wind flux measured at L1, one must check the location of the solar wind monitor and the direction of the solar wind to see if the solar wind striking the magnetosheath is adequately represented by the measurements. If the solar wind monitor is likely to have measured the solar wind that actually strikes the magnetosheath, then the emission is likely to be heliospheric. I am not currently aware of a statistical study characterizing events by their origin, but it would be interesting.

%\clearpage

\section{Comparison of Observations with Models \label{sec:comp}}

\subsection{Heliospheric Emission \label{sec:comp_hs}}

Assuming that the neutral density distribution is relatively well understood, modeling the solar wind velocity, density, abundances, and ionization structure is the primary difficulty. Given the preceding discussion, there are two conditions under which one might expect it to be possible to make an approximate reconstruction: if either the line of sight is entirely within the equatorial flow or if the line of sight passes predominantly through high solar latitudes during solar minimum.

In the first case, with one or more solar wind monitors and interpolation between solar rotations, one can approximate the solar wind along the line of sight. However, the structure of the solar wind evolves as it moves outward, so a simple geometric reconstruction is insufficient. MHD simulations of the solar wind, such as MAS or ENLIL, provide a better approximation. However, the accuracy of these models is not well established (see \S\ref{sec:sw_mod}). Conversely, depending upon the direction, the details of the structure may not matter as much, given that the emission is given by the integral along the line of sight.

\begin{figure}
\center{\includegraphics[width=8cm,angle=0]{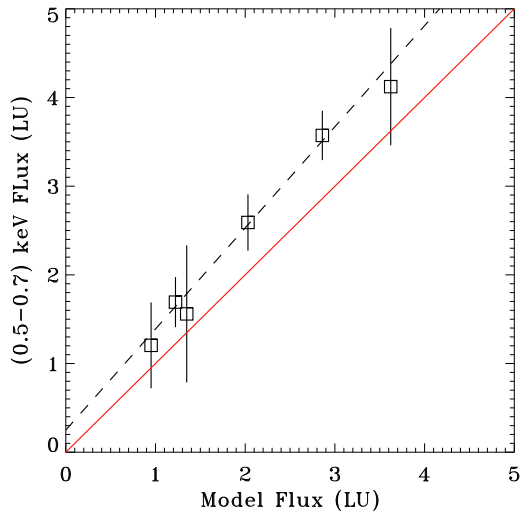}}
\caption{Measured O VII+O VIII flux {\it versus} the modeled SWCX emission in those lines for multiple observations of MBM12, which has an ecliptic latitude of $2.6\arcdeg$. The intercept of the best fit (dashed line) is the emission due to the LHB. The slope of the fit is near unity (red line). The small scatter and near-unity slope suggest that the models accurately reflect reality. This figure has been adapted from \citet{koutroumpa_2012} which, in turn, was adapted from a figure in Galeazzi et al. (in prep).
\label{fig:mbm12}}
\end{figure}

To determine whether a model of the SWCX emission is adequate, one needs to measure the X-ray emission in a single direction over multiple epochs. If the model is correct, a plot of measured flux {\it versus} modeled SWCX flux will be a straight line with an intercept equal to the true cosmic flux. MBM12 lies $\sim2.6\arcdeg$ above the solar equatorial plane and has had six observations in O VII and O VIII between 2000 and 2011. \citet{koutroumpa_2012} showed a preliminary analysis of all of the data to that date using her heliospheric model, a hot Fahr-type model for the neutral distribution and a geometric interpolation of the \ace\ solar wind data. The measured {\it versus} modeled plot show a linear trend with only a small scatter and a slope of nearly unity (Figure~\ref{fig:mbm12}). Thus this relatively simple model is adequate for this line of sight, and perhaps for lines of sight that are totally within the equatorial flow.

A brief filtering of the \xmm\ archive reveals that most pointings that are near the solar rotational plane, have multiple observations over several years, and have observations long enough for adequate signal to noise, are also observations of very bright or extended sources. Thus further testing of the heliospheric SWCX model for the equatorial flow region will require special observations. %\textcolor{red}{[Not for publication: there may actually be a few targets that we could use. I'll look at this more carefully after summer vacation. If there are observations of relatively blank fields with appropriate exposure times and distribution, and they were observed when we have decent \ace\ data, then I'll see if Dimitra is interested. We should be able to do an ADP proposal on this side, and perhaps Dimitra can use it to get funding on her side.]}

If the line of sight passes predominantly through high solar latitudes during solar minimum, one might assume (based on the discussion in \S\ref{sec:sw}) that the high latitude solar wind is uniform and its properties are reasonably well described by the mean abundances. However, this may be contraindicated by the work of Koutroumpa et al. (2019, submitted). 

\begin{figure}
\center{\includegraphics[width=8cm,angle=0]{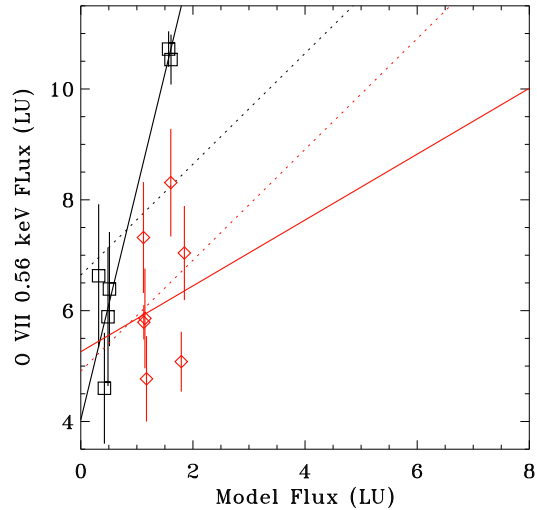}}
\caption{Measured O VII flux {\it versus} the modeled SWCX emission in that line for multiple observations of the North Ecliptic Pole (black boxes) and the HDFN (ecliptic latitude $57\arcdeg$, red diamonds). The intercept of the best fit (solid lines) is the emission due to the LHB and the absorbed Galactic halo. One expects the intercepts to be different as the two lines of sight sample different parts of both the LHB and the Galactic halo. However, one would expect the slopes to be near unity (dshed lines). The strongly divergent slopes and the large scatter suggests that the model of the heliospheric SWCX does not correctly represent the emission in these lines of sight. This figure has been adapted from Koutroumpa et al., (submitted).
\label{fig:nep_hdfn}}
\end{figure}

Koutroumpa and colleagues were searching for the signature of the helium focussing cone by observing the south ecliptic pole (SEP) multiple times as the Earth passed over the focussing cone. In order to track the temporal variation due to changes in the solar wind another series of observations of the north ecliptic pole (or its equivalent) was interleaved. The first such campaign was in 2003 (solar maximum) with \xmm . A second campaign was executed in 2009 (solar minimum) with Suzaku. During the solar maximum campaign, one expects to be observing through equatorial flow (perhaps disrupted by fast flow from coronal holes) and the bulk of the line-of-sight will be through high solar latitudes, containing a mixture of fast and slow solar winds. During the solar minimum campaign, one also expects to observe through the equatorial flow, but the bulk of the line-of-sight will be filled with fast solar wind with a significantly lower X-ray emissivity. The solar minimum campaign should have seen a factor of 2-3 lower emission than the solar maximum campaign. Instead, the signal strengths were comparable; something seems to be seriously amiss with the model. Galeazzi (private communication) found a similar problem in another data set. The issue is summarized by the comparison of \xmm\ observations of the Hubble Deep Field North (2001 and 2003) and \suzaku\ observations of the North Ecliptic Pole (2005, 2006, and 2009) shown in Figure~\ref{fig:nep_hdfn}.

This lack of correlation at high solar latitude may have been anticipated by \citet{yoshitake_etal_2013} who noted that the oxygen line fluxes towards the Lockman Hole did not correlate with the solar cycle. More recently, \citet{henley_shelton_2015} looked at the foreground 0.4-1.0 keV surface brightness towards several targets obtained from shadowing observations and compared those values to the sunspot number. The three lines of sight passing only through low ecliptic latitudes show a clean correlation (as much as three points can) while the four lines of sight (six observations) passing through the polar flow show a large scatter. These results correlate well with those of Koutroumpa (2012, 2019) and Galeazzi (in progress), but there is significant overlap in observations. Small-number statistics remain a major impediment to understanding the heliospheric emission from high solar latitudes.

Thus it is clear that trying to model the heliospheric SWCX emission at high ecliptic latitudes is still fraught with uncertainties. 

\subsection{Magnetosphere \label{sec:comp_ms}}

The discussion in \S\ref{sec:ms_mod} highlighted the difficulties with magnetospheric models; the magnetosheath/magnetosphere is a very complex system with multiple plasmas, multiple current systems, and some important non-MHD physics that must be coupled to the MHD models. The uncertainties are well exemplified by the uncertainty in the model predictions of the magnetopause stand-off distance. The uncertainty in the magnetopause stand-off distance was identified as a prime source of error in the modeling by \citet{wargelin_etal_2014} in his comparison of MHD models (BATS-R-US) with a dozen \chandra\ observations. 

\citet{whittaker_etal_2016} reached similar conclusions in their comparison of MHD models (Gumics) with a set of 19 \xmm\ observations with strong SWCX events. However, they pointed more strongly to the input abundance data. They noted that using the \ace\ measured values of O/H and O$^{+7}$/O produced worse correlations between models and data than did the use of mean values. It is not clear how to start tackling this problem. Both of these sets of comparisons found some events that were well modeled and others that were not. Similarly, direct comparison of different MHD codes find some events that are relatively consistently modeled across the different codes, and many events that are not. 

On a happier note, one test of the magnetospheric models has yielded positive results. \citet{snowden_etal_2009} carefully crafted an observation of an X-ray dark region of the cosmic X-ray background whose line of sight passed through the nose of the magnetosheath. Over the course of an orbit, the motion of \xmm\ caused the line of sight to pass through different parts of the magnetosheath. The results were consistent with the models at the time. Unfortunately, the solar wind was particularly steady and particularly low over the entire observation, so the dynamic range of the emission was lower than expected. An attempt to repeat the experiment in 2013 failed due to soft proton flares. A third attempt is scheduled for early 2019.

\section{Observational/Mitigation Strategies \label{sec:miti}}

From the above, it is clear that modeling the SWCX emission along a particular line of sight is still very difficult and still subject to severe uncertainties. However, despite all the known and unknown problems with the models, we can still use them to construct strategies for mitigating the effects of SWCX emission on an observation or for evaluating the incidence of SWCX emission in a sample of observations. There are multiple types of mitigation: identifying or evaluating the probability that an observation has strong SWCX emission, compensating for the SWCX emission, or minimizing the strength and/or variation in the SWCX emission.

\subsection{Locations}

The discussion in \S\ref{sec:play_enlil} and \S\ref{sec:play_ms} made it clear that the amount and variability of SWCX emission depends upon where one observes. Observations along the Parker spiral, particularly downwind ($\lambda_{as}-43\arcdeg$) near the ecliptic, are subject to large variation on the shortest time scales. Unfortunately, the angle of the Parker spiral with respect to the anti-sun is not constant. It is typically $\sim43\arcdeg$ but, as shown by the ENLIL model in \S\ref{sec:play_enlil}, it can move to large angles and into the region where most X-ray observatories make their observations. 

The extent to which the Parker spiral affects observations at the ecliptic poles is still not clear. Lack of solar data at high solar latitudes restricts the use of MHD models such as ENLIL for the polar regions. However, from the ENLIL simulations and the \ulys\ results, one expects the solar wind to be rather uniform for $r>1$ au in the direction of the poles. There will still be some variation due to the restricted path length through the Parker spiral, so the ecliptic poles are not ``SWCX quiet'' the way the anti-sun is. Indeed, the ecliptic pole analyses by \citet{koutroumpa_2012} and \citet{koutroumpa_etal_2019} demonstrate that the SWCX emission towards the ecliptic poles is quite difficult to model.

Despite the issues of the magnetopause standoff distance, the magnetospheric component is better characterized. LEO missions observe through the flanks of the magnetosheath where the SWCX emission is not strong, but is ubiquitous. Occasionally, observations pass through the bright cusp and experience strong SWCX emission. Although such observations are greatly appreciated by those who study SWCX emission, such observations can usually be avoided by specifying rather weak time constraints. HEO missions generally observe through the flanks, but can sometimes observe through the bright nose of the magnetosheath. Such observations are precious for those who study the magnetosheath but are rare. Scheduling an observation to observe through the nose of the magnetosheath has been surprisingly difficult. Again, such observations can be avoided, if necessary, by specifying rather weak time constraints. 

Of course, most X-ray missions have tight constraints on the allowable Sun angles, so all one can do is attempt to avoid the most problematic directions. New missions focused on diffuse emission, however, can create new strategies.

\subsubsection{The Anti-Sun}

Observations down the magnetotail in the direction of the anti-sun avoid looking through the magnetosheath. Since the anti-sun is also ``SWCX quiet'' for the heliospheric component, such observations are particularly desirable. Observing anywhere within $\sim70\arcdeg$ of the anti-sun direction avoids the bulk of the magnetospheric emission. Observing down the magnetotail is the strategy adopted by the HaloSat mission, which aims to measure O VII and O VIII emission over the entire sky at an angular resolution of $10\arcdeg$.  

One should be aware that the magnetotail experiences ``wind-sock effects'' due to variation in the solar wind velocity direction, ``breathing'' due to substorms, and ``wrenching'' due to changes in the IMF. Thus, observations towards the anti-sun are not guaranteed to be free of magnetospheric SWCX, and the further from the anti-sun the more likely the observation is to experience variation in the SWCX emission. 

There is another caveat to observing in the anti-sun direction; the plasma sheet, which lies horizontally between the north and south lobes of the magnetotail, may contain some O$^{+7}$. The plasma sheet contains O$^{+6}$ whose origin is thought to be the solar wind entering the magnetosheath through any number of mechanisms \citep[See][]{allen_etal_2016a}. The density of O$^{+6}$ in the near magnetotail ($L<20$) is on the order or less than 0.01 cm$^{-3}$ \citep{allen_etal_2017}. The densities at greater $L$ are likely lower \citep{allen_etal_2016b}. Since, in the solar wind, O$^{+7}$/O$^{+6}\sim0.2$ the O$^{+7}$ densities will be $\sim0.002$ cm$^{-3}$. This calculation ignores the charge exchange that will have occurred as the ion enters the magnetosheath, so O$^{+7}$ densities are likely to be even lower. Assuming the Hodges neutral density model and the above density of O$^{+7}$ for $10<L<40$, we find that $Q\lesssim5\times10^{16}$ cm$^{-4}$ s$^{-1}$, which is much smaller than the $Q$ measured through the flanks. Thus, while this value should be verified, SWCX due to the plasma sheet is unlikely to be a significant problem for observations towards the anti-sun. 

\subsection{Time Scales}

The solar wind varies on many time scales, from the rotational period of the Sun, to the size of an active region, to turbulence. The magnetosphere responds rapidly to these changes in the solar wind, so the time scales of variation are mostly those of the solar wind. Of more importance may be the changing observation geometry. The relative importance of those time scales to the heliospheric emission is then modified by the $r^{-2}$ decrease of solar wind density with distance from the Sun, the integration along the line of sight, and the angle of the look direction with respect to the Parker spiral. 

For comparison to X-ray observations, a useful characterization of the time scale can be made by determining the typical change in solar wind flux ($n_pv_p$) between two time periods of length $t$, i.e., the change between two successive observations of length $t$. This quantity is shown in Figure~\ref{fig:delta_q}. It should be kept in mind that that plot is not calculated for the Parker spiral LOS, where we would expect stronger variation. Further, it was calculated for the 95th percentile; values for the 99th percentile are higher by factors of 2-3 for scales less than 10 days.

\subsection{Shadowing Observations}

Shadowing observations are particularly difficult in the current era of small-FOV instruments. Shadowing observations typically require an observation on the densest part of the shadowing cloud and one (or more) observations of nearby regions with minimal absorption. Generally the ``on'' (shadowed) and ``off'' (low absorbing column) regions are close together on the sky so that there will be only a small change in the background and foreground emission between the two. This requirement means that one does not generally have to worry about a change in SWCX emission with location, only in time.

The problem with making a single on-cloud observation and a single off-cloud observation, even if the observations are taken sequentially is that the total observation time, usually more than half a day, is comparable to or longer than the time in which one expects the solar wind (as measured at the Earth) to change significantly. Even for a good observing geometry for the heliospheric emission, the magnetospheric emission will change. (And, as a reminder, we note that not all changes in the solar wind at the Earth are captured by upstream monitors.) Thus, shadowing experiments made with only a single pair of observations have an intrinsic uncertainty.
 
There has been at least one observational program where the on-cloud and off-cloud observations were done close together in time, and the pair of observations was repeated at intervals of several months (Galeazzi, in prep). Such a program is difficult to propose as it typically requires multi-cycle approval. Conversely, one learns about long-term variation in the heliospheric SWCX emission with such a program.

Another technique is to make many short observations on-cloud and many short observations off-cloud, and interleave those sets of observations. Scheduling many short observations is generally inefficient. However, \xmm\ has a ``mosaicking'' mode that allows one to make many short sequential exposures with small slews in an efficient manner. This mode was designed for making shallow mosaics of large regions of the sky, and it has not previously been used to cover the same part of the sky multiple times. The \xmm\ AO-17 contains a program to demonstrate the ability of using the mosaicking mode to make shadowing observations.

A number of proposals have been made for X-ray imagers with FOV having diameters of $\sim1\arcdeg$. With such an imager, many of the most interesting shadowing clouds will be observable. Since both on-cloud and off-cloud are made simultaneously, the temporal variability would no longer be an issue for determining the background spectrum.

\subsection{Identifying SWCX Enhancement in a Particular Observation}

The most convincing demonstration of SWCX emission in an observation is evidence of temporal variation. \citet{carter_etal_2011} implemented the most efficient method for finding SWCX variation; for each time period they plotted the 0.5-0.7 keV band against the 2.5-5.0 keV band. Time periods experiencing soft proton flares would show increases in both bands while time periods experiencing SWCX emission would show an increase in only the soft band. Of course, many observations are too short for such a method to be useful.

The next best method involves solar wind data and MHD models. For LEO missions one should check for periods in which one observes through the cusp, while for HEO missions one should check for periods in which one observes through or near the nose of the magnetosheath. These periods merit further investigation with MHD models, with all of the caveats previously discussed. For lines of sight passing only through the flanks, the solar wind data (as far as it can be trusted to represent the solar wind impinging on the Earth) should indicate whether SWCX enhancements are likely. The efficacy of ENLIL type models to determine the probability of heliospheric emission is not clear. However, use of the solar wind data to determine the location of the Parker spiral with respect to the line of sight and the relative timing of any possible solar wind enhancement should be sufficient to determine the likelihood of a SWCX enhancement.

 %\textcolor{red}{We should probably build a simple tool to do this.}

\subsection{Spectroscopic Methods}

\begin{figure*}
\center{\includegraphics[width=12cm,angle=0]{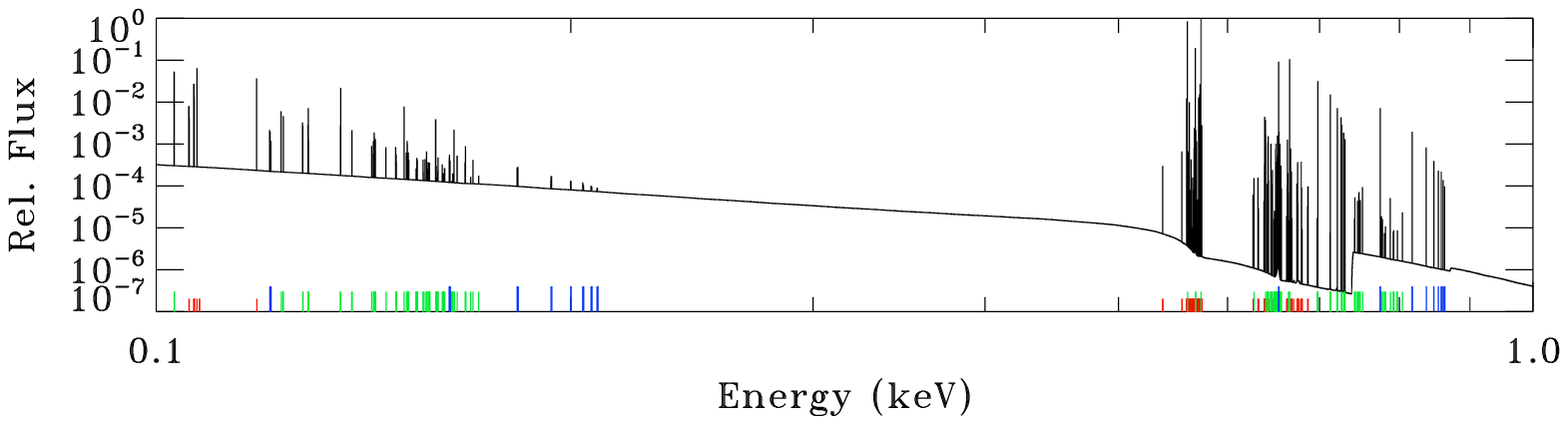}}
\vspace*{-1.5cm}
\center{\includegraphics[width=12cm,angle=0]{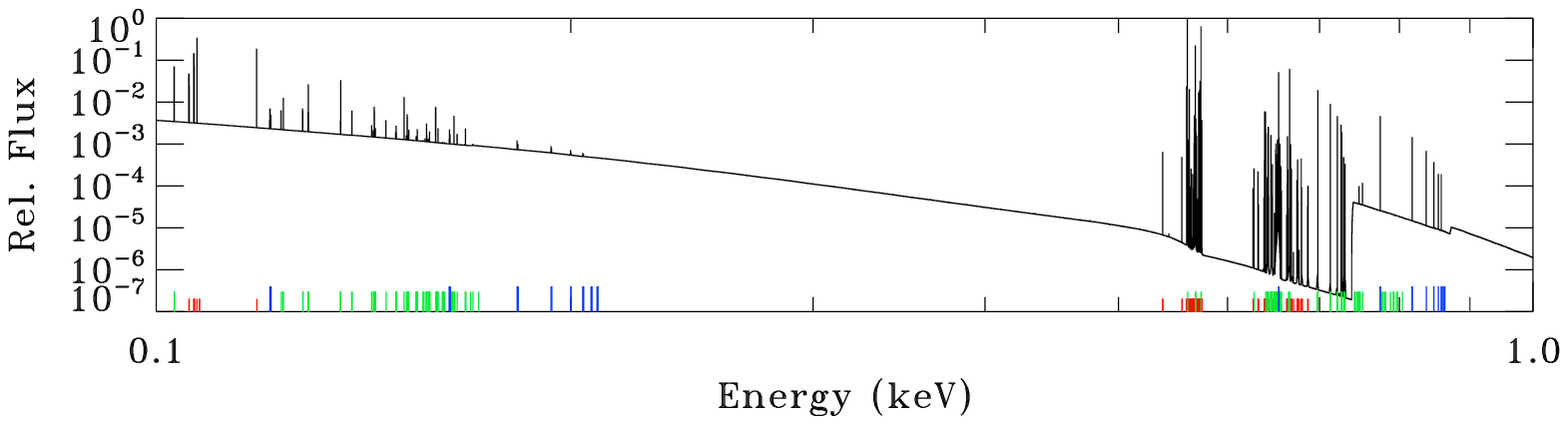}}
\vspace*{-1.5cm}
\center{\includegraphics[width=12cm,angle=0]{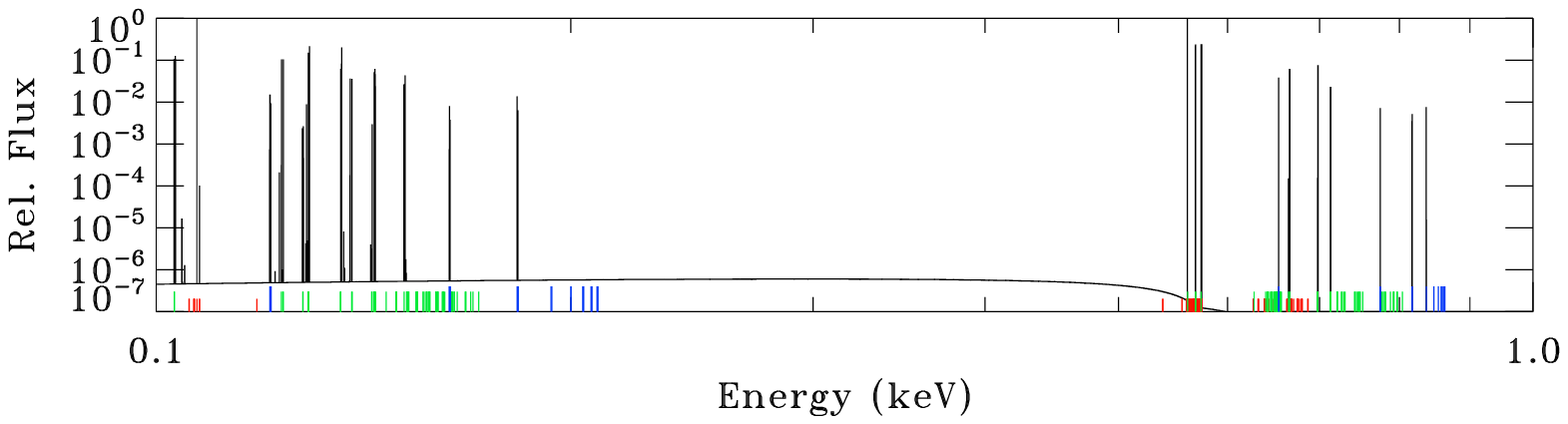}}
\vspace*{-1.5cm}
\center{\includegraphics[width=12cm,angle=0]{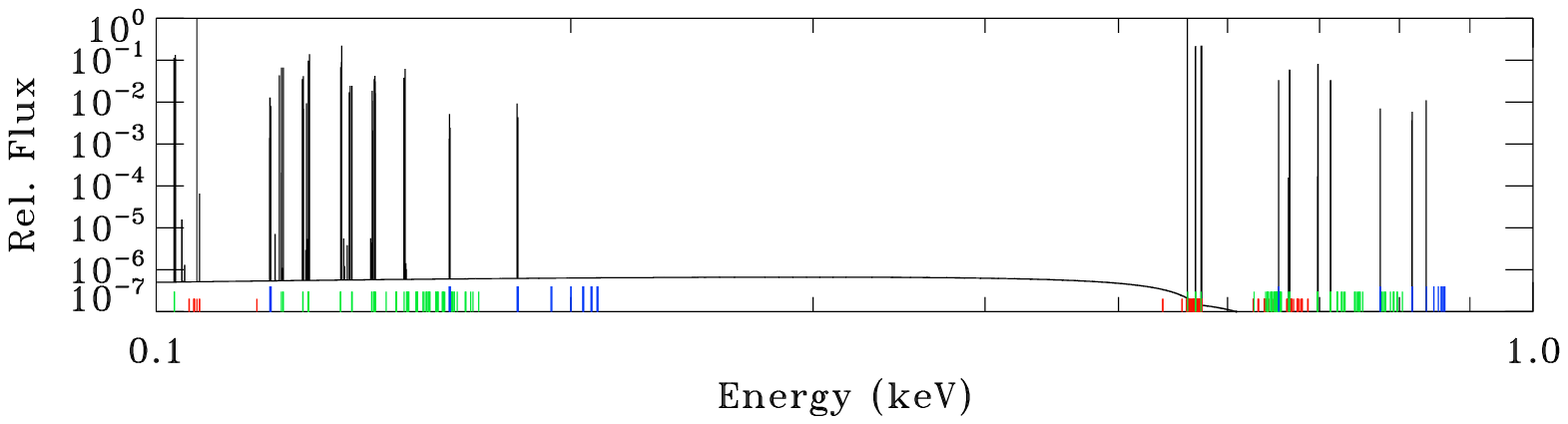}}
\vspace*{-0cm}
\caption{{\it Uppermost: } Spectrum of oxygen in CIE. The abundances of all other elements was set to zero for the vvapec model in xspec. The locations of O VI ({\it red}), O VII ({\it green}), and O VIII ({\it blue}) lines are show by colored ticks. The $kT=0.136$.
%{\it Upper Middle: } Spectrum of oxygen in an over-ionized recombining plasma calculated with vvrnei. The initial $kT=1$, the final $kT=0.136$, and the ionization parameter $\tau=10^8$. 
{\it Upper Middle: } Spectrum of oxygen in an over-ionized recombining plasma calculated with the vvrnei model in xspec. The initial $kT=0.136$, the final $kT=0.08$ (the lowest possible in the model), and the ionization parameter $\tau=10^8$. 
{\it Lower Middle: } Spectrum of oxygen experiencing charge exchange. The freeze in temperature was taken to $kT=0.136$. The spectrum was calculated with the vacx model in xspec using the Landau-Zener distribution.
 {\it Lowest: } Spectrum of oxygen experiencing charge exchange. The freeze-in temperature was taken to $kT=0.136$. The spectrum was calculated with the vacx model in xspec using the ``separable'' distribution. All spectra have been normalized so that the O VIII/O VII 0.561 line has a strength of unity.}
\label{fig:reco_vs_cx}
\end{figure*}

Given the uncertainties in modeling the solar wind over the line of sight, constructing the SWCX spectrum for a given observation is a daunting task. Thus, attention has been turning to understanding the spectral diagnostics that might allow one to fit both thermal and charge-exchange components in order to understand their relative contributions. Of course, given that charge-exchange emission is now expected from many astrophysical objects, such analysis could become very difficult indeed.

It should be kept in mind that charge exchange is, from the point of view of the ion, a recombination without the strong likelihood of a collisional excitation during the electron's journey to its ground state. As a result, charge exchange shares many spectral features with over-ionized recombining plasmas. Thus, for example, the $(x+y+z)/w$ ratio of the O VII triplet is not a signature of charge exchange, but only the signature of recombination in the absence of significant collisional excitation.

That having been said, if one has sufficient resolution and sufficient sensitivity one could distinguish between an over-ionized recombining plasma and charge exchange. For example, the charge-exchange spectrum of a hydrogen-like\footnote{We use ``hydrogen-like'' rather than ``hydrogenic'' to maintain the parallel with ``helium-like'', etc. which have no similar form. It should be noted that ``hydrogenic'', when read by a chemist, would mean ``producing water''.} ion will include series members up to some limiting $n$. The equivalent ``purely recombining'' spectrum will show lines up to the series limit as well as emission due to transitions from the continuum to a bound level. Of course, for a recombination spectrum, the lines grow progressively weaker as $n$ increases, so the higher-$n$ lines will be quite faint. Conversely, in a charge-exchange spectrum, the series line corresponding to $n_i$ will be significantly stronger than those of $n_i-1$ or $n_i-2$, so charge exchange may be positively identified if the series is measured to either $n_i$ or a few steps higher.

Figure~\ref{fig:reco_vs_cx} is intended to show the difference between a CIE spectrum of oxygen, a recombining plasma containing only oxygen, and a charge exchange spectrum for oxygen. The ionization parameter for recombining plasma was made as short as possible in order to accentuate the differences with the CIE spectrum. The charge exchange spectrum was created with the acx\footnote{http://www.atomdb.org/CX/acx\_manual.pdf} model for a freeze-in temperature of $kT=0.136$. The acx model in xspec was run in the ``swcx'' limit, where each ion charge exchanges only once, rather than multiple times to reach the neutral state. One can see that the upper level lines are disproportionally underpopulated. The acx model results are shown for two different $\ell$-state weighting schemes to provide an indication of the uncertainty involved.

This demonstration shows only the emission from oxygen at very high resolution and infinite signal-to-noise, so the differences are apparent. Clearly, if there were contributions from two or more such emission components, even with known metallicities, teasing apart the components could be very difficult even at microcalorimeter resolution.

%\clearpage

\section{Future Efforts\label{sec:future}}

\subsection{Models} 
As noted above, the neutral density within the heliosphere is an active topic of study, and the current hot-Fahr based models do a reasonable job of replicating observations. Models of the exosphere may not be so well tested but can be, in principle, using small Ly$\alpha$ detectors. 

There is active interest in the heliospheric and magnetospheric communities to validate their MHD models. Models of the magnetosphere are particularly important as they are the basis of health and safety plans for satellites. The bulk of the emphasis is on the intensity of the energetic particles that can damage spacecraft. Compressions of the magnetopause can lead to a transient increase in energetic particle intensities due to adiabatic energization followed by loss to the magnetosheath as particles drift towards encounters with the compressed magnetopause. A major solar storm can move the magnetopause within the orbit of geosynchronous satellites, so it is imperative to get these models correct. Significant progress towards validating the magnetospheric models can be expected in the near future. The heliospheric models are less certain, but as they have less application, there is less urgency for their improvement. 

\subsection{Data} 

Solar wind data from L1 monitors are essential for spacecraft operations, so information on the solar wind proton density, velocity, and temperature is assured for the foreseeable future. Unfortunately, there would appear to be no pressing urgency in the solar wind community for the type of abundance and ionization information formerly available from \ace . Nor are there any serious plans for \ulys\ type monitors of the solar wind at high solar latitudes. Both of these would be essential for modeling the SWCX emission.

Laboratory astrophysics continues to produce measurements of charge-exchange cross sections, albeit not as quickly as one might like. However, it appears that techniques are improving and data more directly applicable to SWCX emission will start becoming available in the near future. 

X-ray observations continue to accumulate in the archives. Although the soft response of \chandra\ is no longer sufficient for long term time studies, the \xmm\ soft response remains strong. \citet{yoshino_etal_2009} attempted to measure the difference in mean heliospheric SWCX emission by comparing \suzaku\ observations at solar minimum with the RASS observations at solar maximum, to some success. They found an offset of $(4.1\pm\sim13)\times10^{-6}$ count s$^{-1}$ arcmin$^{-2}$, but here the uncertainty is statistical and does not include the systematic uncertainty of comparing two missions. Since the \xmm\ mission has spanned $\sim$1.5 solar cycles so far, a similar study might be possible using only \xmm , removing much of the systematic uncertainty. Repeated observation of carefully selected archival targets could address both the amplitude of the solar minimum/maximum variation and, through models, perhaps the absolute value of the heliospheric emission. Of course, the {\it caveats} of \S\ref{sec:comp} should be kept in mind.

\subsection{New Missions} 

The magnetospheric SWCX is amenable to observation, and a suite of approved, proposed, or planned missions will make those observations.

 {\it HaloSat} is a 6U cubesat launched in May 2018. It carries three silicon drift detectors sensitive to the 0.23-8.0 keV band. Its purpose is to map the entire sky in O VII and O VIII at a resolution of $\sim10\arcdeg$ \citep{kaaret_2017}. Its SWCX mitigation strategy is to observe within $70\arcdeg$ of the anti-sun whenever possible in order to avoid the magnetospheric SWCX and to minimize the variation in the heliospheric SWCX. Its calibration program will observe the helium focussing cone and make comparative observations of test fields both through the flanks and down the magnetotail in order to test the emissivity in the flanks. 

{\it CuPID}, the {\it Cusp Plasma Imaging Detector} is another 6U cubesat to be launched into LEO in late 2019. It will carry an X-ray sensitive MCP behind a slumped micropore (lobster eye) optic. Its purpose is to measure the size and shape of the cusps as a function of solar wind conditions. It is an outward looking mission which, in its most simple mode, will act as a single element photometer. Although this mission does not have spectroscopic capability, it will parameterize the $\frac{1}{4}$ keV emission quite well.

{\it SMILE}, the {\it Solar wind Magnetosphere Ionosphere Link Explorer} \citep{sembay_sxi_2016} is an approved S-class joint ESA-CAS mission to observe the cusps and the nose of the magnetosheath. The Soft X-ray Imager (SXI) uses CCDs behind an array of slumped micropore optics to achieve  a $15.6\arcdeg\times 27.3\arcdeg$ FOV. By imaging the nose of the magnetosheath, this mission will measure the magnetopause stand-off as a function of solar wind flux. So long as \ace\ continues to function, it will be able correlate the strengths of the oxygen lines with oxygen ion abundances in the solar wind. {\it SMILE} is scheduled to launch in late 2023.

{\it STORM}, the {\it Solar Terrestrial Observer of Reconnection in the Magnetosphere} \citep{sibeck_ssr_2018} is a proposed MIDEX class mission to image the magnetosheath in the $\frac{1}{4}$ keV band. Like {\it SMILE}, its X-Ray Imager (XRI) will use an array of slumped micropore optics to achieve a FOV of $23\arcdeg\times23\arcdeg$. The softer energy band, a longer focal length than {\it SMILE}, and use of MCPs rather than CCDs will allow a factor of 2-7 increase in sensitivity, depending upon the as yet undetermined details of the instrument designs. The requirement for the  (XRI) is to measure the location of the magnetopause to within 0.125 R$_E$ in a 2 minute exposure in order to study features of reconnection events. This resolution will allow unprecedented testing of the MHD models. {\it STORM} will carry a number of other instruments to measure directly the solar wind (light ions only) and the IMF that actually impinges upon the Earth. Although it is primarily a magnetospheric mission, {\it STORM} will provide much of the input data required for robust magnetospheric SWCX models for use in astrophysics.

%As has been seen above, MHD models are not particularly good for predicting the SWCX fluxes for \xmm\ observations due to the observing geometry and the current uncertainty in the location of the magnetopause. Perhaps a more useful study would be a comparison of \suzaku\ observations through the flanks of the magnetosheath, where the emission strength should be far less sensitive to small errors in the MHD models. Given the strength of the O VII and O VI lines, and continued(?) measurement of O$^{+7}$ and O$^{+8}$ by \ace , such a study could help resolve some of the issues surrounding the production factors for those species. 

\subsection{Old Missions with New Capabilities} 

Due to its large grasp, \xmm\ has made many contributions to the study of SWCX emission. With the loss of \suzaku\ it is the only instrument with sufficient soft response to continue these studies. The \xmm\ GOF at GSFC is working diligently on a substantial improvement to the background subtraction below 1 keV. The improvements will affect both the MOS and, to an even greater degree, the pn. The improvements are expected to reduce the discrepancies between the MOS and the pn in just the region where the SWCX spectrum is important. %These improvements will allow new opportunities for archival studies.

\section{Summary}

\rosat\ revealed the existence of solar wind charge exchange, but too late in the mission to allow further study with \rosat . It was not clear to what extent the emission from charge exchange affected individual observations. Simple data reduction methods, given {\it ROSAT's} wide FOV, also minimized the problem in many instances.  Attempts to study the hot ISM in the \xmm\ era stumbled across SWCX in ways that simply could not be ignored. Shortly after the realization that \xmm\  and \chandra\ data could be seriously compromised by SWCX emission, there were a number of proposals to characterize the SWCX emission and to create software tools that would calculate the amount of SWCX expected for a given observation. These projects laid the groundwork for understanding the emission, but found that the problem was too complex for the proposed solutions.

From the foregoing it should be clear that the two different regimes of charge exchange emission have different strengths but a similar amplitude of variation for typical viewing geometries. Thus, with a few exceptions, both regimes must be well characterized in order to remove the SWCX contributions to astrophysical observations. The models are not yet adequate.

For the heliospheric emission the current obstacle is our lack of solar wind data, particularly at high solar latitudes. We will not be getting more data from high solar latitudes any time soon, so it is unclear how we will resolve the problems found by Koutroumpa, Galeazzi, and others. Within the equatorial flow, MHD models of the solar wind are useful in a general way to diagnose the directions in which the emission is more variable, but are not yet capable of timing the enhancements to the degree of accuracy needed. The magnetospheric models are currently more refined. The solar wind data have been adequate for many needs, but the continued aging of \ace\ means that we may soon lack any measurement of the abundance of the minor species that produce the SWCX emission. Even with a fully functional \ace , many of the species that are important to X-ray astronomy were poorly measured. The MHD models are relatively good. It is clear that the models are not consistently reproducing the magnetopause stand-off distance. However, this issue is not so important for the flanks of the magnetosheath, through which the bulk of observations are made. Observations near the cusps or through the nose of the magnetosheath are still problematic.

Modeling both SWCX regimes is hampered by missing interaction cross section data. Some astrophysically important lines, such as O VIII 0.56 keV and O VIII 0.65 keV, have cross sections in the literature, but many other important lines do not. Those cross sections that are in the literature often use empirical formulae to calculate the $n,\ell$ distribution, and experience demonstrates that such formulae can fail spectacularly. 

It is clear that, in the near future, there will not be a general all-purpose tool to calculate SWCX for a given observation. However some problems have been at least partially solved by use of archival data and special observations. \citet{galeazzi_etal_2014} has partially solved the problem of the Local Hot Bubble by determining the band-averaged cross section for SWCX with helium from sounding rocket observations of the helium focussing cone. Other problems, such as shadowing observations, can be accomplished with either wider FOV observations (a hint for the next generation of X-ray instruments) or special observation strategies with the narrow FOV instruments on hand.

Diffuse emission filling the field of view, in the absence of the strongly variable foreground absorption required for shadowing observations, remains a particularly difficult problem. There is some evidence that the SWCX can be modeled within the equatorial flow, within the extended set of caveats discussed in \S\S\ref{sec:comp} and \ref{sec:miti}. However, as has become the common refrain, we do not yet understand the issues of observing through the higher solar latitude solar wind.

Finally, rather than ending on such an uncertain note, we can look to the space physics community to assist in understanding SWCX emission, both in the magnetosheath and in the heliosphere. Many of the issues that are important for modeling the SWCX emission are of interest, though for other reasons, to them. However, continued strong astrophysical involvement in projects such as described in \S\ref{sec:future} will be essential to ensure a productive flow of information and data between fields.

\begin{acknowledgement}
This work would not have been possible without a large number of individuals. When Steve Snowden and I were working on the \xmm\ background, we ran into our first undeniable case of SWCX. In search of illumination, Steve made the dangerous trek across the street, from GSFC building 2 (now defunct) to building 21, and started knocking on doors until he found someone who actually thought that the problem was interesting, Michael Collier. Since then the SWCX family at Goddard has grown to include Scott Porter, David Sibeck, Yari Collado-Vega, Hyunju Connor, and Brian Walsh, all of whom have been pestered by me with naive questions about laboratory astrophysics, space physics and the solar wind for nearly a decade. Also suffering my naive questions have been Robert Allen, Tom Cravens, Renata Cumbee, Adam Foster, Maurice Leutenegger, Ina Robertson, and Randall Smith. Nick Thomas and Lynne Valencic, not being space physicists, have provided the appropriate astrophysics based feedback. The Leicester group, Steve Sembay, Andy Read, Jenny Carter, and Ian Whittaker have been working hard on the magnetospheric emission seen by \xmm\ and have provided invaluable insights as well as great companionship. Massimiliano Galeazzi and the Miami group, whose work is rather underrepresented here, has provided a great deal of information about works in progress. Our late night discussions at Poker Flat helped structure this paper. Renzo Principe red-penned numerous drafts, correcting all the mangetospheres and disambiguating the acronyms.

A very special thanks go to Dimitra Koutroumpa, an integral part of the SWCX family, who is the expert {\it sine qua non} on heliospheric SWCX, whose patience with me has been limitless. Numerous plots have been possible only with the loan of her models.

The Coordinated Community Modeling Center (the CCMC) has provided a large number of simulations for this work and many of the works cited. Their help has been indefatigable. Leila Mays helped set up the special ENLIL simulation used here. Lutz Rastaetter provides great support for the BATS-R-US runs, helping us to set them up properly and to understand the results when they do not meet our expectations. 

My thanks also to OMNI (https://omniweb.gsfc.nasa.gov/) for providing a uniform set of solar wind data, as well as the \ace\ Science Center (http://www.srl.caltech.edu/ACE/ASC/) and the \ulys\ Final Archive (http://ufa.esac.esa.int/ufa/), for providing the solar wind abundance and ionization data required for this work.

I would also thank the several institutions that have hosted me during the work on this article, the Science History Institute (Philadelphia) and the International Space Science Institute (Bern). The first version of this review was a talk constructed for the Chandra X-ray Center and the Smithsonian Astrophysical Observatory in honor of Steve Murray.

Finally, I must thank Joel Bregman for his patience, and Kim Weaver and the \xmm\ GOF for support and forebearance. And thank you, Steve, without you I might not have found this problem.
\end{acknowledgement}
%\clearpage

\bibliographystyle{aps-nameyear} 
\bibliography{apjmnemonic,ms}

\end{document}